\documentclass{vki}
\usepackage{lmodern,pdftexcmds,amsmath}

\usepackage{amsfonts,amsthm,bm} 

\DeclareMathAlphabet{\mathsfbi}{OT1}{\sfdefault}{bx}{sl}
\DeclareMathVersion{sfletters}
\SetSymbolFont{letters}{sfletters}{OML}{ntxsfmi}{b}{it}

\makeatletter
\newcommand{\mathbfsbilow}[1]{%
	\text{\mathversion{sfletters}$\m@th#1$}%
}

\DeclareRobustCommand{\tensor}[1]{%
	\begingroup
	\ifcat\noexpand #1\relax
	\edef\greek@test{\detokenize{#1}}%
	\edef\greek@test{\expandafter\@cdr\greek@test\@nil}%
	\edef\greek@test{\expandafter\@car\greek@test\@nil}%
	\edef\x{\the\lccode\expandafter`\greek@test}%
	\edef\y{\number\expandafter`\greek@test}%
	\ifnum\x=\y\relax
	\mathbfsbilow{#1}%
	\else
	\mathsfbi{#1}%
	\fi
	\else
	\mathsfbi{#1}%
	\fi
	\endgroup
}
\makeatother

\usepackage[most]{tcolorbox}
\usepackage{booktabs}
\usepackage{amsmath}
\usepackage{wrapfig}
\usepackage{amsthm}
\usepackage[toc,page]{appendix}
\usepackage{subfig}
\usepackage{fullpage}
\usepackage{array}
\usepackage{pbox}
\usepackage{tikz}
\usepackage{url}
\usepackage{listings}
\usepackage[colorlinks=true]{hyperref}

\usepackage{color,soul}
\usepackage{mathtools}
\usepackage{booktabs}

\usepackage[colorlinks=true]{hyperref}

\usepackage{forest}	
\usetikzlibrary{arrows.meta, shapes.geometric, calc, shadows}
\usepackage{multicol}
\usepackage{imakeidx}
\definecolor{lightblue}{rgb}{.90,.95,1}
\definecolor{lightgreen}{rgb}{.90,1,.95}
\definecolor{darkgreen}{rgb}{0,.5,0.5}

\usepackage[linesnumbered,ruled,vlined]{algorithm2e}

\usepackage{pdflscape}

\usepackage{comment}

\usepackage{mathtools}
\tcbset{enhanced,colback=cyan!2!white,colframe=cyan!90!white,coltitle=blue!20!black,fonttitle=\bfseries}

\usepackage{listings}
\usepackage{hyperref}
\hypersetup{
	colorlinks=true,
	linkcolor=blue,
	filecolor=magenta,      
	citecolor=blue,
	urlcolor=cyan,
}

\graphicspath{ {./figures/}}
\makeindex
\makeatother

\begin{document}


\pagenumbering{roman}
\title{Data-driven turbulence modeling}
\author{Paola Cinnella\\Institut Jean Le Rond D'Alembert, Sorbonne Université 
\thanks{paola.cinnella@sorbonne-universite.fr}}

\date{29 January 2024}
\maketitle

\begin{abstract}
This chapter provides an introduction to data-driven techniques for the development and calibration of closure models for the Reynolds-Averaged Navier--Stokes (RANS) equations. RANS models are the workhorse for engineering applications of computational fluid dynamics (CFD) and are expected to play an important role for decades to come. However, RANS model inadequacies for complex, non-equilibrium flows and uncertainties in modeling assumptions and calibration data are still a major obstacle to the predictive capability of RANS simulations. In the following, we briefly recall the origin and limitations of RANS models, and then review their shortcomings and uncertainties. Then, we provide an introduction to data-driven approaches to RANS turbulence modeling. The latter can range from simple model parameter inference to sophisticated machine learning techniques. We conclude with some perspectives on current and future research trends.
\end{abstract}

\pagenumbering{arabic}
\setcounter{page}{1}
\clearpage{\pagestyle{empty}} 

\tableofcontents
\clearpage{\pagestyle{empty}} 

\section{Introduction}
\label{sec:introduction}

\begin{figure}[!htbp]
  \centering
    \includegraphics[width=0.5\textwidth]{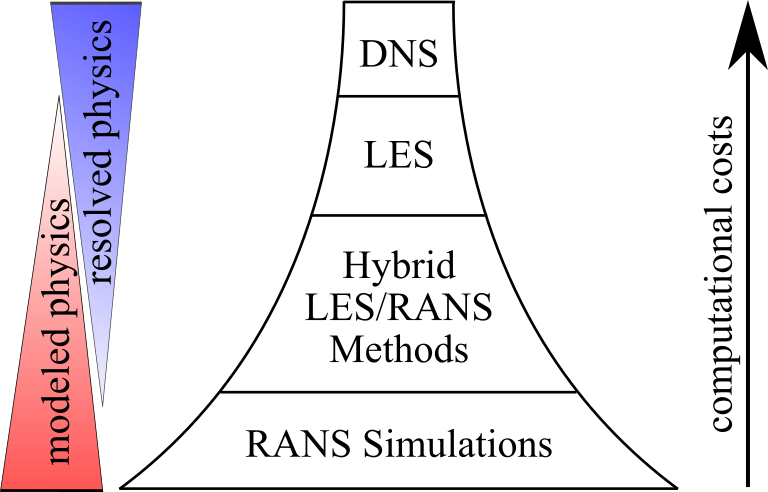}
   \caption{A schematic representation of the hierarchy of turbulence modeling approaches based on computational costs and the amounts of resolved versus modeled physics. From \citet{xiao2019review}.}
  \label{fig:classification}
\end{figure}

Accurate predictions of turbulent flows are of vital importance of natural and engineering systems, including climate, weather forecast, ocean dynamics, astrophysics, aerospace applications, energy conversion systems, civil engineering, and many other.
The dynamics of  fluid flows is described through conservation for mass, momentum, and energy.  While  many of the above-mentioned application may involve compressible flows, reacting flows or multi-phase flows, hereafter we restrict our attention to incompressible, single-phase, Newtonian fluids with constant properties, described by the celebrated Navier-Stokes (NS) equations:
  \begin{subequations} \label{eq:ns}
  \begin{align}
    \quad \frac{\partial u_i}{\partial x_i} & = 0 \label{eq:ns-mass} \\
    \frac{\partial u_i}{\partial t}+\frac{\partial \left( u_i u_j \right)}{\partial x_j} & =
    -\frac{\partial {p}}{\partial x_i} +\frac{1}{Re}\frac{\partial^2 u_i}{\partial x_j \partial
      x_j} ,
\label{eq:ns-momentum}
\end{align}
\end{subequations}
where $u_i$, $p$, $x_i$ and $t$ are, respectively, the flow velocity, pressure, and spatial and temporal coordinates. Although simpler in form than the partial differential equations governing the above-mentioned problems, incompressible NS equations cover a very wide variety of flow configurations and bear the key difficulty that leads to the turbulence modeling dilemma, i.e., the nonlinear convective  term in Eq.~(\ref{eq:ns-momentum}). Equation~(\ref{eq:ns}) is normalized with respect to a reference length $L_{\text{ref}}$, a reference velocity $U_{\text{ref}}$, and the density $\rho$ and viscosity $\mu$ of the fluid.  The parameter $Re=\rho U_{\text{ref}} L_{\text{ref}}/\mu$ is the Reynolds number, a measure of the relative importance of inertia to viscous forces.  Because of the nonlinearity of the convection terms $\frac{\partial \left( u_i u_j \right)}{\partial x_j}$, the NS equations admit chaotic solutions when the Reynolds number is beyond some flow-dependent critical value. As the Reynolds number increases, eventually the flow reaches a state of motion characterized by strong three-dimensional and unsteady chaotic fluctuations of the velocity and pressure fields, which is referred to as the turbulent regime.
The nonlinear and multi-scale character of the governing equations for fluid flows makes the numerical solution of fluid problems particularly challenging, typically requiring an extremely high number of degrees of freedom (DOF). Precisely, the number of DOF required to resolve all spatial scales in turbulent flow grows as $Re^{9/4}$, making so called Direct Numerical Simulation (DNS) of turbulent flow at all relevant scales impossible for most practical engineering problems, for which the Reynolds number is typically very large (see \citep{pope2000turbulent}, for a detailed discussion).
This has favored the development of so-called low-fidelity models, obtained by applying a coarse-graining operator to the NS equations. 
Denoting ${\cal N}(w)=0$ the exact Navier-Stokes operator and $w=(u_1,u_2,u_3,p)$ the exact state vector, the coarse-grained equations are of the form:
$${\cal N}_\text{CG} (w_\text{CG} )=0, \quad \text{with} \quad {\cal N}_{CG}=\langle{\cal N}\rangle$$
$\langle\cdot\rangle$  being the coarse-graining operator, and $w_\text{CG}=\langle{w}\rangle$ the coarse-grained solution.  Due to the nonlinearity of the NS equations, ${\cal N}_\text{CG}(w_\text{CG} )\ne {\cal N}(w_\text{CG} )$ so that new, unclosed term, appear in the coarse-grained equations. To fix ideas, $\langle\cdot\rangle$ could be for instance:
\begin{itemize}
\item	A filtering operator: ${\cal N}_\text{CG}$ corresponds to the filtered NS equations, and the corresponding simulations are referred-to as Large Eddy Simulations (LES). The coarsening procedure generates unclosed terms, known as subgrid scale (SGS) stresses, expressing the interactions between the unresolved subgrid scales and the resolved ones, which need to be modelled~\citep{sagaut2006large}. 
\item	The Reynolds averaging operator: ${\cal N}_\text{CG}$ then corresponds to the Reynolds-Averaged Navier-Stokes (RANS) equations. Such equations are alike the exact equations, except for the appearance of unclosed term expressing the contribution of unresolved turbulent scales (the main one being the Reynolds stress tensor), which need to be modeled to close the system.
\item	A combination of the preceding ones: ${\cal N}_\text{CG}$ then corresponds to a hybrid RANS/LES operator, degenerating to an averaging operator in the limit where no scales are resolved.
This strategy has been introduce to palliate to the prohibitive computational cost of LES for wall bounded flows at high Reynolds number due to the small yet energetic scales dominating the dynamics in the near-wall regions~\citep{spalart2009detached-eddy}. RANS/LES methods then combine LES in free shear regions with RANS models or other simplified models (e.g., boundary layer equation or law of the wall) in the under-resolved near-wall regions~\citep{frohlich2008hybrid,chaouat2017state}. Wall-modeled LES is also used with similar purpose~\cite{cabot2000approximate,piomelli2002wall,kawai2012wall,yang2015integral}. 
\end{itemize}
The hierarchy of turbulence modeling approaches is illustrated in Figure~\ref{fig:classification}, with the top represented by the most physics-resolving and computationally expensive approach (DNS) and the bottom by the most empirical and computationally affordable approach (RANS).  Lower fidelity models toward the bottom of the pyramid involve more flow-dependent, uncertain closures than the higher-fidelity, scale-resolving approaches towards the top of the pyramid. On the other hand, high-fidelity, scale-resolving models are more susceptible to influences from numerical uncertainties and boundary conditions.

For flows dominated by the dynamics of large-scale turbulent structures, LES simulations are the reference high-fidelity approach for low to moderate Reynolds numbers and mildly complex geometries. Unfortunately, they become computationally unfeasible for high-Reynolds-number wall-bounded flows encountered in industrial configurations. Hybrid approaches combining RANS and LES have been proposed to alleviate the computational burden, ranging from wall-modeled LES (WMLES) to zonal or global RANS/LES models \citep{sagaut2006multiscale}. The latter are the most attractive for industrial applications due to their ability to switch automatically from RANS to LES based on the local mesh resolution and turbulence characteristics. However, they are sensitive to the numerical set up and computational grid, with the appearance of “grey zones” where the model behaves in between the RANS and LES, possibly leading to aberrant solutions \citep{spalart2009detached-eddy}. Efforts for improving RANS/LES models often rely on empirical shielding functions or forcing terms. Also note that RANS/LES are inherently unsteady and 3D, hence much more costly than RANS. Methods for drastically reducing their computational cost are then highly desirable. 

The RANS equations are the workhorse model in CFD, because of their robustness and relatively low computational cost for complex, high-Reynolds-number configurations of practical engineering interest compared to higher-fidelity models. The most widely used RANS models, referred-to as Linear-Eddy-Viscosity Models (LEVM), rely on the so-called Boussinesq assumption of alignment between the Reynolds stress and the mean strain rate tensor. Such condition is however not verified even for relatively simple flows (see \cite{wilcox2006turbulence,schmitt2007boussinesq}). Besides, the eddy viscosity coefficient is computed via auxiliary relations (e.g. transport equations for turbulent properties) introducing supplementary modeling hypotheses and closure coefficients. More advanced models than LEVM exist, but they also rely on modeling approximations and uncertain closure parameters (see \cite{xiao2019review}). Whatever the closure assumptions, RANS models suffer from uncertainties associated with~\citep{duraisamy2019turbulence} :   
\begin{enumerate}
\item the applicability of a RANS-type description of turbulence for a given flow (often challenged for flows that are not statistically steady and do not exhibit large scale separation between macroscopic unsteadiness and turbulent motions); 
\item the way of representing the Reynolds stress as a functional form of the problem unknowns (the mean fields)
\item the choice of a suitable mathematical structure for constitutive relations and auxiliary equations used to link turbulent quantities to the mean field, referred-to as structural or model-form uncertainty; 
\item the calibration of the model closure parameters, known as parametric uncertainty.
\end{enumerate}
While scale-resolving simulations such as DNS, LES, and hybrid RANS/LES provide more insights of fluid flow physics, their cost is still too high for routine use in aerodynamic design and optimization, especially when the flow field dynamics is not of concern, and only knowledge of mean flow behavior is required.  The choice of the appropriate modeling level remains a matter of expert judgment. In particular, it will always involve a compromise between computational cost and predictive accuracy.  Even after a given fidelity level is selected (e.g., RANS or LES), several possible closure models may be designed for relating the unclosed terms to the resolved variables. These closure models differ both by their mathematical structure and by the associated model parameters.  The common practice in turbulence modeling is to leave the choice of a specific closure model to user judgment and to treat model parameters as adjustable coefficients that are generally calibrated to reproduce simple, canonical flows.  Both of the preceding aspects, however, represent sources of uncertainty in the prediction of new flows.  Recent development of turbulence modeling in RANS, LES, and hybrid approaches has been reviewed by~\citet{durbin2018some}.  Despite considerable progress recently made in LES and hybrid RANS/LES models (e.g., \cite{sagaut2006large,girimaji2006partially-averaged,frohlich2008hybrid,spalart2009detached-eddy,xiao2012consistent}), RANS models are expected to remain the workhorse in engineering practice for decades to come, due to their much lower computational costs and superior robustness.  

For this reason, in the following we will focus only on the RANS modeling approach. 

\section{RANS models: derivation and uncertainties}\label{sec:origin-rans-uq}
To derive the RANS equations, the instantaneous velocity $u_i$ and pressure $p$ are decomposed into the sum of the mean components $U_i$ and $P$ and the fluctuations~$u'_i$ and~$p'$, respectively. Note that several definitions exist for the mean or average quantities~\citep[see, e.g.,][]{wilcox2006turbulence}. The most general one is the statistical ensemble average, which however is rarely used in current practice due to the large number of independent flow realizations required for convergence. For statistically steady flow, time average is used instead based on an ergodicity hypothesis. The same is also used for unsteady flows, although its validity is still controversial. Substituting the decomposition into the Navier--Stokes equations and taking the ensemble-average leads to the RANS equations:
\begin{subequations}
  \label{eq:rans}
  \begin{align}
    \quad \frac{\partial U_i}{\partial x_i} & = 0  \\
    \frac{\partial U_i}{\partial t}+\frac{\partial \left(U_i U_j \right)}{\partial x_j}
                                            & =  -\frac{\partial {P}}{\partial x_i}
                                              +\frac{1}{Re}\frac{\partial^2 U_i}{\partial x_j \partial  x_j}
                                              - \frac{\partial \overline{  u_i'  u_j'}}{\partial x_j}
 \label{eq:rans-momentum}
  \end{align}
  \end{subequations}
  The RANS equations are  similar in form to the Navier--Stokes equations except for the term involving the tensor $-\overline{u_i' u_j'}$. As with the SGS stress term in the filtered NS equations for LES, this term stems from the nonlinear convection term in the NS equation and represents the cross-component covariance among the velocity fluctuations. It is often referred to as Reynolds stress due to its formal similarity to the viscous stresses and is denoted as 
\begin{equation}
\tau_{ij} = -\overline{u_i' u_j'}  .  
\end{equation}
Since the velocity fluctuations are not available in RANS simulations, one must resort to closure models to supply the Reynolds stress tensor (RST), which lies at the root of most efforts of turbulence modeling.
As an  covariance tensor, the RST must be symmetric positive semi-definite. This is referred to as \emph{realizability requirement}. We also observe that $\tau_{ij}$ appears in the RANS momentum equation through its divergence $\nabla \cdot \boldsymbol{\tau}$. While the RST as a symmetric rank-two tensor has six independent components, the divergence $\nabla \cdot \boldsymbol{\tau}$ as a forcing term only has three components.  The majority of existing turbulence models use the RST as the target of modeling. The rationale behind this choice is that the divergence form ensures conservation of momentum for any arbitrary volume of fluid. A few authors have however attempted to directly model $\nabla \cdot \boldsymbol{\tau}$, called the Reynolds force vector (RFV) through a conservative forcing term, although this is not straightforward~\citep{perot1996new}.  Therefore, in the following we discuss only turbulence models based on the Reynolds stress $\boldsymbol{\tau}$.

A plethora of turbulence models has been proposed through the years. Despite continuous research efforts though several decades,  "no class of models has emerged as clearly superior, or clearly hopeless" until now \cite{spalart2000} . Clearly, RANS models suffer from several shortcomings for complex flow configurations involving turbulence nonequilibrium, strong gradients, separations, shocks, 3D effects, etc.  Flow-specific tuning and fudge functions are still an indispensable part of RANS simulations~\citep{spalart2015philosophies}.  Current development of improved turbulence models faces the dilemma of conserving the low computational costs and high robustness of RANS approaches while approaching the predictive accuracy of high-fidelity models as much as possible.

As mentioned in the above~\cite{duraisamy2019turbulence} classified the model uncertainties in RANS simulations into four levels, including uncertainties due to information loss in the Reynolds-averaging process, uncertainties in representing the Reynolds stress as a functional form of the mean fields,  uncertainties in the choice of the specific function, and uncertainties in the parameters of a given model.  Hereafter we focus mostly on the two latter sources of uncertainties.

RST based turbulence models require prescribing a constitutive relation for~$\boldsymbol{\tau}$ as a function of the mean flow fields.  The most widely used class of models, generally known as linear eddy viscosity models (LEVM), relies on the Boussinesq analogy (see, e.g., \cite{wilcox2006turbulence}).  This consists in assuming that the anisotropic part of~$\boldsymbol{\tau}$  behaves similarly to the viscous stress tensor of a Newtonian fluid, i.e. it is a linear function of the local mean flow rate-of-strain~$S_{ij}$:
\begin{subequations}
\label{eq:evm}
\begin{align}
 \tau_{ij} + \frac{2k}{3} \delta_{ij}& =  2 \nu_t S_{ij} , \qquad \textrm{with}   \label{eq:evm-tau} \\
S_{ij} & =\frac{1}{2}\left(\frac{\partial U_i}{\partial x_j}+\frac{\partial U_j}{\partial x_i}\right)
 \label{eq:evm-s}
\end{align}
\end{subequations}
where $\tau_{ij} + \frac{2k}{3} \delta_{ij}$ is the Reynolds stress anisotropy, $k = \frac{1}{2} \overline{u'_i u'_i} = - \frac{1}{2}\tau_{ii}$ is the turbulent kinetic with a summation over index $i$ implied, $\delta_{ij}$ is the Kronecker delta (or the second order identity tensor in its vector form $\mathbf{I}$), and the eddy viscosity $\nu_t$ is the proportionality scalar.

The limitations of the Boussinesq assumption have been widely recognized in the literature. Specifically, the LEVM postulate of  a linear relationship between the Reynolds stress tensor and the mean strain rate tensor is not fully verified even for relatively simple flows \citep{wilcox2006turbulence,schmitt2007boussinesq}. A large part of the turbulence modeling literature during the last three decades reports attempts to upgrade the baseline LEVM by adding nonlinear terms suited to sensitize the model to curvature effects or to improve its anisotropy.
Examples are given by the SARC (Spalart--Allmaras with Rotation and Curvature, \cite{spalart_shur1997}), non-linear models \citep{speziale1987on}, elliptic relaxation models \citep{durbin1991}, algebraic Reynolds Stress models \citep{rodi1976algebraic} or explicit algebraic Reynolds Stress models \citep{pope1975more,gatski1993on,wallin2000explicit}, and full Reynolds Stress Models \citep{speziale1995,gerolymos2012term-by-term}. The latter require the solution of additional transport equations for the Reynolds stress components plus an equation for a quantity allowing to determine a turbulent scale. 
Unfortunately, the balance accuracy / robustness / computational cost of such more complex models has prevented a widespread use in CFD applications, despite their theoretical superiority~\citep{spalart2015philosophies} and anyway, no turbulence models (even the most sophisticated ones) are able to accurately predict the flow physics in all circumstances. Of note, more complex models typically involve a larger number of adjustable closure coefficients, typically calibrated against experimental or numerical data for so-called "canonical flows", \emph{i.e.}, simple turbulent flows representative of some elementary turbulent dynamics. However, 1) it is not always possible to determine closure coefficients that are simultaneously optimal for all canonical flows (as exemplified by, e.g., the so-called round/plane jet anomaly \cite{wilcox2006turbulence}); 2) the calibration data are affected by observational uncertainties that propagate to the closure coefficients; and 3) the final values retained in some models are not even the best fit to the data, but rather a compromise with respect to other requirements, e.g. numerical robustness. A discussion of uncertainties associated with turbulence models can be found in \cite{xiao2019review}.

\begin{figure}[!htbp]
  \centering
    \subfloat[Effects of turbulence model]{\includegraphics[width=0.45\textwidth]{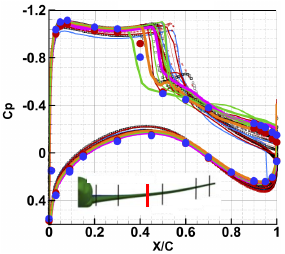}}
    \subfloat[Effects of model coefficients]{\includegraphics[width=0.5\textwidth]{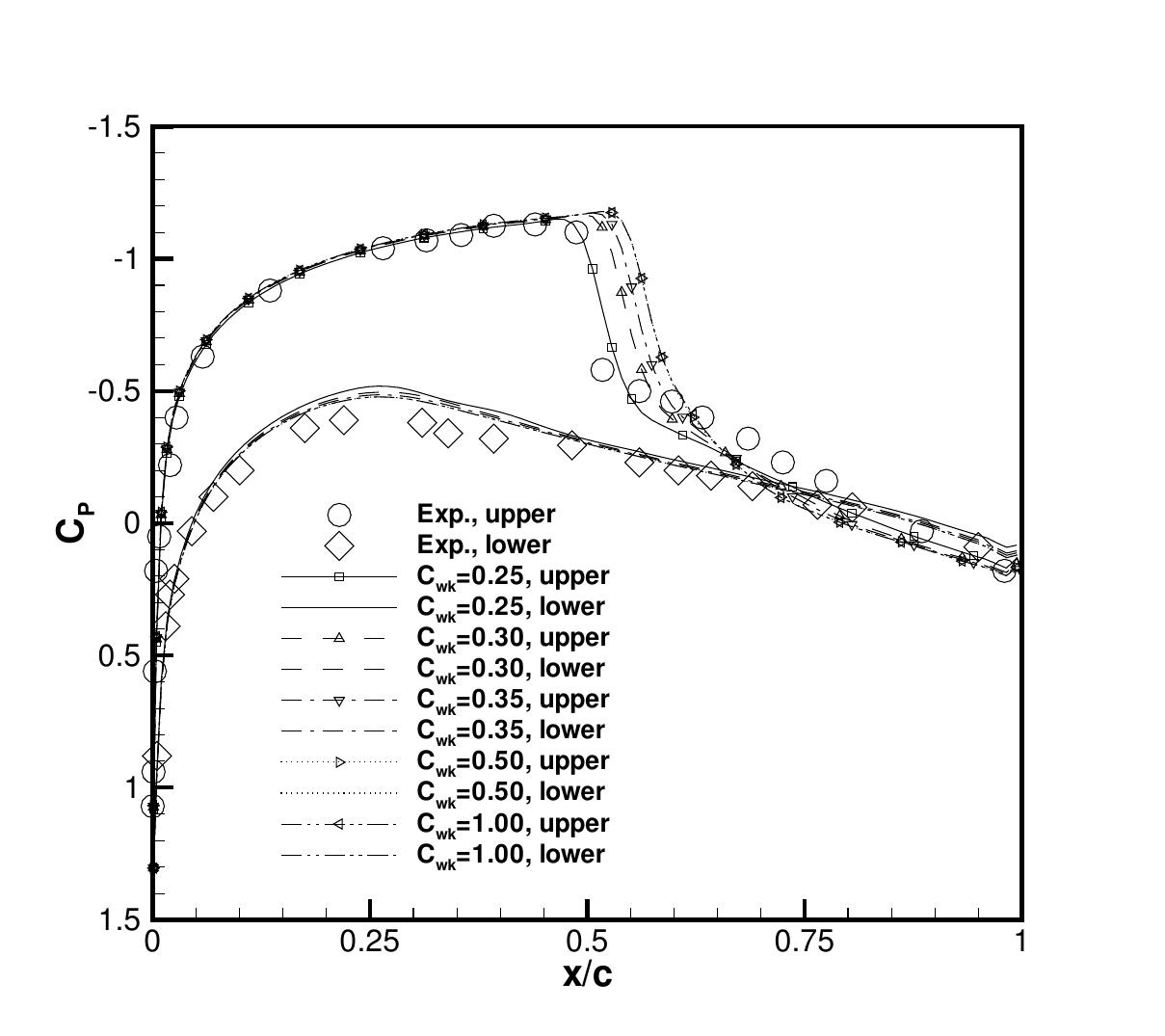}}
    \caption{Examples of uncertainties in RANS predictions of pressure coefficient $C_p$ distribution on wings and airfoils due to (a) model form and (b) model coefficients. Panel (a) shows the $C_p$ profile on a CRM wing-body configuration at $4.0^\circ$ angle of attack.  Results are from the $6^{\text{th}}$ AIAA CFD drag prediction workshop based on different RANS models, including $k$--$\varepsilon$ model, $k$--$\omega$ model, SA model, SA with quadratic constitutive relation (QCR), and EARSM. The location of the presented pressure distribution is indicated by the red/solid line on the wing (see inset; showing the port half of the fuselage and the wing only).  Figure reprinted from~\citet{tinoco2018summary} with permission. Panel (b) shows the $C_p$ profile on a NACA0012 airfoil in a transonic flow with freestream Mach number 0.8 and Reynolds number $9\times 10^6$, obtained from RANS simulations with the algebraic model of Baldwin and Lomax~\citep{baldwin1978thin}.  The figure shows the effect of varying $C_{wk}$, one of the seven model parameters, from 0.25 to 1, adopted from an unpublished report of the second author~\citep{cinnella2016review}.
      \label{fig:CRM-RANS}
    }
\end{figure}

The importance of model uncertainty is clearly illustrated in Figure~\ref{fig:CRM-RANS}a, which shows the predicted pressure distribution on the wing section of a Common Research Model (CRM) predicted by a number of turbulence models. A large scattering of the predictions is observed, particularly downstream of the shock wave generated at the upper wing surface.
Figure~\ref{fig:CRM-RANS}b illustrates the effect of varying only one of the seven parameters in the algebraic model of \citet{baldwin1978thin}. In particular, the location of the shock wave at the airfoil upper surface and the post-shock pressure are very sensitive to the model coefficient~\citep{cinnella2016review}.

Both the structural and the parametric uncertainties mentioned above are of an \emph{epistemic} nature, i.e. theoretically they could be reduced when better knowledge of turbulent flow physics and/or more abundant or more accurate data become available.  This is in contrast to \emph{aleatory} uncertainties, which arise from intrinsic variability of a process, e.g., uncertainties in manufactured geometries~\citep{parussini2007fictitious,liu2017quantification}, operation conditions of turbines or aircraft~\citep{avdonin2018quantification} or inflow conditions~\citep{gorle2015quantifying,mariotti2016freestream}.  In practice, reducing epistemic uncertainties by leveraging additional knowledge (e.g., by developing more advanced models to incorporate such knowledge) is far from straightforward.  Additionally, sophisticated models may lack numerical robustness or incur excessive computational costs.  Except for a few canonical examples, it is challenging, if not impossible, to identify the dominant source of uncertainty with definitive evidence, even for a given flow and a specific turbulence model. For instance, in many cases it is possible to improve the results of a model flawed by structural inadequacy by over-tuning its closure parameters. However, such over-tuning typically leads to poor predictions when applying the model to different flows from the calibration flows.  Such a phenomenon is referred to as over-fitting in statistics and machine learning~\citep{james2013introduction}.

\section{Data-driven approaches in turbulence modeling: state of the art}

In recent years, there has been a growing interest in applying Machine Learning (ML) techniques for turbulence modeling. ML can be used to analyze large amounts of data and discover non-trivial patterns. Since all RANS models involve some degree of empiricism \citep{spalart2000,spalart2015philosophies}, including those that were initially derived from exact manipulations of the Navier--Stokes equations, the use of data-driven has emerged during the last decade as a natural way to systematize the development and calibration of RANS models, and to discover improved RANS formulations for more complex flows (see the reviews of \cite{duraisamy2019turbulence,xiao2019review,duraisamy2020perspectives,sandberg2022review}).

Early studies of so-called data-driven RANS models focused on quantifying and reducing model uncertainties using interval analysis or statistical inference tools. This was achieved either by directly perturbing the Reynolds stress anisotropy tensor computed with a basic LEVM model \citep{emory2013modeling,gorle2013framework,thompson2019eigenvector} or by treating the model closure coefficients as random variables endowed with probability distributions \citep{platteeuw2008uncertainty,cheung2011bayesian,edeling2014bayesian,margheri2014epistemic}.
The first approach accounts for structural uncertainties in the constitutive relation of the Reynolds stresses, while the second considers only parametric uncertainties. On the other hand, the tensor perturbation approach is intrusive, as it implies modifications of the Reynolds stress representation in the RANS solver, while the parametric approach only requires a modification of the turbulence closure coefficients introduced in the CFD solver. A non-intrusive procedure to account for model-form uncertainties was proposed by \cite{edeling2014predictive}, who used Bayesian Model-Scenario Averaging (BMSA) to combine the predictions obtained from a set of competing baseline LEVM models calibrated on different data sets (scenarios). Since Bayesian Model Averaging builds a convex linear combination of the underlying models, its prediction accuracy cannot be better than the best model in the considered set, even if it outperforms the worst model. 

With the aim of reducing modeling inadequacies, data-driven  methods for turbulence modeling have  been introduced in recent years, mostly relying on supervised ML. Examples of early contributions can be found in \cite{tracey2013application,parish2016paradigm}, who proposed field inversion to learn corrective terms for the turbulent transport equations, along with ML to express the correction as a black-box function of selected flow features and to extrapolate it to new flows. Other contributions to the field-inversion and ML approach can be found, e.g., in \cite{ferrero2020field,volpiani2021machinePRF,volpiani2022neuralIJHFF}. One of the advantages of field inversion is that the (goal-oriented) correction can be inferred from sparse data or even global performance parameters. On the other hand, the ML step uses flow features estimated with a baseline RANS model to infer the corresponding correction. This may cause a feature mismatch for severe flow cases, since the baseline RANS flow field may completely miss flow features expected in the high-fidelity solution. In order to improve the consistency of the data-driven correction, iterative procedure have been proposed by \cite{taghizadeh2020turbulence,duraisamy2021perspectives}.

Another important question in data-driven turbulence modeling is the selection of input features that satisfy  known properties of turbulent flows. One of them is invariance in all Galilean reference frames, and under arbitrary rotations of the frame axes. The seminal work of \cite{ling2015evaluation} introduced a novel neural network architecture (Tensor Basis Neural Network, TBNN) that allows frame invariance constraints to be incorporated into the learned Reynolds stress anisotropy correction. The idea is to project the correction term onto a minimal integrity basis, as in the extended eddy viscosity model of \cite{pope1975more}, leading to a form of generalized Explicit Algebraic Reynolds Stress model, whose function coefficients are regressed from high-fidelity data using ML. 
 \cite{xiao2016quantifying,wu2018data-driven} combined ML classification techniques for identifying regions of high RANS modeling uncertainty with ML regression techniques for inferring model corrections from data and  predicting new configurations.
Other recent contributions are given by \cite{kaandorp2020data} who used tensor-basis Random Forests to learn data-driven corrections of the Reynolds stress tensor, and \cite{jiang2021interpretable} who proposed a general principled framework for deriving deep learning turbulence model corrections using deep neural network (DNN) while embedding physical constraints and symmetries. 

The aforementioned approaches to modeling turbulence from data use so-called black-box ML, such as neural networks or Random Forests. They allow a flexible approximation of complex functional relationships, but do not provide an explicit, physically interpretable mathematical expression for the learned correction. Recent attempts to interpret ML-augmented turbulence models rely on nonlinear sensitivity analysis tools, such as Shapley factor analyses \citep{he2022explainability}. Another drawback of black-box methods is that they are difficult to implement in CFD solvers and may lack numerical stability.
An interesting alternative is represented by so-called open-box ML approaches, which consist in selecting explicit mathematical expressions and/or operators from a large pre-defined dictionary to build a suitable regressor for the data. 
Examples of open-box ML include Gene Expression Programming (GEP) \citep{weatheritt2016novel} and symbolic identification \citep{schmelzer2020discovery}. 
Although less expressive than black-box ML, due to the rapidly escalating complexity of the search procedure in large mathematical operator dictionaries, the open-box models have the merit of providing tangible mathematical expressions that can be easily integrated into existing CFD solvers and interpreted in the light of physical considerations. 

Regardless of their formulation and training procedure, both black-box and open-box suffer from common drawbacks. First, the learned corrections are generally non-local, which means that they can alter the predictions of RANS models even where the baseline LEVM already gives good results. Second, such models tend to perform well only for narrow classes of flows and operating conditions, which means that they often need to be retrained whenever a new flow configuration needs to be addressed. This has recently led to attempts to merge or combine models trained for different settings to make accurate predictions for a wider range of flows. 
For instance, \cite{MZB2023} (to which we refer for a more complete literature review on ensemble models) proposed a method for spatially combining the predictions of a set of well-known LEVM taken from the literature, called space-dependent Model Aggregation (XMA). For that purpose, a cost function is introduced to evaluate the local model performance with respect to some training data, which is used to build the model weights. To make predictions, the weights are regressed in a space of flow features (representative of different flow phenomena).
Recently, \cite{cherroud2023space} replaced the set of  "on-the-shelf" LEVMs with a set of data-driven models. The latter are obtained by augmenting a baseline LEVM with stochastic corrective terms learned from high-fidelity data using open-box ML, namely, the Symbolic Bayesian Learning Sparse Regression of Turbulent stress Anisotropy (SBL-SpaRTA) of \cite{cherroud2022sparse}. 

In the following we briefly illustrate some data-driven RANS modeling techniques. We start with simple model coefficient calibration from data, then we present the main families of data-driven RANS modeling approaches, and finally we discuss their validity as physical "models", in terms of compliance with some general modeling rules. We also provide evidence of the difficulty of such model to generalize, and we introduce the above-mentioned mixture-of-experts technique as a possible improvement.

\section{Parametric approaches}
We call parametric approaches the data-driven procedures consisting in adjusting the closure parameters of RANS models with given (physics-based) mathematical form. Such approaches are non intrusive and can be applied to existing RANS solvers, provided that the model closure parameters can be prescribed as free inputs to the solver.
\subsection{Data-driven update of turbulence model closure coefficients}
\label{sec:prob-theory-da}
Initial attempts to improve turbulence models using data originate from research on the quantification and reduction of turbulence modeling uncertainties (see \cite{xiao2019review} for a comprehensive review). Early works  by  \cite{cheung2011bayesian,oliver2011bayesian,emory2013modeling,edeling2014bayesian}  treat  turbulence   modeling  uncertainties   in  a
probabilistic framework:  instead of producing a  single deterministic prediction associated  with a
model form  and a  set of parameters,  they try  to estimate the  probability distribution  of model
outputs, conditioned on some random inputs.  The analysis is conducted either by perturbing directly
the     Reynolds    stress     anisotropy     tensor    computed     with     a    baseline     LEVM
\citep{emory2013modeling,gorle2013framework,thompson2019eigenvector} or by treating the turbulence model
closure    parameters   as    random   variables    with   associated    probability   distributions
\citep{platteeuw2008uncertainty,cheung2011bayesian,edeling2014bayesian,margheri2014epistemic}.   

Hereafter, we recall the Bayesian framework for model calibration, which corresponds to an inverse statistical problem whereby, given some data, one tries to update the values of some input parameters: in our case, the closure coefficients of a turbulence model.
Let us consider a physical model of the form:
\begin{equation}
\Delta = M(\theta)
\end{equation}
with $\Delta  = (\Delta_1 ,..., \Delta_N)$  a vector of Quantities  of Interest (QoI) computed  by a
model $M$ given a set of parameters $\theta$ of dimension $P$.

In the  deterministic framework, the  components of  $\theta$ are perfectly  known and have  a fixed
value. In a Bayesian statistical framework,  the unknown  parameters vector $\theta$  is treated as  a random
vector,  characterised  by a  joint  probability  density function  (pdf),  noted  $f$. Due  to  the
uncertainty on $\theta$, $\Delta$ is also a random vector.

The scope of Bayesian inference is to gain  new knowledge about $\theta$ by constructing an improved
representation of its  pdf, based on prior  knowledge and on the observed  data.  For that
purpose, let  us note $D$  the random vector  of observed high-fidelity  data. Bayes
rules states that :
\begin{equation}
p(\theta|D = \overline{D}) = \dfrac{p(D = \overline{D}|\theta)}{p(D = \overline{D})} p(\theta)
\label{equ_Bayes}
\end{equation} 
Here,  $p(\theta)$ is  the  prior pdf  and  represents the  initial belief  about  $\theta$, $p(D  =
\overline{D}|\theta)$   is  the   likelihood  and   corresponds  to   the  probability   to  observe
$\overline{D}$,  a realization  of the  random  variable $D$,  if  $\theta$ is  known exactly.   The
posterior pdf $p(\theta|D  = \overline{D})$ represents the updated knowledge  of $\theta$ given
the observed data  vector $\overline{D}$, of size $N$.  In  practice, calibration  compares the model prediction  and the observations
to extract  the pdf of the  parameters vector $\theta$ that  is the most likely  to capture the
data. In our  case, $\theta$ is the set of closure parameters  associated with a given RANS
model, and $D$ is some observable quantity that can be obtained as an output $y$ of a RANS solver.

From Eq. (\ref{equ_Bayes}), it appears that the posterior distribution is entirely determined by the
prior  and likelihood  function.   Several strategies can be used to choose the prior pdf, including expert judgement or information theory.
In cases where little is known about the parameters,  Arnst \citep{Arnst2010}  suggests to  use uninformative  priors,
\textit{e.g.} large uniform  priors, for each component of  $\theta$ (supposed as independent). As  RANS models  have been
carefully designed, one may assume that  the standard values are included within the
range of  the prior. This is done for instance in \cite{cheung2011bayesian,edeling2014bayesian}. 
The likelihood function $p(D = \overline{D}|\theta)$ is a statistical model for
observation errors  (discrepancies between the  data and their  true, unobserved, values)  and model
inadequacies. The latter accounts for  the fact that part of the physics is  missed by the model due
to  any approximation  introduced in  its construction,  so that  the true  phenomenon can  never be
exactly captured, even with the best possible model coefficients. Possible choices for the prior are discussed in \cite{cheung2011bayesian,oliver2011bayesian}

As an example, the observation error can be modeled as  an additive noise, so that the data $\overline{D}$ at a given location
$x_i$ are related to the observation error by:
\begin{equation}
\overline{D}(x_i) = \widehat{D}(x_i) + e_i(x_i)
\end{equation}
with $e_i$ the observation noise at position $x_i$  and $\widehat{D}(x_i)$ the (unobserved) true value of the QoI
vector.  A simple choice consists in assuming that  the components  of  the  observation  noise  are  independent  and  normally
distributed, with zero mean and a standard deviation equal to some percentage of the observed value.

Because the model also suffers from structural inadequacy stemming from modeling assumptions, the model output is likely not to capture the data even with the best possible choice of the parameters. This can be taken into account by introducing a statistical term representative of model inadequacy, i.e. of the gap between the optimally calibrated model and the true data \citep{kennedy2001bayesian}.  For instance, \cite{cheung2011bayesian} treat model inadequacy as a
multiplicative term $\eta_i$, so that the data are related to the model output for $D$, $\Delta(x_i , \theta)$, at some space location $x_i$ by
\begin{equation}
\widehat{D}(x_i) = \eta_i  \Delta(x_i , \theta)
\end{equation}
Once again, a simple choice consists in assuming
the      model      errors     as     independent      and     Gaussian,      \textit{i.e.}
$ \eta_i \sim  \mathcal{N}(1 ,\sigma_{\eta}^{2}) $ where $\sigma_{\eta}$ is  an additional uncertain
hyper-parameter  that needs  to  be calibrated,  and  therefore  is concatenated  to  the vector  of
parameters $\theta$. The hyperparameter  $\sigma_{\eta}$ is a measure of the  magnitude of the model
inadequacy and thus can be taken as an indicator  of the accuracy of a given model, calibrated for a
given  scenario.   The preceding choices for $\eta_i$ and $e_i$  lead to a likelihood function of the form: 
\begin{equation}
p(\overline{D}|\Delta,\theta) = \dfrac{1}{\sqrt{(2\pi)^N |K|}} exp \left[ -\dfrac{1}{2} (\overline{D} - \Delta(\theta))^T K^{-1} (\overline{D} - \Delta(\theta)) \right]
\label{equ_likelihood}
\end{equation}
with $K =  K_e + K_M$ where $K_e$  is a diagonal matrix representing the  observational error vector
and $K_M=\sigma_{\eta}^2I$ a diagonal matrix reflecting model inadequacy.

For  complex models, such as RANS flow solvers,  the term  $\Delta(\theta)$ cannot  be
computed  analytically,   and  the  posterior   distribution  for  $\theta$  must   be  approximated
numerically.  Specifically,  we use  a  Markov-Chain  Monte-Carlo method  to  draw  sample from  the
posterior   pdf, and namely  the   Metropolis-Hastings   algorithm
  \citep{hastings1970MCMC}. 
Typically, $\mathcal{O}(10^5)$ samples  are needed to reach convergence, which  is unacceptably high
for costly RANS models.   To reduce the computational effort to an  amenable level, the calibrations
presented   in   the   following   are   based   on   surrogate   models. These are approximate models based, e.g., on polynomial interpolation, radial basis functions, gaussian processes or neural networks that emulate the response of the CFD solver to a given choice of the $\theta$ parameters, at a much lower computational expense.

When the exact probability is not critical and only the low order moments such as the mean and the variance are important, various approximate Bayesian inference methods can be used~\citep[e.g.,][]{mons2016reconstruction,iglesias2013ensemble-kalman}. These methods use maximum a posteriori (MAP) probability estimate to obtain the mode (peak) of the posterior and not the full posterior distribution.The MAP estimate can be computed in several ways, among which the most commonly used are variational methods and ensemble methods.  Both methods are used in data assimilation with a wide range of applications ranging from numerical weather forecasting to subsurface flow characterization.  Both variational methods and ensemble methods have been adopted for parameter inferences.  Obtaining the MAP estimate is equivalent to maximizing an error function to the reference data under the constraint imposed by the models describing the physical system (i.e., RANS equations in case of turbulent flows), by finding the set of parameters minimizing the discrepancies between the prediction and the observation data.  In variational methods the minimization problem is often solved by using gradient descent methods, with the gradient obtained with adjoint methods.  In contrast, ensemble methods use samples to estimate the covariance of the state vector, which is further used to solve the optimization problem.  
Ensemble Kalman filtering (EnKF)~\cite{evensen2003ensemble,evensen2009data} has been widely used in inverse modeling to estimate model uncertainties~\citep{iglesias2013ensemble-kalman,xiao2016quantifying}.  In EnKF-based inverse modeling, one starts with an ensemble of model parameter values drawn from their prior distribution. The filtering algorithm uses a Bayesian approach to assimilate observation data (e.g., data from experiments and high-fidelity simulations) and produces a new ensemble that represents the posterior distribution. In parametric or field inference of concern here, the EnKF method is used in an iterative manner to find the states that optimally fits the model and data with uncertainties of both accounted for, which is essentially a derivative-free optimization. As such, it is referred to as the \emph{iterative ensemble Kalman method}. This is in contrast to the EnKF-based data assimilation as used in numerical weather forecasting, where the observation data arrive sequentially. EnKF has some well known limitations due to its assumptions of linear models and Gaussian distributions, and theoretically they would perform poorly for non-Gaussian priors and highly nonlinear forward models.  However, despite the above-mentioned limitations, EnKF methods have been successfully used in a wide range of applications.

The first application of Bayesian uncertainty quantification techniques to the process of calibrating a turbulence model and making stochastic predictions of a new flow was done by \cite{cheung2011bayesian}.  A Bayesian calibration of the Spalart-Allmaras model from velocity and skin friction data for three boundary layers characterized by zero, adverse, and favorable pressure gradients was carried out by using MCMC sampling. \cite{kato2013approach} used ensemble Kalman filtering to determine the values of the parameters of the Spalart--Allmaras turbulence model for a zero-pressure gradient flat plate boundary layer. The results show the ability of the EnKF method to identify the correct model parameters for a relatively low computational cost (ensembles of 100 function evaluations, i.e. CFD calculations). Bayesian strategies similar to that of \citet{cheung2011bayesian} can also provide estimates of the uncertainty associated with the model form, grounded in uncertainties in the space of model closure coefficients.
This can be achieved by calibrating the model separately against several sets of data.
The spread in the posterior estimates of closure coefficients across calibration scenarios provides a measure of the need for readjusting the model coefficients to compensate for the inadequacy.
An example of  such a sensitivity study is given by \citet{edeling2014bayesian}, where the Launder--Sharma model was calibrated separately against 13 sets of flat-plate boundary layer profiles from \citet{kline1969computation}.
The results showed a significant variation in the most-likely closure-coefficients values for the different pressure gradients, despite the relatively restricted class of flows (flat plate boundary layers) considered for the calibrations.

\begin{figure}[!htbp]
  \centering
   \includegraphics[scale = 1.2]{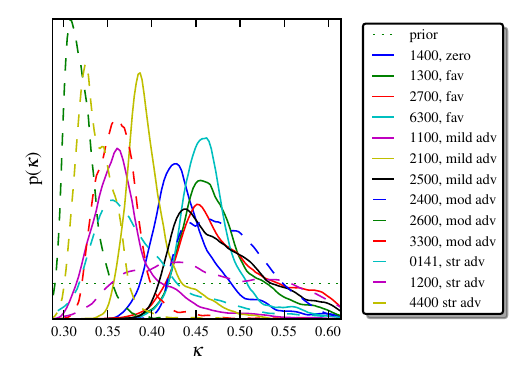} \\
     \caption{Sample posterior distributions from separate calibrations of the $k$--$\varepsilon$ model \citep{edeling2014bayesian}. The figure shows the wide spread of the Karman constant $\kappa$ calibrated from data for boundary layers with various pressure gradients}
  \label{fig:edeling2014a}
\end{figure}

The main lessons learned from the preceding exercise are: 
\begin{enumerate}
\item there are no universal values for the closure parameters of the turbulence models; 
\item the parameters need to adjust continuously when changing the dataset to compensate the intrinsic inadequacy (simplifying modeling assumptions) of the chosen model (see, e.g., the variation of the marginal posterior pdf for $\kappa$, reported in Figure~\ref{fig:edeling2014a}a); 
\item closure coefficients obtained by calibrating the model against a given boundary layer are generally not valid for the prediction of a different one.
\end{enumerate}

\subsection{Accounting for uncertain model structure: Bayesian model selection and averaging}
\label{sec:bsma}

An interesting outcome of Bayesian calibration is the possibility of deriving statistical criteria for model selection, i.e., for choosing the best model in some statistical sense among a class of competing models.  This consists in providing estimates of the posterior probability of each model in the considered set of models $\mathcal{M} = \{M_1, M_2, \cdots, M_I\}$ given the observed data. The ``model'' here should be interpreted in a broader sense, including not only physical models (e.g., $k$--$\varepsilon$, $k$--$\omega$, and Reynolds stress models) with associated coefficients but also statistical models (e.g., covariance kernel used to construct likelihood functions~\citep{cheung2011bayesian,oliver2011bayesian}.  Model probabilities are obtained as an outcome of parameter calibration introduced above. First, each model in the set $\mathcal{M}$ is assigned a probability $\mathbb{P}(M_j)$, $j=1, \cdots, I$, based on prior knowledge (e.g. from expert elicitation) or the lack thereof, in which case a noninformative, uniform distribution is chosen.  Additionally, the prior distributions for the closure coefficients $\theta$ or statistical hyperparameters associated with each model are also specified.  If data $\mathcal{D}$ are available, the prior probability mass function (pmf) can be updated according to Bayes' theorem, leading to the {\it posterior pmf} of model $M_j$:
\begin{align}
\label{eq:model-probability}
\mathbb{P}(M_j|\mathcal{D})=\frac{p(\mathcal{D}|M_j) \, \mathbb{P}(M_j)}{\sum_{i=1}^I p(\mathcal{D}|M_i) \, \mathbb{P}(M_i)} \qquad j = 1, \cdots, I
\end{align}
where $p(\mathcal{D}|M_j)$ is the evidence for model $M_j$ that normalizes the posterior pdf of the model parameters $\theta$, as in Equation~(\ref{equ_Bayes}).  The evidence can be computed at the end of the calibration by numerically integrating the numerator of~Eq.~(\ref{equ_Bayes}), using the posterior samples of $\theta$. This can be a challenging process requiring special techniques \citep[e.g.,][]{prudencio2012parallel}). The estimated pmf of the models can subsequently be used for predictions by choosing the model with maximum posterior probability in the case of \emph{model selection}, or alternatively by weighting the various posterior predictive distributions for the QoI with the posterior pmf in the case of \emph{model averaging}.

\begin{figure}[!htbp]
  \centering
    \includegraphics[width=0.6\textwidth]{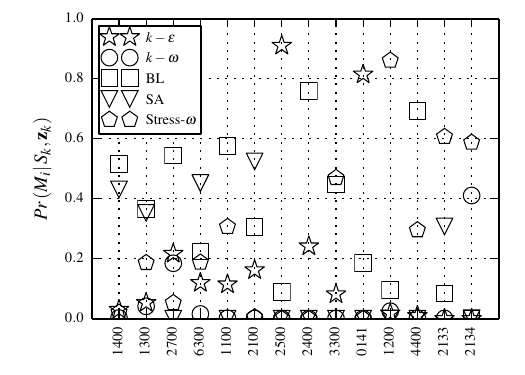}
    \caption{Posterior probabilities $\mathbb{P}(M_i | \mathcal{D}_k)$ of five turbulence models for 13 calibration datasets (boundary layers of various external pressure gradients). The set of models includes a simple algebraic model (Baldwin and Lomax~\citep{baldwin1978thin}), one-equation and two-equation eddy viscosity models (SA model~\citep{spalart1992one-equation}, $k$--$\varepsilon$, and $k$--$\omega$~\citep{wilcox2006turbulence}), and a Reynolds-stress model (stress--$\omega$ model \citep{wilcox2006turbulence}). Numbers on the horizontal axis denote identification codes for datasets (flow configurations). From \cite{edeling2014predictive}.}
  \label{fig:posterior}
\end{figure}

For instance \cite{oliver2011bayesian} calibrated the combination of four eddy viscosity models and three statistical models by using DNS data of plane channel flows and compared the posterior probabilities and predictive capabilities. The results showed that the considered data slightly favored Chien's $k$--$\varepsilon$ model~\citep{chien1982predictions} with an inhomogeneous stochastic model for the inadequacy, but no clear winner emerged with a dominantly high posterior probability.  \citet{edeling2014predictive} systematically demonstrated the difficulty of identifying a single best model without ambiguity. They used Bayesian inference to compute the posterior probabilities of five turbulence models ranging from simple algebraic eddy viscosity models to sophisticated Reynolds stress models by using DNS data of 13 boundary layer flows of various configurations. The posterior pmf for each dataset are presented in Fig.~\ref{fig:posterior}, which suggests that none of the models has a consistently higher probability than other models for all datasets, and the probabilities of all models are highly flow-dependent. As a consequence, it was not possible to select a single best model valid for all flow configurations.  Moreover, somewhat surprisingly, the Reynolds stress model was not the most plausible one for all flows despite its theoretical superiority; on the other hand, after calibration the algebraic model performed rather well over a wide range of flow configurations.

The difficulty of making predictions with a single calibrated model clearly calls for a framework based on multi-model ensembles. Multi-model approaches have been used in aerodynamics~\citep{poroseva2006improving} and many other applications~\citep{diomede2008discharge,duan2007multi-model,tebaldi2007use}.  Bayesian model averaging (BMA) is the among the most widely used multi-model approach, where the posterior of any predicted quantity $d$ is~\citep{draper1995assessment,hoeting1999bayesian}:
\begin{align}
  p(d \mid {\cal D}, {\cal M}) = \sum_{i=1}^I  p(d\mid M_i) \;\, \mathbb{P}(M_i \mid \mathcal{D}) ,
 \label{eq:bma-post}
\end{align}
given calibration data $\mathcal{D}$ and a set of models $\mathcal{M}$.  In this framework the posterior of $d$ is an average of $I$ posterior predictive distributions corresponding to $I$ competing models weighted by their respective model posterior as computed from Equation~(\ref{eq:model-probability}).

\cite{edeling2014predictive} proposed the Bayesian model--scenario averaging (BMSA), which is an extension of the classical Bayesian model averaging as shown in Equations~(\ref{eq:model-probability}) and (\ref{eq:bma-post}) above.  BMSA accounts for uncertainties on the choice of the calibration flow configuration (referred to as \emph{scenario}).  It predicts the QoI for a new scenario $\tilde{S}$ (not used for model calibration) as a weighted average of the predictions provided by a set of models ${\cal M} = \{M_i\}_{i=1}^I$, each model being previously calibrated against a set of scenarios $\mathcal{S} = \{S_k\}_{k=1}^K$ with corresponding datasets ${\cal D} = \{\mathcal{D}_k\}_{k=1}^K$.  Specifically, BMSA yields the posterior distribution of $d$ as follows:
\begin{align}
  p(d \mid \tilde S; {\cal D}, {\cal M}, {\cal S}) = \sum_{k=1}^K \sum_{i=1}^I p(d\mid \tilde{S};   M_i,  S_k, \mathcal{D}_k) \; \underbrace{\mathbb{P}(M_i\mid \mathcal{D}_k, S_k)}_{\text{model posterior}}
  \; \underbrace{ \mathbb{P}(S_k)}_{\text{ scenario prior}}
 \label{eq:pbmsa}
\end{align}
which is an average of the $I\times K$ posterior predictive distributions $p(d\mid \tilde S; \mathcal{D}_k, M_i, S_k)$,  each corresponding to the forward propagation of the parameter posterior obtained by calibration of model $I$ against scenario $K$ through the new prediction scenario $\tilde{S}$. The average is weighted by the corresponding posterior model probability $\mathbb{P}(M_i \mid \mathcal{D}_k, S_k)$ and prior scenario probability $\mathbb{P}(S_k)$. It is important to stress here that, for nonlinear systems, averaging the posterior predictive distributions of the QoI obtained by propagating the posterior pdf of the parameters for various scenarios through each model, as in Eq.~(\ref{eq:pbmsa}), is radically different than creating a mixture of the $K$ pdfs for the closure coefficients and then propagating it through the model. Specifically, \citet{ray2018robust} showed that latter provided unsatisfactory predictions, albeit being less expensive computationally.

In the BMSA prediction, the posterior probability of model $M_i$ is the outcome of the multiple calibration process after application of Eq.~(\ref{eq:model-probability}). On the other hand, the scenario probability $\mathbb{P}(S_k)$ needs to be specified \textit{a priori} and represents the user's belief about the similarity between calibration scenario $S_k$ to prediction scenario $\tilde{S}$ when the prediction of $\Psi$ is concerned.  When a physically justified prior is not available, a non-informative, uniform pmf can be used, implying equal probabilities for all scenarios. However, this may overestimate the posterior variance for $\Psi$, which leads to an overly pessimistic estimate of the prediction uncertainty~\citep{edeling2014predictive}. To address this issue, \citet{edeling2014bayesian} proposed an empirical scheme for choosing the scenario prior, with $\mathbb{P}(S_k)$ being inversely proportional to the scattering of all models trained on scenario $S_k$ when predicting the QoI for $\tilde{S}$.  The rationale is that if a calibration scenario $S_k$ is similar to the prediction scenario $\tilde S$, the models would give similar predictions of the QoI. Alternative models for the scenario pmf are discussed in \cite{MZB_2022}.

A major drawback of BMSA is its high computational cost, since it requires $I\times K$ stochastic calculations, each requiring forward propagation of a posterior parameter pdf (corresponding to a model/scenario combination) through the CFD model. The computational cost can be drastically reduced to~$I\times K$ deterministic CFD simulations by propagating though $\tilde{S}$ only the set of parameters  with maximum posterior probability (MAP) for each model and calibration scenario~\citep{edeling2018bayesian}, instead of the full pdf. With this simplification, the BMSA approach was applied to complex flow configurations such as the transonic three-dimensional flow around the ONERA M6 wing.
BMSA based on MAP estimates of turbulence model parameters has been also applied successfully to predict and reduce turbulence modeling uncertainty in turbomachinery configurations \citep{MZB_2020,MZB_2022}.

As noted by \citet{draper1995assessment}, multi-model approaches still introduce biases in the prediction because of the subjective selection of a finite set of models. However, they play a useful role in reducing the bias compared to predictions based on a single model. An averaged model is a way of obtaining a conservative prediction for an unseen configuration.  Indeed, the result will not be as good as the (\textit{a priori} unknown) best model but will not be as bad as the worst one. Additionally, BMSA provides an estimate of the solution variance based on the solution variability among the competing models.

\section{Non-parametric approaches to data-driven \\turbulence modeling}
\label{sec:nonparametric}
The preceding parametric and multi-model approaches account for uncertainties in the model coefficients and in the model choices. However, it is possible that the true solution lies outside the region in the solution space reachable by the parametric approaches. For example, it is well-known that linear eddy viscosity models are intrinsically not capable of predicting the secondary flows in a square duct.  In order to go beyond these limitations, an active track of research consist in learning directly from data the unclosed terms in the RANS equations and/or in the auxiliary transport equations for turbulent quantities or, more frequently, corrective terms to be applied on top of some baseline (inadequate) turbulence model. The latter option is generally preferred because it is easier to learn corrective terms to models that have been fine-tuned at least for some classes of flows over years, than learning all unclosed terms from scratch using limited high-fidelity data.

\begin{figure}[!htbp]
  \centering
    \includegraphics[width=0.45\textwidth]{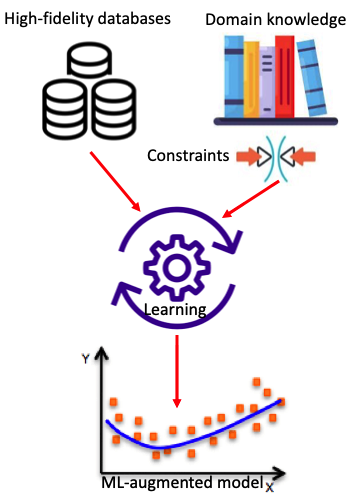}
    \caption{Sketch of a data-driven turbulence model: high-fidelity data are collected from numerical simulations (with higher hierarchical level than RANS) or experiments. Domain knowledge, i.e. available information about turbulence properties, invariances, symmetries, etc., are enforced as constraints to the learning process. The model is trained against data, i.e. model parameters as fit as to minimize the deviation to the observed data, under the given constraints. The resulting data-augmented RANS model is finally assessed again data not used for training (test set).}
  \label{fig:datadriven_model}
\end{figure}

The general philosophy of data-driven turbulence models can be summarized as follows (see also figure \ref{fig:datadriven_model}):
\begin{itemize}
\item The idea of developing a single « universal » model is generally abandoned;
\item Specialized models ("experts") are directly learned from high-fidelity data available for a given class of flows;
\item For that purpose, the procedure consists in:
\begin{itemize}
\item Collecting and curating data;
\item Choosing a representation (function dictionary or machine learning architecture) for the unknown fields to be modeled;
\item Enforce physical constraints (whenever possible)
\item Train the model against data
\item Validate (in the CFD sense) for a "test set" (in the machine learning sense)
\end{itemize}
\end{itemize}
%

A number of data-driven turbulence modeling strategies have been proposed during the last decade. These can be broadly classified as follows
\begin{enumerate}
\item models using data-assimilation to identify corrective fields (to the RANS transport equations and/or to the Reynolds stresses), which are subsequently parametrized as functions of a set of flow features to allow predictions of new flows~\citep{singh2016using,parish2016paradigm};
\item models directly learning corrections of the Reynolds stress field from high-fidelity data~\citep{emory2011modeling,emory2013modeling,xiao2016quantifying,ling2016reynolds,weatheritt2016novel,schmelzer2020discovery,cherroud2022sparse};
\item model-consistent approaches, whereby the corrective fields and their parametrization are directly learned from data using backpropagation of errors through the RANS solver [REFS] or gradient-free algorithms~\citep{zhao2020rans,duraisamy2021perspectives,saidi2022}
\end{enumerate}

Data-driven turbulence modeling methods can be also classified according to the regression method used to learn the input/output relationships between selected flow features and the corrective fields: 
\begin{enumerate}
\item Black-box models, were the unclosed quantities (e.g., the Reynolds stresses) are completely learned from data using machine learning methods such as Neural Networks (NN), Random Forests (RF), Gaussian Processes (GP) or other, which only provide input/output relationships and do not have any straightforward physical interpretation, hence the name of "black box". An example is provided in \cite{ling2016reynolds}, where a standard fully-connected neural network (multi-layer perceptron) is used to learn the Reynolds stresses as a function of the pointwise components of the mean strain rate and mean rotation tensors. Such models do not naturally account for flow invariances and symmetries, and a very large amount of data or data-inflation techniques are needed for the model to "learn" such properties.
\item Grey-box models, i.e. models where unknown subparts of a model are learned from data. These are often expressed as corrections to an existing baseline RANS model, and can be formulated in different ways: corrective terms in the model transpot equations, corrective terms for the Reynolds stresses, corrected eddy viscosity, etc... In all cases, the corrective term, which are often formulated to respect known physical constraints, are subsequently learned using black-box machine learning methods, hence the denomination of "grey-box". Some examples are discussed in the following. 
\item Open-box or white-box models, i.e. models where corrective terms for the Reynolds stresses are learned under the form of tangible mathematical expressions. Sparsity constraints and feature engineering can be used to make the resulting expressions as simple as possible, and physically interpretable. Additionally, such methods can easily incorporate invariance and symmetry properties. The downside of such methods is their lesser expressivity compared to black-box models, and their difficulty in scaling up to very high-dimensional problems. Some examples are also discussed in the following.
\end{enumerate}

\begin{figure}[!htbp]
  \centering
    \includegraphics[width=0.8\textwidth]{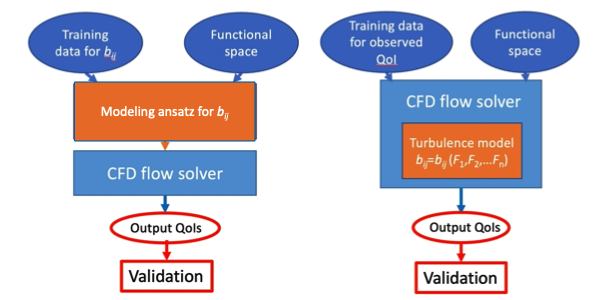}
    \caption{Generic workflows of a priori (left) and model-consistent (right) training ap-
proaches. In a priori training, data are gathered for the quantity to be modeled, e.g. the
Reynolds stress anisotropy $b_{ij}$, and for a set of features $F_1, ...F_n$. The model parameters are trained, and the resulting data-driven RANS model is then plugged in
a RANS solver and used to predict quantities of interest (QoI). In model-consistent training the
search space is the same, but the data can be any QoI computable as an output of the RANS solver; the
unknown model is embedded in the solver, and evaluations of "candidate" models involve
CFD solves.}
  \label{fig:apriori_vs_consistent}
\end{figure}

Finally, yet another classification of data-driven RANS models can be operated based on the training strategy.
\begin{enumerate}
\item "A priori" training: this consists in first obtaining full fields of turbulent properties (Reynolds stresses, terms in the transport equations, etc.) either directly \cite{weatheritt2016novel,wu2018data-driven,schmelzer2020discovery,cherroud2022sparse} from high-fidelity experiments/simulations or indirectly, via a preliminary data assimilation step \citep[e.g.]{parish2016paradigm,volpiani2021machinePRF}
\item "A posteriori", "Model-consistent" or "CFD-driven" training, where a modeling ansatz for the unclosed terms is directly trained by solving a "CFD-in-the-loop" optimization problem, whereby the target data are compared to output of the RANS solver given a set of model parameters. Based on the outcome of the comparisons, the parameters are adjusted and/or the model form/architecture is modified, and updated RANS output are computed. Such an approach is more costly than "a priori" strategies, because each training episode involves a costly CFD calculation, but it mitigates inconsistencies between the input features (obtained with the baseline model or the high-fidelity data) used for training and those used for prediction (uniquely based on the imperfect RANS model)~\citep{zhao2020rans,duraisamy2021perspectives,saidi2022}.
\end{enumerate}
The "a priori" and "model-consistent" training procedures are illustrated schematically in Fig. \ref{fig:apriori_vs_consistent}.
%
\emph{A priori} training is naturally cheap and nonintrusive (i.e., the solver is not involved), and provides
opportunities to directly impose physics-based constraints on the Reynolds stresses. However, it suffers a consistency problem 
of the ML model. The latter being trained using only DNS quantities, it is consistent with the DNS field, but does
not guarantee consistency with the RANS or LES environment. One reason is the so-called feature mismatch between training and prediction. 
During the training process features computed from DNS data are used as inputs to the learning model, whereas in the
prediction process, features predicted by the model are used. Unless the model is very accurate, the latter will be very different from the DNS ones. 
This is difficult
to achieve because RANS models are typically constructed to
predict certain quantities (e.g., mean flow, Reynolds shear stress, etc.) well, but not all quantities.
This is especially true in RANS where several secondary quantities are used to provide length and
time scales which should not be interpreted to have a one-on-one correspondence with the DNS
quantities \citep{duraisamy2021perspectives}. 
Another drawback of \emph{a priori} training is that it requires full field data of turbulent quantities, and especially the Reynolds stresses, which may be no always easily available.

On the other hand, model-consistent training involves large scale inverse problems, typically
necessitating the use of adjoint-driven optimization techniques. Due to high nonlinearities of the RANS equations
and highly nonconvex loss function landscapes, gradient-based algorithms are easily trapped into local minima, while gradient-free methods can be extremely slow and expensive to converge.
This has fostered the development of "iterative" or "closed-loop" training techniques \citep{taghizadeh2020turbulence,liu2021iterative}, whereby the corrected model is used as the baseline for subsequent steps of correction.

In the following of this Section, we first discuss possible representations for the corrective terms and physical constraints to be considered for the selection of input flow features used to express such terms.
Afterwards, we illustrate into some detail examples of some widespread data-driven turbulence modeling approaches among those mentioned in the above.

\subsection{Representation of the data-driven corrections}
The data-driven corrections can be applied either directly on the Reynolds stress tensor (RST) $\tau_{ij}$ or on the model transport equations.
As previously mentioned, the former is a symmetric, semi-positive definite tensor. Thus, any model for $\tau_{ij}$, before or after application of data-driven corrections, should respect such constraints.

A list of additional hard or soft constraints that turbulence models (whether data-driven or not) should respect, has been recently given by Spalart \citep{spalart2023oldfashioned}, to support the development of well-founded data-driven turbulence models.
The latter are classified as hard or soft constraints, and include the use of input properties that satisfy frame invariance, well-posedness of the resulting governing equations, and universality, as well as practical considerations about implementation in complex geometries.

For instance, using corrective fields of the form $\beta(\bm{x})$ is not acceptable, since it violates  the “universality principle” and the requirement for turbulence models to use local quantities unrelated to the geometry.  Hence the necessity of finding a set of local flow quantities relevant to the evolution of turbulent quantities, and respectful of frame invariance.

For cases where the RST is the target of the learning process, three main representations have emerged in the literature. 

The first one, initially considered in uncertainty quantification studies by \cite{emory2013modeling,gorle2013framework,tracey2013application}, and then in the works about machine learning of turbulence model corrections  by Xiao and co-workers \citep{wu2016bayesian,wu2018data-driven}, expresses the RST through the eigenvalues of its anisotropic part:
\begin{equation}
  \label{eq:tau-decomp}
   -\boldsymbol{\tau} = 2 k \left( \frac{1}{3} \mathbf{I} +  \mathbf{b} \right)
  = 2 k \left( \frac{1}{3} \mathbf{I} + \mathbf{V} (\Lambda+\Delta\Lambda(\bm{\eta})) \mathbf{V}^\top \right)
\end{equation}
where $k$ is the turbulent kinetic energy, equal to one half the trace of $ \boldsymbol{\tau}$; $\mathbf{I}$ is the second-order identity tensor; $\mathbf{b}$ is the anisotropy tensor; $\mathbf{V} = [\mathbf{v}_1, \mathbf{v}_2, \mathbf{v}_3]$ and $\Lambda = \textrm{diag}[{\lambda}_1, {\lambda}_2, {\lambda}_3]$ where ${\lambda}_1+ {\lambda}_2 + {\lambda}_3=0$ are the orthonormal eigenvectors and eigenvalues of $\mathbf{b}$, respectively, indicating the \emph{shape} (aspect ratio) and \emph{orientation} of $\bm{\tau}$, if the latter is visualized as an ellipsoid~\cite{simonsen2005turbulent}.
In the above, $\Delta\Lambda(\bm{\eta})$ is a corrective term that models the error in anisotropy via supervised learning, generally
on a DNS dataset for $\tau$. The features $\bm{\eta}$ may include the eigenvalues of the anisotropy tensor, the ratio of
the production-to-dissipation rate (of turbulent kinetic energy), or a more comprehensive list of features including not only the
eigenvalues, but also the eigenvectors and the turbulent kinetic energy \citep{wu2016bayesian,wu2018data-driven}.

In their seminal work, \cite{ling2016machine,ling2016reynolds} proposed a neural network architecture, named "Tensor Basis Neural Network" (TBNN) to learn the coefficients of a tensor basis
expansion for the anisotropic Reynolds stresses in the form:
\begin{equation}\label{bdelta}
 b_{ij} = \sum_{n=1}^{10} T_{ij}^{(n)}(S_{ij}^*,\Omega_{ij}^*) \alpha_{n} (\bm{\eta}) 
\end{equation}
where $S_{ij}^*=S_{ij}^*/\omega$ and $\Omega_{ij}^*=\Omega_{ij}/\omega$ are  a  non-dimensional strain rate tensor and a non-dimensional rotation rate tensor, respectively, with respect to the turbulent time scale $\omega^{-1}$, and 
where $T_{ij}^{(n)}(S_{ij}^*,\Omega_{ij}^*)$ are a minimal integrity tensorial basis \citep{pope1975more} (with respect to the strain rate and vorticity tensors)
and $\bm{\eta}$ are the coefficients which, in the original formulation of \cite{pope1975more} are taken to be five invariants based on $S_{ij}^*,\Omega_{ij}^*$, and obey frame invariance by definition. Such coefficients are learned under the form of a neural network in  \cite{ling2016machine,ling2016reynolds}, but many alternatives have since then been explored in the literature:
\cite{kaandorp2020data} leverage the above integrity basis,
but use random forest regression instead of neural networks; \cite{weatheritt2016novel,zhao2020rans} use
symbolic regression with genetic programming; \cite{schmelzer2020discovery,beetham2020formulating,cherroud2022sparse} employ sparse deterministic regression or sparse Bayesian learning
on a library of candidate functions which are written as tensor polynomials of the aforementioned invariants. 
For 2D flows, equation (\ref{bdelta}) can be simplified, and reduces to a combination of the first three tensors and two invariants, noted $I_1$ and $I_2$.
Note that Pope's representation relies on an hypothesis of local equilibrium of turbulence: in such conditions, the RST can be represented as a function of local mean quantities, and specifically the local mean velocity gradient. This assumption is no longer true for flows with strong nonequilibrium (see also \cite{taghizadeh2020turbulence}).
For this reasons, various authors have tried to enrich the representation with additional features, e.g. accounting for near-wall turbulent scales \citep{jiang2021interpretable} or for turbulent kinetic energy and pressure gradients, projected onto a local invariant frame \citep{brener2024highly}.

The above-mentioned representations deal with the RST. However, using the Reynolds force vector (RFV) has also been considered by some authors.  
\citet{cruz2019use} introduced a new propagation method in which they only used the mean velocity field in RANS equations to calculate the term associated with the RST (i.e. the divergence of RST). They compared the explicit treatment of both RST and RFV and showed that using RFV results in a lower error for both propagation and prediction. In a more comprehensive comparison, \citet{brener2021conditioning} studied both implicit and explicit treatment techniques with both RST and RFV values obtained from high-fidelity data of channel flow, the square-duct flow, and the periodic-hills flow. They reported that the propagation of the RFV results in lower error than the propagation of the RST. This has been recently confirmed by \cite{amarloo2022frozen}. Regarding the vector-based framework, \citet{berrone2022invariances}  introduced a vector-basis neural network (VBNN) for the prediction of the divergence of the RST. The VBNN is derived and simplified from the TBNN \citep{ling2016reynolds}. \cite{amarloo2023data} also derived a VBNN from the TBNN to predict the RFV discrepancy.
Taking the divergence of the Reynolds stress representation in eq. (\ref{bdelta}), the following RFV representation is obtained:

\begin{equation}
\label{eq:tiPope}
t_{i} = \partial_j \left(b_{ij}\right)= \partial_j \left(\Sigma_{n=1}^{3}\left(\alpha_nT^{(n)}_{ij}\right)\right),
\end{equation}
which can be extended into 
\begin{equation}
\label{eq:tiEqPope}
t_{i} = \Sigma_{n=1}^{3}\left(\alpha_n q^{(n)}_{i} + \frac{\partial \alpha_n}{\partial I_1}r^{(n)}_{i} + \frac{\partial \alpha_n}{\partial I_2} s^{(n)}_{i}\right),
\end{equation}
\begin{equation}
\label{eq:vectorsPope}
\begin{split}
    q^n_{i} = & \partial_j(T^{(n)}_{ij}) + \partial_j(k) T^{(n)}_{ij},\\
    r^n_{i} = & \partial_j(I_1) T^{(n)}_{ij},\\
    s^n_{i} = & \partial_j(I_2) T^{(n)}_{ij}.
\end{split}
\end{equation}
Similarly to the tensor base in Eq.~(\ref{bdelta}),  the $\alpha_n=\alpha_n(\bm{\eta})$ ($\bm{\eta}=(I_1,I_2)$) should be learned from data using some machine learning technique.


A wide range of approaches do not use the RST or RFV as the modeling targets, but other terms in the RANS or auxiliary transport equations. For instance
Duraisamy et al. [13,26,27] applied similar feature-based augmentations to transport
equations, rather than directly to the Reynolds stress. \cite{volpiani2022neuralIJHFF} proposed a neural-network correction to the eddy viscosity field predicted by the Spalart Allmaras model.
In such cases, the choice of representation is much freer, and several sets of flow features have been proposed in the literature.
\cite{schmelzer2020discovery} proposed an additive corrective term for the turbulent kinetic energy equations, which was modeled as an "extra production": in practice the term was written as the product of a tensor written in a form similar to Eq. (\ref{bdelta}) by the mean velocity vector, and the coefficients of the tensor expansion are learned from data using sparse symbolic regression.
Specifically, the term writes:
\begin{equation}\label{extraprod}
R = a_{ij}^{R} \frac{\partial U_i}{\partial x_j}
\end{equation}
with
\begin{equation}\label{bR}
b^{R}_{ij}= \sum_{n=1}^{3} T_{ij}^{(n)} \alpha_{n}^{R} (I_{1},I_{2}) 
\end{equation}
Other choices are possible. For instance \cite{stocker2023dns} reformulated $R$ by using a "dissipation analogy" instead of the preceding "production analogy". This has the advantage of simplifying the learning problem, because only a scalar term has to be searched. However, the representation (and hence the choice of the features) is no longer based on a minimal integrity basis.

\subsection{Field inversion and Machine Learning (FIML) approach}
\label{sec:nonpara-beta}

Model calibration cannot overcome the structural inadequacies inherent with the chosen baseline models, e.g., those associated with the Boussinesq constitutive model or with drastic simplifying assumptions used to close auxiliary transport  equations for turbulent quantities. A way for overcoming such limitations is inferring from data corrections to the model structure, e.g. modifying the constitutive equation or the source terms in the turbulent transport equations (e.g., for $k$, $\omega$, and $\bm{\tau}$).  

The first attempts to use data-driven approaches where related to uncertainty quantification \citep{singh2016using,parish2016paradigm}.
It was assumed that errors in the turbulent transport equations rather than the constitutive equations, are the dominant source of the prediction errors in RANS simulations.  The learned corrections depend on the specific form of the baseline turbulence model.  Taking the $k$--$\omega$ equation for example, a multiplicative discrepancy field $\beta(\bm{x})$ is introduced to the source terms of the $\omega$ transport equation by:
\begin{equation}
    \label{eq:omega}
    \frac{D \omega }{D t} = \beta(\bm{x}) \mathcal{P}_\omega(k, \omega, U_i) - D_\omega(k, \omega, U_i) + T_\omega(k, \omega, U_i)
\end{equation}
where $\omega$ is the turbulent frequency; $ \mathcal{P}_\omega$, $D_\omega$, and $T_\omega$ indicate production, dissipation, and transport, respectively, of $\omega$.  The optimal discrepancy field $\beta_\text{opt}(\bm{x})$ minimizing the deviation between the RANS solver output and the data can be inferred by using DNS or experimental data of velocities or other quantities of interest, e.g.,  drag, lift,  pressure coefficient, and surface friction.  Assuming the velocity is the data to be used,  the inference can be cast as the following optimization problem:
\begin{equation}
\label{eq:beta-opt}
  \beta^{\text{opt}} = \arg \min_{\beta} J \; ,  \qquad \text{with} \quad J = \| U(\beta) - U^{\text{dns}}  \|_{L^2}
\end{equation}
where $\| \cdot \|_{L^2}$ indicates $L2$ norm. In cases where other derived quantities (e.g., drag and lift) are used in the optimization, an observation operator $\mathsf{H}$ is needed to map the solution to these quantities, i.e., $g = \mathsf{H}[u]$, and the cost function would be $J = \| g(\beta) - g^{\text{dns}} \|_{L^2}$.  The inferred discrepancy $\beta^{\text{opt}}(\bm{x})$ is a correction that allows the baseline $k$--$\omega$ model to agree with the data.  The discrepancy field $\beta$ resides in a space of very high dimensions with a dimension equal to the number of cells in the CFD mesh, and thus the optimal solution is not unique. In the terminology of inverse modeling, this problem is ill-posed and needs to be regularized.  The deviation of $\beta$ from $1$ is used as a penalty to regularize the problem, which leads to the following cost function~\cite{singh2016using}:
\begin{equation}
  \qquad \text{with} \quad J = \| U(\beta) - U^{\text{dns}}  \|_{L^2} + \gamma \| \beta(\bm{x}) - 1 \|_{L^2}
\end{equation}
where $\gamma$ is a regularization parameter.
The second term, $\beta(\bm{x}) - 1$, prevents the corrected model from deviating too much from the baseline model. With such a regularization, the corrected model is constrained to explore only the \emph{vicinity} of the baseline solution, which greatly reduces the dimension of the search in the high-dimensional space of possible discrepancy fields $\beta$.

\begin{figure}[!htbp]
  \centering
    \includegraphics[width=0.9\textwidth]{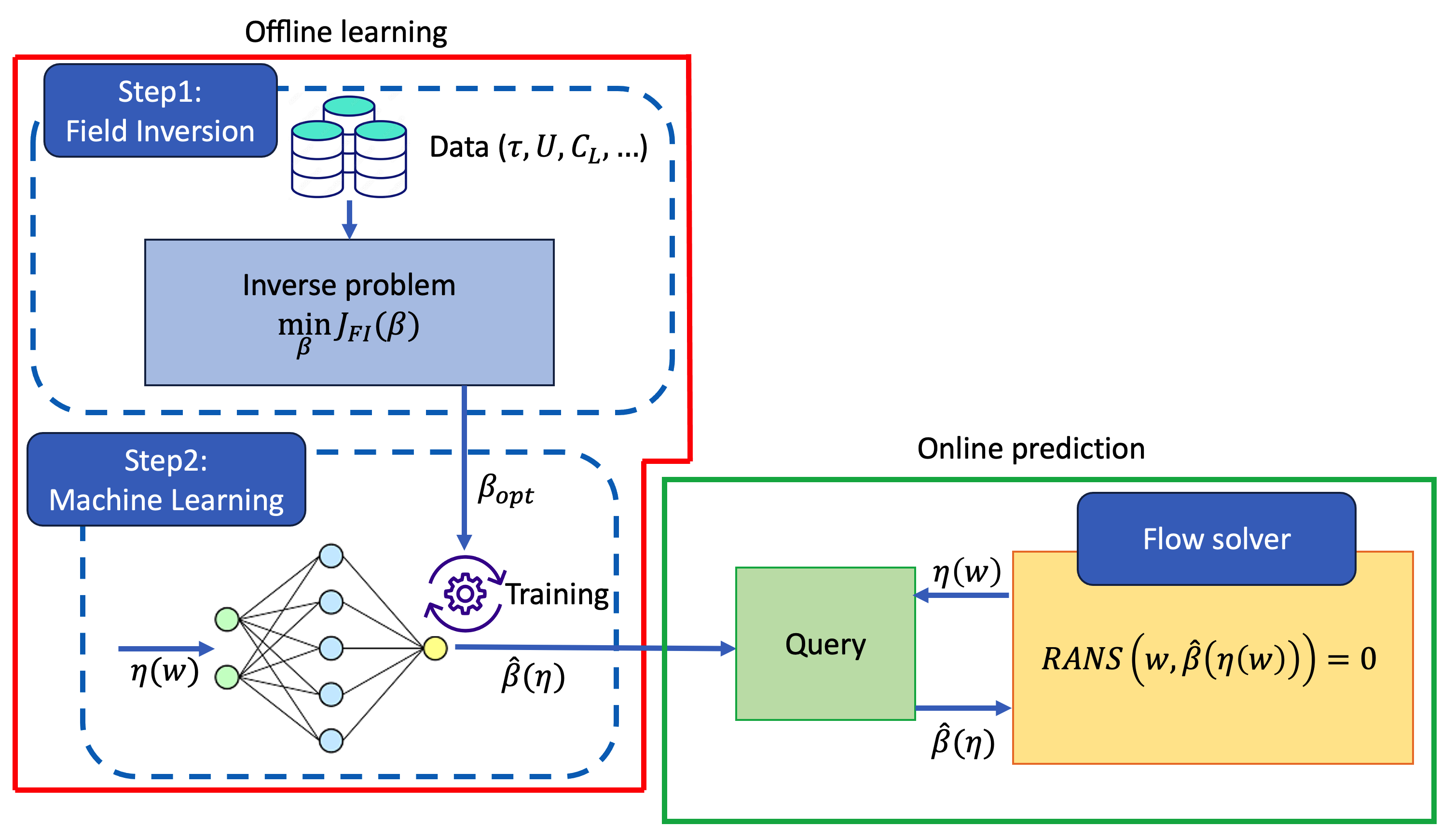}
    \caption{Sketch of the classic Field Inversion and Machine Learning (FIML) procedure.}
  \label{fig:fiml}
\end{figure}

The inferred discrepancy field can be subsequently used to guide the improvement of the baseline model and to develop data-driven correction schemes. For that purpose, the correction function $\beta(\bm{x})$ is first reformulated as a function of a set of well-chosen \emph{flow features} $\bm{\eta}$: $\beta(\bm{\eta})=\beta(\bm{\eta}(\bm{x}))$.
For instance, $\bm{\eta}$ can include nondimensional counterparts of flow quantities such as the mean strain rate $\bm{S}$ or rotation rate $\bm{\Omega}$~\citep{ling2015evaluation}, as well as the ratio $\mathcal{P}/D$ between production and dissipation~\cite{singh2017machine-learning-augmented}). Choosing flow variables $\bm{\eta}$ rather than spatial coordinate $\bm{x}$ as the input of the regression enables generalization of the learned function in different flows, possibly at different spatial scales. Afterwards, a supervised machine learning method is used to train an approximant $\hat{\beta}(\bm{\eta})$ to $\beta^{\text{opt}}$. Various approaches can be used for that purpose, including shallow neural networks, random forests, and Gaussian processes.
A sketch of the FIML procedure is provided in figure \ref{fig:fiml}.

Although the correction scheme is applied on a few specific models ($k$--$\omega$ or SA model), generalization to additional models (e.g., $k$--$\varepsilon$ model or Reynolds stress transport model) is straightforward. On the other hand, since the corrected model is obtained by perturbing the transport equations in the baseline model, it is still constrained by the limitation of the latter.  For example, if a linear eddy viscosity model is chosen as baseline, the corrected model would still be limited by the Boussinesq assumption.

Several variants and applications of the FIML approach have been proposed in the literature, see for instance \cite{volpiani2021machinePRF,volpiani2022neuralIJHFF,mandler2022realizable,rumsey2022search}. 
Recently, a so-called "model consistent" formulation of FIML to overcome the feature mismatch problem, involving strongly coupled inference of the corrective fields and model coefficients, whereby the candidate models are propagated through the RANS solver to compute modeled output quantities to be used in the cost function \citep{holland2019towards}.

\subsection{Physics-informed correction of Reynolds stress discrepancies}
\cite{wang2017physics-informed,wu2018data-driven}  proposed an alternative approach that directly utilizes training data to construct functions of the discrepancies
(compared to the DNS data) in the RANS-predicted Reynolds stresses and use these functions to predict Reynolds stresses in new flows. The procedure is summarized as follows:
\begin{enumerate}
\item Perform baseline RANS simulations on both the training flows and the test flow.
\item Compute a vector field of well-chosen flow features $\bm{\eta}(\bm{x})$, based
on the RANS-predicted mean flow fields for all flows.
\item Compute a discrepancy tensor field $\Delta\tau(\bm{x})$ in the RANS modeled Reynolds stresses for the
training flows based on the high-fidelity data.
\item Construct regression functions $\widehat{\Delta\tau} : \bm{\eta}\rightarrow \Delta \tau$ for the discrepancies based on the training data
prepared in Step 3.
\item Compute the Reynolds stress discrepancies for the test flow by querying the regression
functions. The Reynolds stresses can subsequently be obtained by correcting the baseline RANS
predictions with the evaluated discrepancies.
\end{enumerate}
In \cite{wang2017physics-informed,wu2018data-driven} , the functions $\widehat{\Delta\tau}$ are obtained by means of Random Forest regressors (RFR), but other models are possible, e.g. Gaussian Processes or Neural Networks. The search space is constrained using prior physical knowledge, in particular about realizability constraints for the RST.

\begin{figure}[!htbp]
  \centering
    \includegraphics[width=0.6\textwidth]{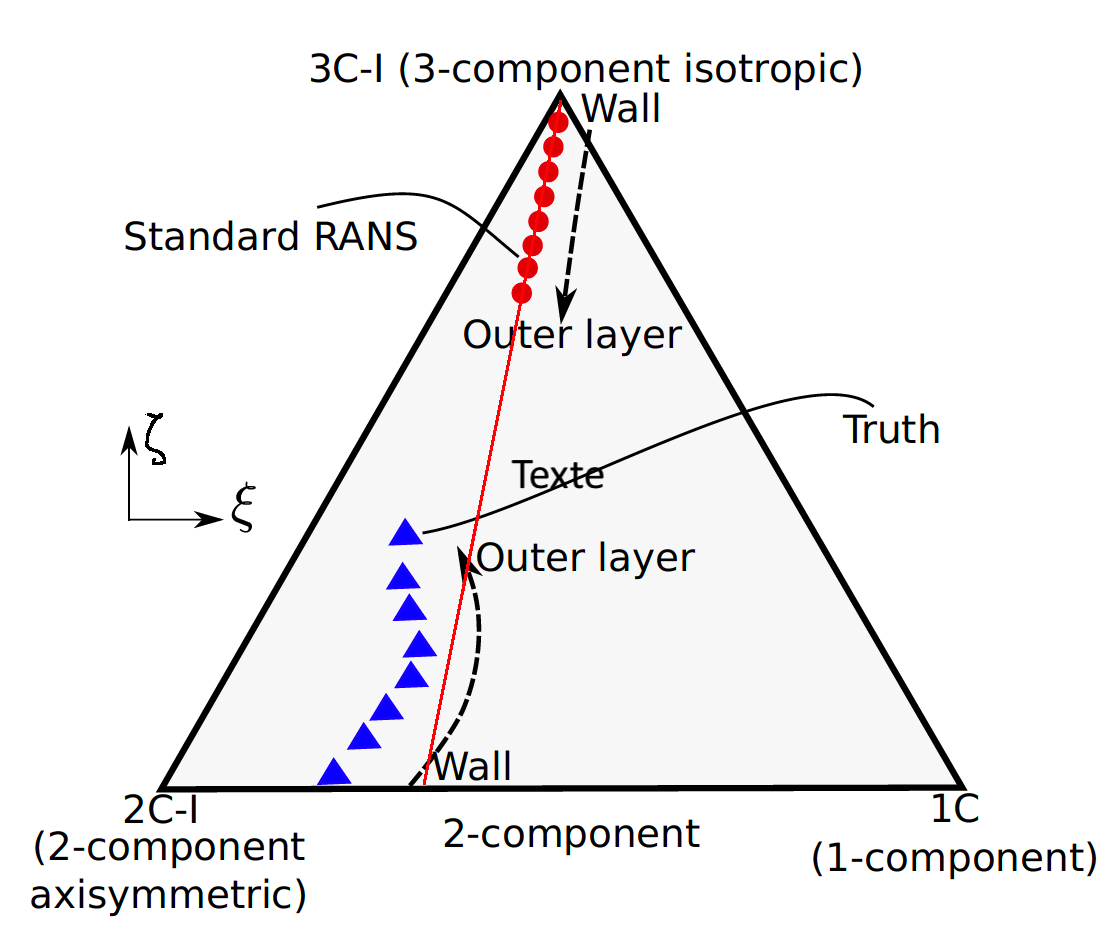}
    \caption{Illustration of the barycentric triangle of physically realizable states of Reynolds stress \citep{banerjee2007presentation}. Turbulence states predicted by Boussinesq 
    RANS models lie near the
isotropic state (vertex 3C-I), and along the plane-strain line (in red), while the actual locations from DNS are near the bottom edge (2C-I) for wall-bounded turbulence (nearly 2D turbulence). The typical
RANS-predicted trend of spatial variation from the wall to shear flow and the corresponding actual trend are
indicated with arrows. Adapted from \cite{wang2017physics-informed}.}
  \label{fig:bary}
\end{figure}

For that purpose, the decomposition scheme in Equation~(\ref{eq:tau-decomp}), can be used as a parameterization scheme for correcting RANS-predicted Reynolds stresses by using machine learning and training data. \citep{wang2017physics-informed,wu2018data-driven} developed a systematic strategy to predict discrepancies in the magnitude, anisotropy, and orientation of the Reynolds stress tensor in terms of an invariant feature set for a set of tensor variables of the mean flow (e.g., $\mathbf{S}$, $\bm{\Omega}$, $\nabla p$, $\nabla k$). 
Transformation of the eigenvalues leads to invariants that can be mapped to the well-known Lumley triangle~\citep{lumley1978computational} or the barycentric triangle~\citep{banerjee2007presentation}, both of which provide a map for all realizable states of turbulence.  Any realizable turbulence state can be mapped to a point within or on the edge of the triangles after the respective transformations.  In the case of the barycentric map, the following linear transformation from eigenvalues $(\lambda_1, \lambda_2, \lambda_3)$ of the anisotropy $\mathbf{a}$ to the barycentric coordinates $(c_1, c_2, c_3)$ is adopted:
\begin{subequations}
  \label{eq:lambda2c}
\begin{align}
  c_1 & = \lambda_1 - \lambda_2 \\
  c_2 & = 2(\lambda_2 - \lambda_3) \\
  c_3 & = 3 \lambda_3 + 1 \ .
\end{align}
\end{subequations}
with $\lambda_1 + \lambda_2 + \lambda_3=0$ for the anisotropy (deviatoric) tensor.
Placing the triangle in a Cartesian coordinate system $\bm{\xi}=(\xi,\zeta)$, the location of any point within
the triangle is a convex combination of those of the three vertices, i.e.,
\begin{equation}
\bm{\xi}=\bm{\xi}_{1c} c_1 + \bm{\xi}_{2c} c_2 + \bm{\xi}_{3c} c_3
\end{equation}
where $\bm{\xi}_{1c}$, $\bm{\xi}_{2c}$ and $\bm{\xi}_{3c}$ denote coordinates of the three vertices of the triangle. An advantage
of representing the anisotropy of Reynolds stress in the barycentric coordinates is that it has a
clear physical interpretation, i.e., the dimensionality of the turbulence state~\citep{banerjee2007presentation}. Typically, the
standard-RANS-predicted Reynolds stress at a near-wall location is located close to the isotropic,
three-component state (vertex 3C-I) in the barycentric triangle, while the true stress is near the
two-component limiting states (bottom edge). The barycentric triangle representation of RST eigenvalues is illustrated in Fig. \ref{fig:bary}.

The next step consists in finding representations of the discrepancies.
Perturbations on $k$ and $\Lambda$ with realizability constraints are represented in terms of their logarithmic discrepancy. Parameterizing the perturbations on the eigenvectors is more challenging due to the need to maintain their \emph{orthonormal} property, which is necessary to ensure that the perturbed Reynolds stresses remain symmetric positive semidefinite tensors. To this end, the perturbation from $\mathbf{V}$ to $\mathbf{V}^\star$ is mapped as a rigid-body rotation, i.e., $\mathbf{V}^\star = Q^\delta \mathbf{V}$ with $Q^\delta$ being an orthonormal rotation matrix representing the perturbation. In fact, a rotation can be represented more compactly by using a set of Euler angles ($\varphi_1, \varphi_2, \varphi_3$). That is, any rigid-body rotation in a three-dimensional space (with a few rare exceptions) can be achieved by the following three consecutive intrinsic rotations about the axes of the local coordinate system. 
In summary, the Reynolds stress tensor is projected to six physically interpretable, Galilean
invariant quantities representing the magnitude ($k$), shape $(\xi,\zeta)$, and orientation ($\varphi_1, \varphi_2, \varphi_3$). They
are collectively denoted as $\Delta \tau$. The actual values of these quantities can be written as baseline RANS predictions corrected by the corresponding discrepancy terms, i.e.,
\begin{eqnarray}
\log k = \log k^\text{RANS}+\Delta(\log k)\\
\xi=\xi^\text{RANS}+\Delta\xi\\
\zeta=\zeta^\text{RANS}+\Delta\zeta\\
\varphi_i=\varphi_i^\text{RANS}+\Delta\varphi_i\quad i=1,2,3
\end{eqnarray}

The target discrepancies $\Delta \tau=(\log k, \xi, \zeta, \varphi_1, \varphi_2, \varphi_3)$, are postprocessed from high-fidelity full-field data. This means that DNS data are needed.
Then the approximant $\widehat{\Delta \tau}=\widehat{\Delta \tau}(\bm{\eta})$ is identified through supervised machine learning.
Due to the \emph{a priori} training and the use of pure DNS data at the training stage, the methodology suffers from the previously mentioned feature mismatch problem. In the attempt of improving model consistency, \citet{wu2018data-driven} proposed learning the linear and nonlinear parts of the Reynolds stress separately, with the linear part treated implicitly to improve model conditioning. 
The reader is referred to that reference, as well as to \cite{wu2018rans} for more details about the ill-conditioning problem.

\subsection{Data-driven RANS modeling through Symbolic Regression}
An interesting alternative is represented by so-called open-box ML approaches, which consist in selecting explicit mathematical expressions and/or operators from a large pre-defined dictionary to build a suitable regressor for the data. The process of systematically determining symbolic equations that fit certain data with an
unknown underlying function, i.e. to find explicit mathematical models that best explain
the relationship between the inputs and the outputs, is called Symbolic Regression (SR).
Unlike standard regression, where data is fit to a pre-defined function, SR attempts to find
both the model structure and the model parameters simultaneously. 
For that purpose, SR searches a space of mathematical expressions to find the model that best fits a given dataset, both in terms of accuracy and simplicity. Since such a problem is NP-hard \citep{virgolin2022symbolic},
the search is restricted to a reduced functional space, possibly making SR less flexible than other machine learning techniques, and specifically ANNs.
However, SR offers
the possibility of not only modeling the data and being able to predict new outputs from new inputs, but also actually grasping why the variables are related in a given way, i.e. it provides more naturally interpretable models. 
Additionally, if the obtained equation is physically sound, the model is likely to extrapolate better than most other mathematical models, especially those based on ANNs, which are often excellent to interpolate but can barely extrapolate outside of their training domains.
In addition to being more interpretable, less of a black box, and likely more generalizable, SR is also attractive in terms of
computational efficiency, ease of integration with existing RANS solvers, and portability, because the discovered analytical expressions can be easily hard-coded within existing numerical solvers by using any computer language at hand.

In the following, we describe two examples of open-box ML for data-driven RANS modeling, based on Genetic Programming (GEP) \citep{weatheritt2016novel} and sparse symbolic identification \citep{schmelzer2020discovery,cherroud2022sparse}. 

\subsubsection{Gene expression programming}
\begin{figure}[t]
\centering
\includegraphics[width=.8\textwidth]{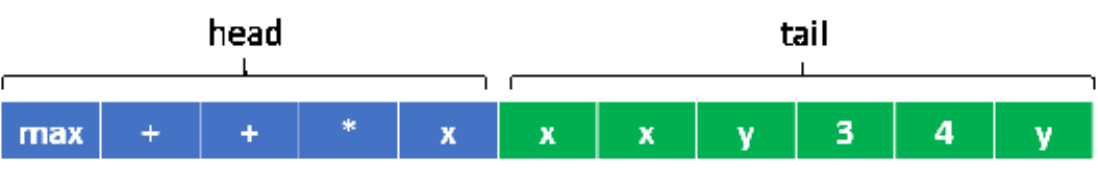}
\caption{The blue section, a string of length 5, contains 4 function operators (here, the max function and the elementary operations of addition and multiplication) and one terminal (the variable $x$). The green section is the tail, and it contains only "terminals", i.e. variables and constants, to which the operators are applied. From \url{https://github.com/ShuhuaGao/geppy}}
\label{fig:gep1}
\end{figure}

\begin{figure}[t]
\centering
\includegraphics[width=.3\textwidth]{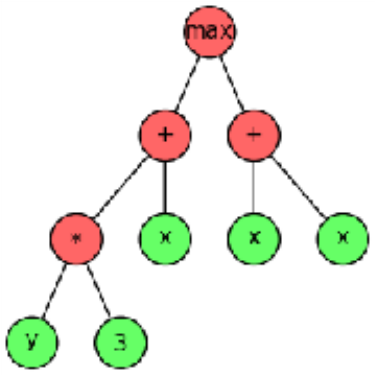}
\caption{Decoding the gene from left to right, we find a max function that takes as inputs the result of two additions. One takes as inputs two terminals, and the second one the result of a multiplication of two other terminals and the head terminal $x$. The last two elements in the string are not used here. From \url{https://github.com/ShuhuaGao/geppy}}
\label{fig:gep2}
\end{figure}

Gene Expression Programming (GEP) was initially introduced by \cite{ferreira2001gep} and coworkers. A recent review can be found in \cite{zhong2017gep}, and  through the online resource  of \url{https://github.com/ShuhuaGao/geppy}.
GEP belongs to the family of genetic algorithms (GAs), since
it uses populations of individuals, selects them according to fitness, and introduces genetic variation using one or more
genetic operators. However, in typical GAs the individuals are encoded as fixed-length strings
(chromosomes) of binary digits or real numbers; in GP the individuals are non-linear entities
of different sizes and shapes (parse trees); and in GEP the
individuals are encoded as symbolic strings of fixed length
(chromosomes) which are then expressed as non-linear entities of different sizes and shapes (expression trees).
Figure \ref{fig:gep1} illustrates a gene of head length 5 and tail length.
The strings contained in a gene constitute his "genotype", and must be translated into a "fenotype", i.e. a mathematical expression.
For that purpose, the gene is treated as a syntax tree,  called an expression tree (ET) in GEP. Specifically, starting from the first position of a gene, an expression tree is constructed according to the number of arguments that the various functions accept.) The expression tree of the gene in the preceding example is shown in figure \ref{fig:gep2}.

Each gene-expression tree is evaluated for its "fitness", i.e. its performance at predicting the desired output value given the data set.
The GEP algorithm starts by randomly creating
$N_\text{pop}$ candidate solutions to form the initial population (of guess functions).
The fitness of each individual is determined, and the fittest individuals are most likely to be selected for reproduction according to a custom probability.
Their chromosomes are then randomly split and recombined. 
A prescribed amount of individuals is modified through different mechanisms, such as mutation.
In this way, the algorithm explores new solutions that combine certain aspects of existing solutions. As a final
step, poor candidate solutions are filtered out in the selection process.
The fitness of the individuals in the new generation is evaluated and  the process is looped until a stopping criterion is satisfied.
\subsubsection{Sparse identification methods}
Sparse regression was introduced by Tibshirani et al. \cite{tibshirani1996regression} in the context of identification of linear models, to mitigate overfitting problems often encountered in high-dimensional regression problems when using ordinary least mean squares (OLMS). The Least Absolute Shrinkage and Selection Operator (LASSO) algorithm shrinks most coefficients in the linear regressor to zero, thus selecting a small subset of relevant terms to explain the data.
The approach can be extended to generalized linear models, i.e. linear combinations of functions $\Phi_m(\mathbf{x})$:
\begin{equation} \label{genlinear}
\mathbf{y}=M(\mathbf{x};\boldsymbol{\theta})=\sum_{m=1}^M \theta_m\Phi_m(\mathbf{x})
\end{equation}
where $\boldsymbol{\theta}=(\theta_1,\theta_2,...,\theta_M)$ is a vector of parameters.
The data $\mathbf{Y}=\left(y(\mathbf{x}_1),\dots,y(\mathbf{x}_N)\right)$ for the problem are regressed as a linear combination of candidate functions (\ref{genlinear}), the coefficients of which are searched for by minimizing the mean square error as in OMLS with a sparsifying regularization term:
\begin{equation} \label{sparse:regression}
{\arg\,\min}_{\boldsymbol{\theta}} \left(\text{MSE}(\boldsymbol{\theta})+\lambda\Omega(\boldsymbol{\theta})\right)
\end{equation}
where $\text{MSE}=||\mathbf{Y}-\boldsymbol{\Phi}\boldsymbol{\theta}||_2^2$ is the mean square error to the data, $\lambda$ is a regularization parameter, $\Omega$ denotes a regularization
term, and the design matrix $\boldsymbol{\Phi}$ corresponds to the collection of the basis functions evaluated at each data point:
\begin{equation}\label{designmatrix}
\boldsymbol{\Phi}=
\begin{bmatrix}
\Phi_1(\mathbf{x}_1) & \ldots & \Phi_M(\mathbf{x}_1)\\
\vdots & \ddots & \vdots\\
\Phi_1(\mathbf{x}_N) & \ldots & \Phi_M(\mathbf{x}_N)
\end{bmatrix}
\end{equation}

Using only the OLMS estimates can lead to a prediction in
which the bias is low and the variance high \cite{tibshirani1996regression}, leading to a poor performance in prediction and
 interpretation \cite{zou2005regularization}. The variance can be reduced by shrinking the coefficients
with the goal of improving the predictive capability and avoid overfitting. However, this tends to increase
the bias, i.e. the MSE. 

There are two reasons to introduce a regularization term, namely, an improvement in prediction accuracy
and interpretation. Using only the OLMS estimates can lead to a prediction in
which the bias is low and the variance high \cite{tibshirani1996regression}, leading to a poor performance in prediction and
 interpretation \cite{zou2005regularization}. The variance can be reduced by shrinking the coefficients
with the goal of improving the predictive capability and avoid overfitting. However, this tends to increase
the bias, i.e. the MSE. 
Several regularization strategies are possible, including LASSO, Ridge, and Elastic net penalizations among others. Another option consists in using a sparse Bayesian formulation \citep{tipping2001sparse}.The interested reader is referred to \cite{bishop2006} for more details. 

\subsubsection{Reynolds stress representation and frozen-RANS procedure}
Both the SR approaches discussed in the following use Pope's representation in Eq. (\ref{bdelta}) for formulating a non-dimensional correction to the RST. Such a quantity, noted $b^{\Delta}_{ij}=(\tau_{ij}^\text{DNS}-\tau_{ij}^\text{BL})/(2k)$ is hereafter expressed for 2D flows. In such cases, only the first three tensors of the minimal integrity bases are linearly independent, and only two
invariants are nonzero:
\begin{equation}\label{bdelta2D}
  b^{\Delta}_{ij}    =   \sum_{n=1}^{3} T_{ij}^{(n)} \alpha_{n}^{\Delta} (I_{1},I_{2}) 
\end{equation}
where :
\begin{equation}
\begin{cases}
\label{tensors}
    \displaystyle T_{ij}^{(1)}  = S_{ij}^* \\
    \displaystyle T_{ij}^{(2)}  = S_{ik}^*\Omega_{kj}^* - \Omega_{ik}^* S_{kj}^* \\
    \displaystyle T_{ij}^{(3)} = S_{ik}^* S_{kj}^* - \frac{1}{3} \delta_{ij} S_{mn}^* S_{mn}^* \\
    \displaystyle I_{1} = S_{mn}^* S_{mn}^* \\
    \displaystyle I_{2} = \Omega_{mn}^* \Omega_{mn}^*
\end{cases}
\end{equation}

The preceding models additionally require the specification of the eddy viscosity coefficient $\nu_t$ and of the turbulent length scales. These are computed by solving the auxiliary transport equations for the RANS model at stake.
To fix ideas, in the following we focus on the $k-\omega$ SST model \citep{menter1992improved}.

Several studies have shown that when Reynolds-stress corrections learned from DNS data are propagated through the RANS equations, output quantities such as the velocity fields are not error-free \cite{duraisamy2021perspectives}. The reason for that is the above-mentioned ill-conditioning of the Navier-Stokes operator.
To improve such ill-conditioning, \cite{schmelzer2020discovery} proposed to correct the linear eddy viscosity and the turbulent time scale $\omega^{-1}$ indirectly, by accounting for model-form errors in the auxiliary turbulent transport equations. This favors the discovery of a model formulation that is more consistent with the RANS solver, thus reducing propagation errors. 
More precisely, the so-called $k$-corrective frozen RANS approach of  \cite{schmelzer2020discovery}, was initially inspired from the frozen-RANS procedure of \cite{weatheritt2016novel} for estimating the turbulent time scale from LES and RANS/LES data, which consists in solving the turbulent transport equations with 
frozen high-fidelity values for all quantities except $\omega$. In the $k$-corrective-frozen RANS, however, an additive corrective term $R$ is added to the transport equation for the turbulent kinetic energy (with its counterpart also added to the equation for $\omega$):
\begin{equation}\label{frozen}
  \begin{cases}
    \displaystyle \frac{\partial k}{\partial t}  +U_{j} \frac{\partial k}{\partial x_{j}} = P_{k}  + R -\beta^{*} k\omega  + \frac{\partial }{\partial x_{j}} \left((\nu
      +\sigma_{k} \nu_{t}) \frac{\partial k}{\partial x_j} \right) \\ 
    \displaystyle  \frac{\partial \omega}{\partial t} +U_{j}  \frac{\partial \omega}{\partial x_j} =  \frac{\gamma}{\nu_{t}}(P_{k} +  R) -\beta^{*}\omega^{2}  +
    \frac{\partial}{\partial x_j} \left((\nu +\sigma_{w} \nu_{t}) \frac{\partial \omega}{\partial x_j} \right) 
  \end{cases}
\end{equation}
In the preceding equations, the production of turbulent kinetic energy is computed by adding to the Boussinesq Reynolds tensor the high-fidelity extra anisotropy:
\begin{equation}
     P_k = \min{\left(  2k\left(-\frac{\nu_t}{k} S_{ij} + b_{ij}^{\Delta,hf}\right) \frac{\partial U_i}{\partial x_j},
      10 \beta^{*} \omega k \right)} 
\end{equation}
with the model constants $\beta^*$,  $\sigma_k$ and  $\sigma_{\omega}$  given in \cite{menter1992improved}. 

In (\ref{frozen}), $k$, $U$, etc. are evaluated using high-fidelity data if an \emph{a priori} training approach is considered; if a \emph{model-consistent} (also \emph{coupled} or \emph{CFD-driven}) approach is used, the equations are iteratively solved alongside the meanflow equations. The two training strategies are discussed in the next Section.

A modeling ansatz for the residual $R$ can obtained by rewriting it in a form similar to the turbulent kinetic energy production \cite{schmelzer2020discovery}:
\begin{equation}
  R\approx 2k b_{ij}^{R} \frac{\partial U_i}{\partial x_j}
\end{equation}
with the fundamental difference that it can take both positive (extra production) and negative (under-production) values. The tensor
$\mathbf{b}^R$ is projected onto the integrity basis previously used for $\mathbf{b}^{\Delta}$ :
\begin{equation}
  b^{R}_{ij} =  \sum_{n=1}^{3} T_{ij}^{(n)} \alpha_{n}^{R} (I_{1},I_{2})
\end{equation}
thus introducing a new set of unknown functions  $\alpha_n^R$  that  are  sought  by  some SR procedure.

\subsection{GEP regression of data-driven RANS closures}
In GEP regression, tangible
mathematical expressions of the tensors $T^{(n)}_{ij}$, and the invariants
$I_k$ are obtained through the GEP evolutionary algorithm. The framework described in the following is based on \cite{weatheritt2016novel,weatheritt2017development}. 

\begin{figure}[t]
\centering
\includegraphics[width=.6\textwidth]{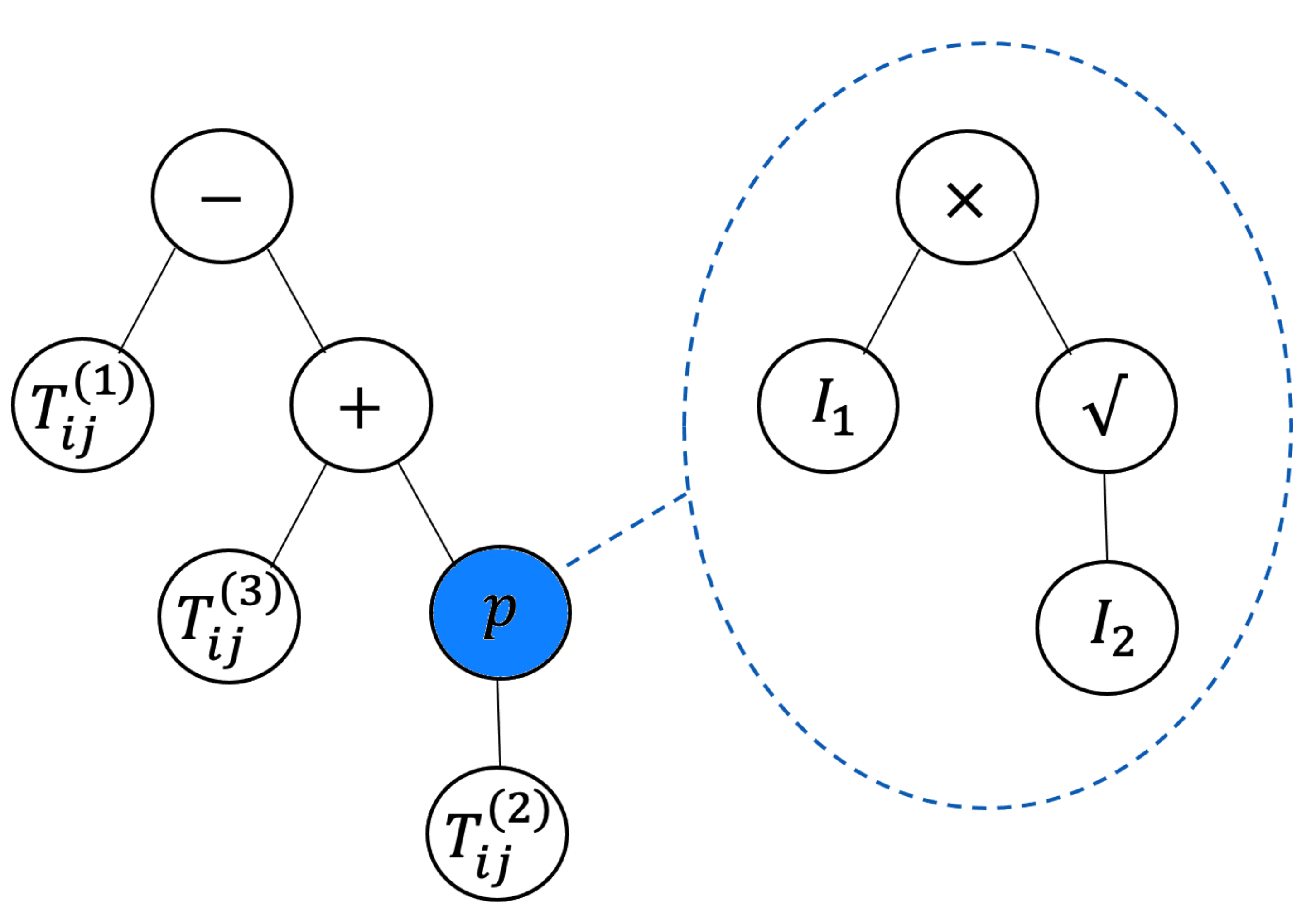}
\caption{Example of expression tree for GEP turbulence modeling.}
\label{gep:tree}
\end{figure}

The first step consists in encoding combinations of the integrity basis tensors and invariants as genes.
To fix ideas, consider an algebraic expression of the form:
\begin{equation}\label{gep:expression}
b^\Delta_{ij}=T^{(1)}_{ij}-\left(T^{(3)}_{ij}+(I_1I_2^{0.5}T^{(2)}_{ij}) \right)
\end{equation}
An expression tree for (\ref{gep:expression}) is represented in Fig. \ref{gep:tree}, which in turn is constructed from two chromosomes of the form:
$$\left(-T^{(1)}_{ij}+T^{(3)}_{ij} p | T^{(2)}_{ij} T^{(3)}_{ij}  T^{(2)}_{ij}  T^{(3)}_{ij}  T^{(2)}_{ij}  T^{(1)}_{ij}  \right)$$
$$\left(*  I_1 Q | I_2 I_2 I_2 I_2 \right)$$
As explained previously, the $|$ separates the head and the tail, while $Q$ represents the square root.
Of note, not all symbols are expressed in the final equation in this case. Specifically, only the first symbols in the tails of both chromosomes are used here.
The first chromosome is a tensor expression, while the second is a scalar, i.e. they do not have the same tensorial order.
The $p$ symbol, introduced and named the plasmid in \cite{weatheritt2016novel} is a symbiotic join between the two chromosomes, corresponding to the scalar/tensor product.
In the algorithm, the populations of tensorial terms and scalar expressions (functions of the invariants) are evolved separately, and joined together via the plasmid.
On initialization of the algorithm, populations of chromosomes are randomly generated, then
evolved according to survival of the fittest. After a given number of generations, the best expression
from the population is taken as the output.
Various options can be considered to express the fitness, which expresses the distance between the output of the candidate model and the training data.
A similar procedure can be used to encode the $R$ correction in the turbulent kinetic energy equation, if any, or the turbulent heat flux model.

\begin{figure}[t]
\centering
\includegraphics[width=.6\textwidth]{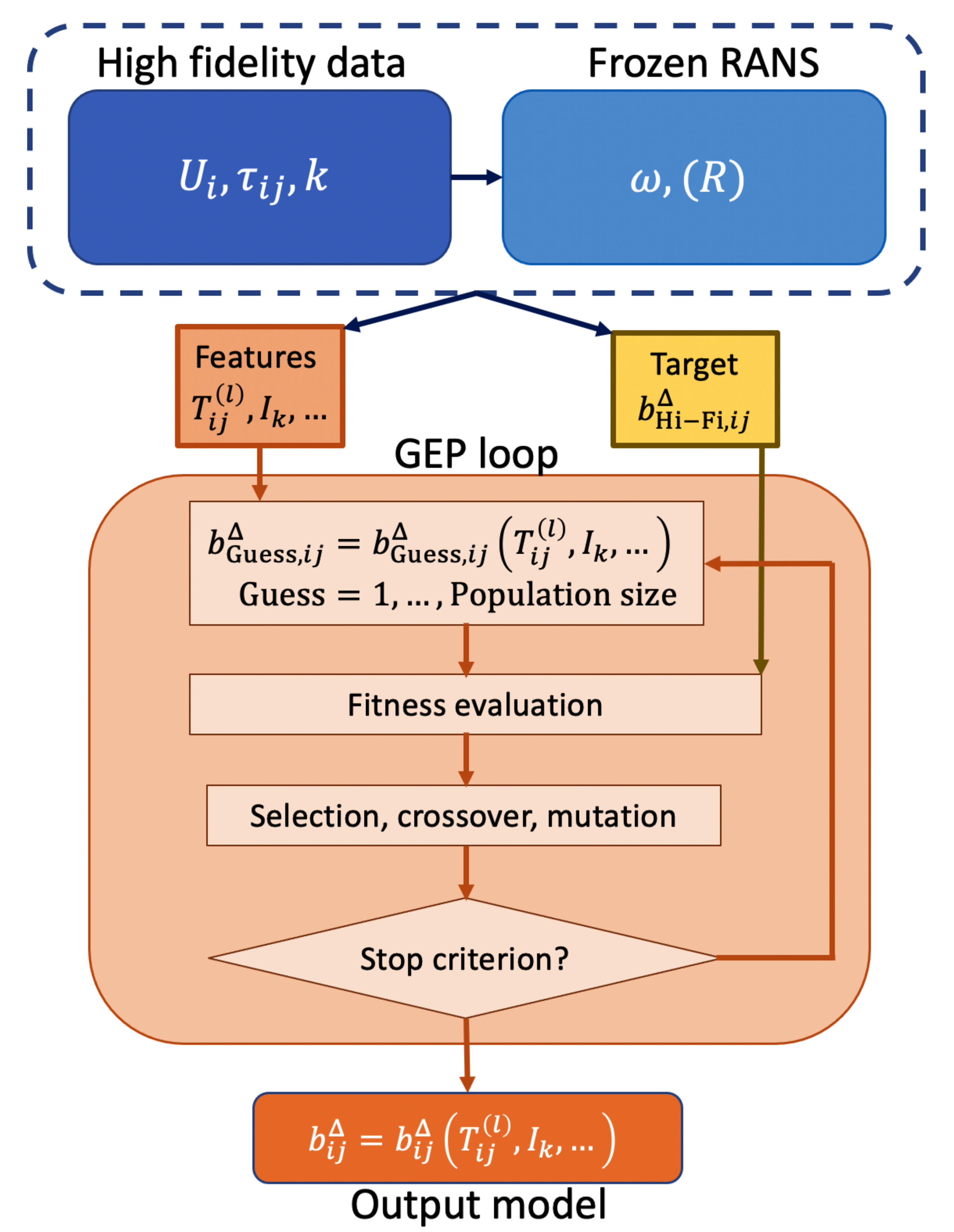}
\caption{Workflow of the GEP procedure for data-driven turbulence modeling. High fidelity data are used to compute the modeled time scale $\omega^{-1}$ by solving the frozen-RANS system. High-fidelity data and the time scale are then used to compute the tensor and invariant features, as well as the target Reynolds stress anisotropy. A population of candidate models is randomly initialized and evolved using genetic operators. When the stop criterion is met, the training loop outputs the optimal model for $\mathbf{b}_{ij}^\Delta$
\label{gep:loop}}
\end{figure}

Figure \ref{gep:loop} summarizes the GEP procedure for learning the extra anisotropy $\mathbf{b}_{ij}^\Delta$.

\subsection{SpaRTA: Sparse regression of Reynolds sTress Anisotropy}
Due to its non-deterministic nature, GEP discovers a different model at each run, as the random initial population changes. The discovered models terms and/or other values for coefficients, with varying
complexity. Weatheritt and Sandberg \cite{weatheritt2017development} observe that the models using only a few nonlinear terms show lower
training and prediction errors and higher numerical robustness that more complex models. 
Additionally, evolutionary algorithms tend to have slower convergence than convex optimization approaches, due to the random nature of the search.

With this in mind, the deterministic SR algorithm SpaRTA (\emph{Sparse Regression of Turbulent Stress Anisotropy}) \cite{schmelzer2020discovery} was introduced
to constrain the search towards sparse algebraic models using sparsity-promoting regression
techniques. SpaRTA combines functions from a predefined library of candidates
without any random recombination. 
The algorithm is based on the following steps:
\begin{enumerate}
\item building a library of candidate
functions,
\item model selection using sparse-regression techniques, 
\item inference of
model coefficients
\item cross-validation of the resulting models
\end{enumerate}
The first three steps are offline, and hend computationally very cheap.

In the following we briefly recall the various steps, referring the reader to \cite{schmelzer2020discovery} for more details.

\paragraph{Function dictionaries}
In SpaRTA, the object of the learning procedure are the function coefficients $to$ and $to$ in the tensor polynomial expansions for 
$\mathbf{b}^\Delta$ and $\mathbf{b}^3$ .

To identify expressions for the $\alpha_n^{\Delta}$  and $\alpha_n^{R}$, a dictionary $\boldsymbol{\mathcal{B}}$ of  functions of the invariants $I_m$  is constructed.
A convenient choice is a dictionary of monomials. A possible choice could be:
\begin{equation}
  \boldsymbol{\mathcal{B}} = \{ I_1^{l},\ I_2^{m},\ I_1^{p}I_2^{q} \ \|  \ 0 \leq l,m \leq 9, \ 2\leq p+q \leq 4 \} 
\end{equation}
leading in this case to   25     candidate terms     for     each     function
($\mathbf{b}^{\Delta}=\{\alpha_n^{\Delta}\}$ and $\mathbf{b}^{R}=\{\alpha_n^R\}$), \emph{i.e.} a total of $25\times 3 \times 2=150$ function candidates.
Other choices are possible, including more terms or different kernel functions (such as radial basis functions).
With such a large number of functional terms used to represent the learning targets, an efficient learning procedure is needed to select a parsimonious model (i.e. a sparse model involving a small subset of functions selected from the initial redundant dictionary) and limit the risk of overfitting the data.

Sparse symbolic regression constructs a large library of nonlinear candidate functions from which a few optimal terms are extracted to
to regress data. Specifically, SpaRTA relies on the fast function extraction (FFX) 
algorithm \cite{McConaghy2011fast}, for which a library is constructed using a
set of raw input variables and mathematical operations. 
The library to regress models for $\mathbf{b}^\Delta$ is constructed by multiplying each
candidate function of $I_m$ in $\boldsymbol{\mathcal{B}}$  with each base tensor $\mathbf{T}^{(n)}$, leading (for the example at hand) to the dictionary of tensorial candidate functions:
\begin{equation}\label{lib:delta}
\mathbf{C}_{b^\Delta}=
\begin{bmatrix}
      T_{11|k=0}^{(1)} & I_1 T_{11|k=0}^{(1)} &  ... & I_1^{2} I_{2}^{2}T_{11|k=0}^{(3)} \\
      T_{11|k=1}^{(1)} & I_1 T_{11|k=1}^{(1)} & ... & I_1^{2} I_{2}^{2}T_{11|k=1}^{(3)} \\
      ... & ... & ... & ... \\
      T_{33|k=K}^{(1)} & I_1 T_{33|k=K}^{(1)} & ... & I_1^{2} I_{2}^{2}T_{33|k=K}^{(3)}
    \end{bmatrix} 
\end{equation}

In order to regress models for $R$, we compute the double dot product of each function in $\boldsymbol{\mathcal{B}}$
with the
mean velocity gradient tensor $\frac{\partial U_i}{\partial x_j}$ leading to the following dictionary of scalar functions:
\begin{equation}\label{lib:R}
\mathbf{C}_{b^R}=
    \begin{bmatrix} 
    T_{ij}^{(1)}  \left.\frac{\partial U_i}{\partial x_j}\right |_{k=0}   &   I_1 T_{ij}^{(1)}  \left.\frac{\partial U_i}{\partial x_j}\right |_{k=0}   &   ...   &
    I_1^{2}I_{2}^{2}T_{ij}^{(3)} \left.\frac{\partial U_i}{\partial x_j}\right |_{k=0} \\
    ... & ... & ... & ... \\
    T_{ij}^{(1)}  \left.\frac{\partial U_i}{\partial x_j}\right |_{k=K}   &   I_1 T_{ij}^{(1)}  \left.\frac{\partial U_i}{\partial x_j}\right |_{k=K}   &   ...   &
    I_1^{2} I_{2}^{2}T_{ij}^{(3)} \left.\frac{\partial U_i}{\partial x_j}\right |_{k=K}
  \end{bmatrix}
\end{equation}

The two libraries $\mathbf{C}_{b^\Delta}$ and $\mathbf{C}_{b^R}$ are evaluated using the high-fidelity data. 
The target data $\mathbf{b}^\Delta$ and $R$ are stacked to vectors. For instance, high-fidelity values of the second order tensor $\mathbf{b}^\Delta$ at each data point are written as:
\begin{equation}\label{lib:R}
\mathbf{B^\Delta}=
    \begin{bmatrix} 
    b_{11|k=0}^\Delta,&
    b_{12|k=0}^\Delta,&
    b_{22|k=0}^\Delta,&
    ...  ,&
     b_{11|k=K}^\Delta,&
    b_{12|k=K}^\Delta,&
    b_{22|k=K}^\Delta
  \end{bmatrix}^T
\end{equation}
and similarly for $\mathbf{B}^R$.
The learning problem can than be written in the compact notation:
\begin{align}\label{sparta:learning}
    \displaystyle \mathbf{C}_{b^{\square}}  \boldsymbol{\theta}_{b^{\square}}= \mathbf{B^{\square}};\quad
    \square = \Delta,R
\end{align}

\paragraph{Model selection}
Sparse symbolic identification  automatically select from the dictionaries $\mathbf{C}_{b^{\square}}$, $\square = \Delta,R$ a small subset of functions that best approximate the data.
This is achieved by performing an elastic net regression, corresponding to the solution of the following optimization problem:
\begin{equation}\label{sparta:enet}
\hat{\boldsymbol{\Theta}}_{b^{\square}}=\underset{\boldsymbol{\theta}_{b^{\square}}}{\arg \,\min} \left(\left|\left|  \mathbf{C}_{b^{\square}}  \boldsymbol{\theta}_{b^{\square}} - \mathbf{B^{\square}} \right|\right|_2^2 +\lambda\rho ||\boldsymbol{\theta}_{b^{\square}}||_1 + \frac{1}{2}\lambda(1-\rho)||\boldsymbol{\theta}_{b^{\square}} ||_2^2\right)
\end{equation}
The elastic net regularization is carried out for a set of candidate values for the blending parameter $\rho$:
$$\rho = [0.01, 0.1, 0.2, 0.5, 0.7, 0.9, 0.95, 0.99, 1.0]^T$$
and
$$\lambda = [\lambda_0, ..., \lambda_{max}]^T$$
(see \cite{schmelzer2020discovery} for more details).
At each grid point of the net, $(\lambda_m, \rho_n )$ a vector
$\hat{\boldsymbol{\theta}}_{b^{\square}|mn}$ of parameter values is discovered.
It may happen that different $(\lambda, \rho)$ combinations might produce parameter vectors with the same null elements.
The selection step then concludes by 
filtering out the set of ${\cal D}$ unique abstract model forms:
$${\cal D}_{\square}=\left\{
\hat{\boldsymbol{\theta}}^d_{b^{\square}}|d=1,...D
\right\}$$

\paragraph{Model inference}
The parameters of the abstract models ${\cal D}_{\square}$ are recalibrated by performing an additional Ridge regression, which guarantees convexity while ensuring small-sized parameters that best fit to the data.
Large coefficients are more likely to induce numerical robustness problems when propagated through the CFD solver.
More precisely the inference step consists in solving:
\begin{equation}\label{sparta:ridge}
\hat{\boldsymbol{\theta}}^{s,d}_{b^{\square}}=\underset{\boldsymbol{\theta}^{s,d}_{b^{\square}}}{\arg \,\min} \left(\left|\left|  \mathbf{C}_{b^{\square}}  \boldsymbol{\theta}^{s,d}_{b^{\square}} - \mathbf{B^{\square}} \right|\right|_2^2 +\lambda_r||\boldsymbol{\theta}^{s,d}_{b^{\square}} ||_2^2\right)
\end{equation}
with $\lambda_r$ is the Tikhonov-regularisation parameter. The index $s$ denotes the submatrix of
$\mathbf{C}_{b^{\square}}$ and the subvector of $\boldsymbol{\theta}^{d}_{b^{\square}}$ consisting of the selected columns or elements respectively as
defined in ${\cal D}_{\square}$. 
The optimal value of $\lambda_r$ is found through cross validation, i.e. by solving (\ref{sparta:ridge}) for a set of values of $\lambda_r$ and comparing the corresponding MSE on some test dataset. 
This could be done by splitting the $\mathbf{B^\Delta}$ and $\mathbf{B^R}$ data sets into a training and a test subset. However, small deviations of the discovered model with respect to the data can translate in large errors in output quantities such as the velocity or pressure fields, due to the nonlinear nature of the RANS equations. Additionally the \emph{a priori} training does not provide any guarantee that the discovered model is numerically robust when propagated through the RANS system.
In order not to overcharge the training data and to select models that are robust when propagated through the CFD solver, 
the cross validation step in SpaRTA is performed \emph{a posteriori}, by propagating the first the candidate models through the RANS equations, and by computing the MSE on a set of output QoIs.
This procedure is of course more expensive that \emph{a priori} cross validation, because of the CFD solves required. However, it allows to select more accurate and robust models.

\begin{figure}[t]
\centering
\includegraphics[width=.6\textwidth]{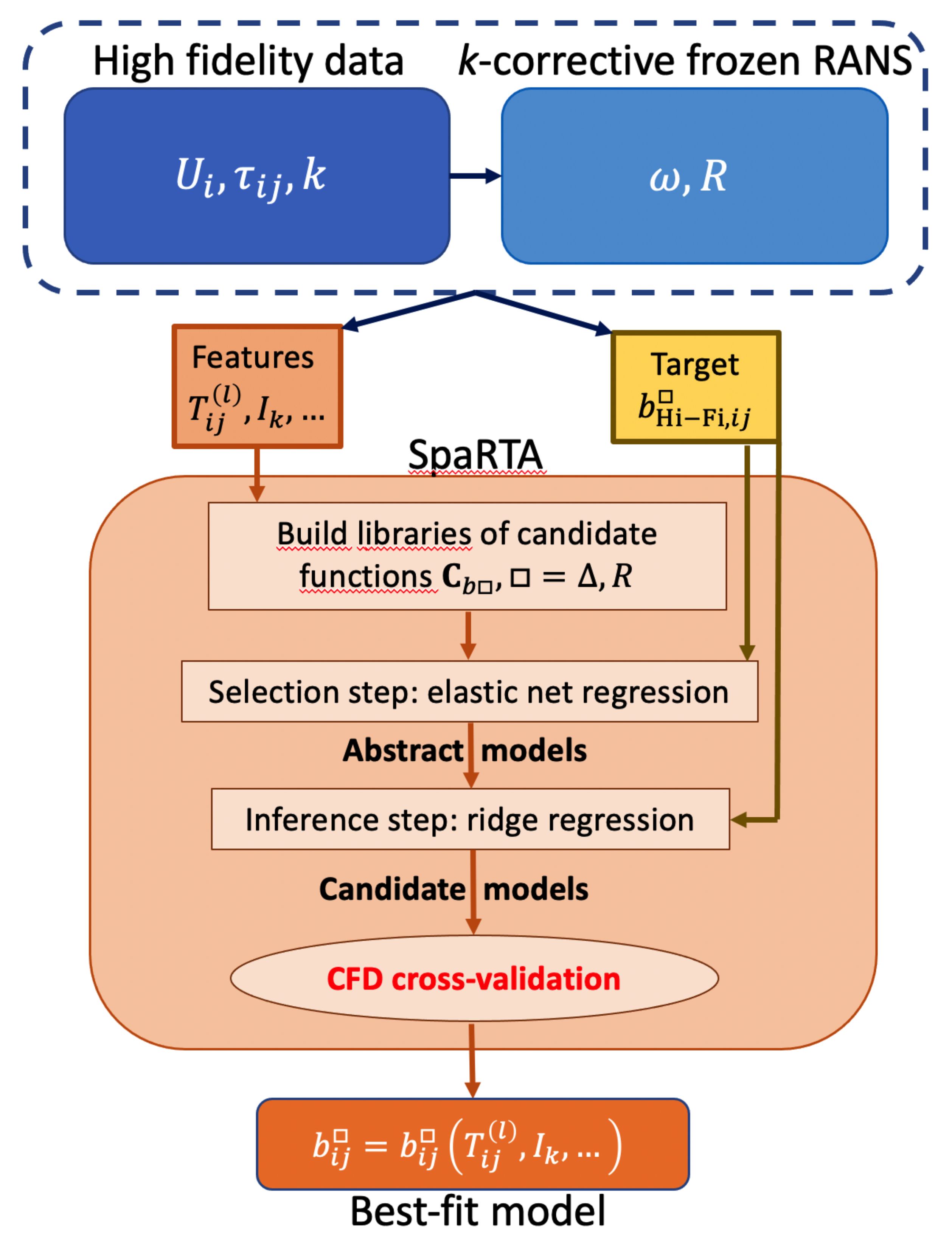}
\caption{Summary of the SpaRTA procedure for data-driven turbulence modeling.}
\label{fig:sparta}
\end{figure}

Figure \ref{fig:sparta} summarizes the various steps of the SpaRTA algorithm.

\paragraph{SBL-SpaRTA}
The classic SpaRTA approach requires separate selection and inference step. Although each step is performed \emph{a priori} and hence is inexpensive, the algorithm may lead to a large number of candidate models for the cross-validation step.
Recently, Cherroud et al. \cite{cherroud2022sparse} introduced a Bayesian SR approach for data-driven turbulence modeling, based on the Sparse Bayesian Learning algorithm initially introduced by Tipping \cite{tipping2001sparse}
for support vector machines. The latter basically consists in the Bayesian counterpart of SpaRTA (SBL-SpaRTA) and, similarly to SpaRTA, it provides sparse and interpretable models.
However, the Bayesian formulation has several advantages:
\begin{itemize}
\item It potentially better accounts for noisy/sparse data due to the regularizing effect of Bayesian priors
\item It allows for model selection and inference simultaneously, thanks to the Bayesian "Occam razor", i.e. model selection during inference, based on model evidence \cite{tipping2001sparse} 
\item It provides full posterior distributions of the inferred model parameters, thus enabling uncertainty quantification
\end{itemize}
We refer the reader to  \cite{cherroud2022sparse} for more information.

\section{Sample results}
In the following we report some results exemplifying the performance of a data-driven RANS model. We chose the case of SBL-SpaRTA, but rather similar behaviors can be expected from other data-driven strategies.

The SBL-SpaRTA is trained for different datasets, listed in Tab. \ref{tab:training_cases}. Different symbolic expressions for the RST and turbulent kinetic energy equation discrepancies are discovered, that equal or outperform the baseline $k-\omega$ SST model on the training set. Specifically,
\begin{itemize}
\item The discovered model coincides with the baseline for channel flow configurations, showing that the latter has  been trained properly to capture velocity profiles for such kind of simple wall-bounded shear flow. Rather similar results are obtained for flat plates.
\item For jet flows, the discovered model tends to decrease the turbulent kinetic production and the eddy viscosity, allowing to palliate the so-called round/plane jet anomaly of LEVM.
\item For separated flows, the discovered model introduces some nonlinear corrections (extra anisotropy) of the RST while increasing turbulent kinetic energy production. The extra production contributes to reduce the size of the recirculation bubble, which tends to be overestimated by the baseline model.
\end{itemize}

\begin{figure}
        \centering
        \includegraphics[width=0.48\linewidth]{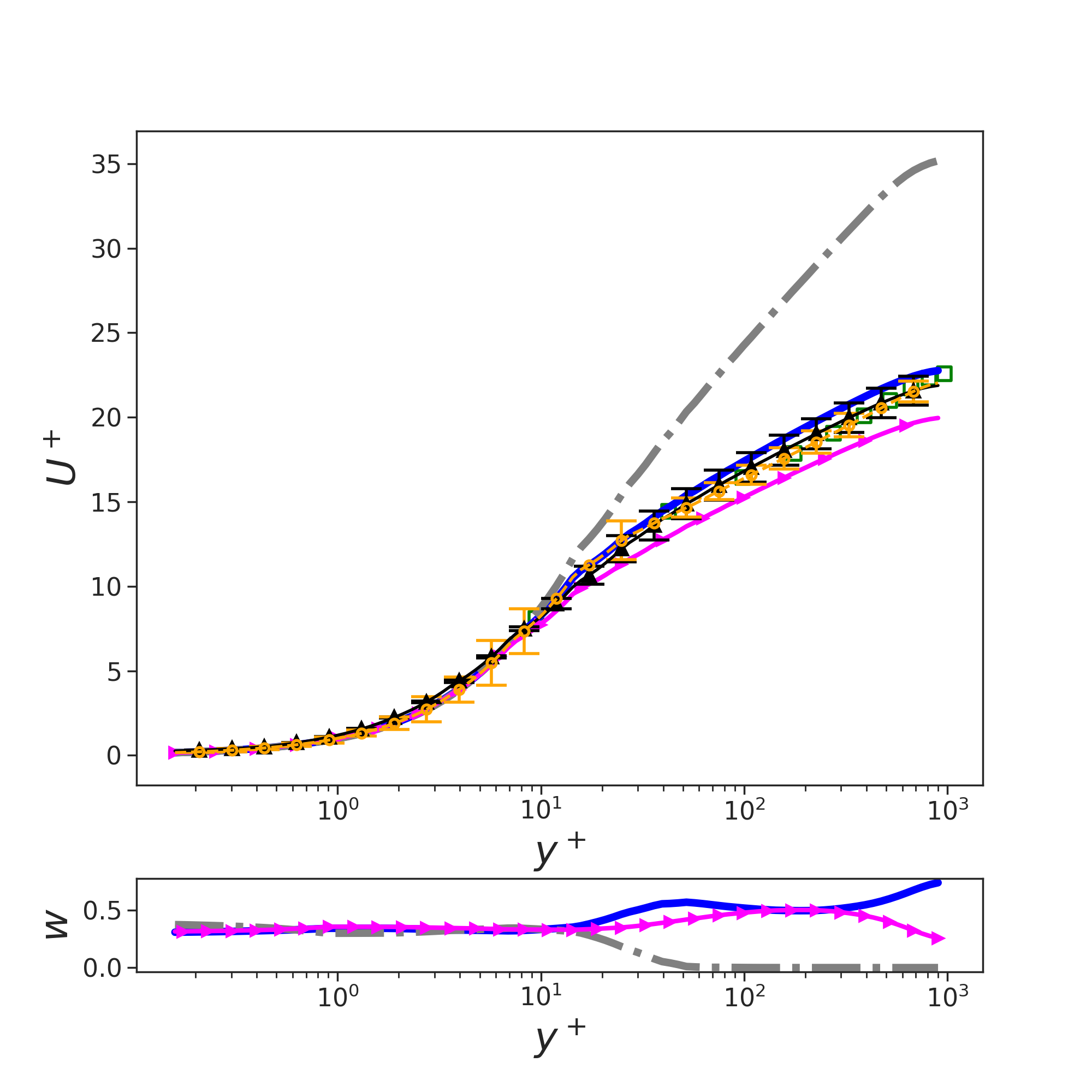}\vspace{-0.5cm}
        \includegraphics[width=0.48\linewidth]{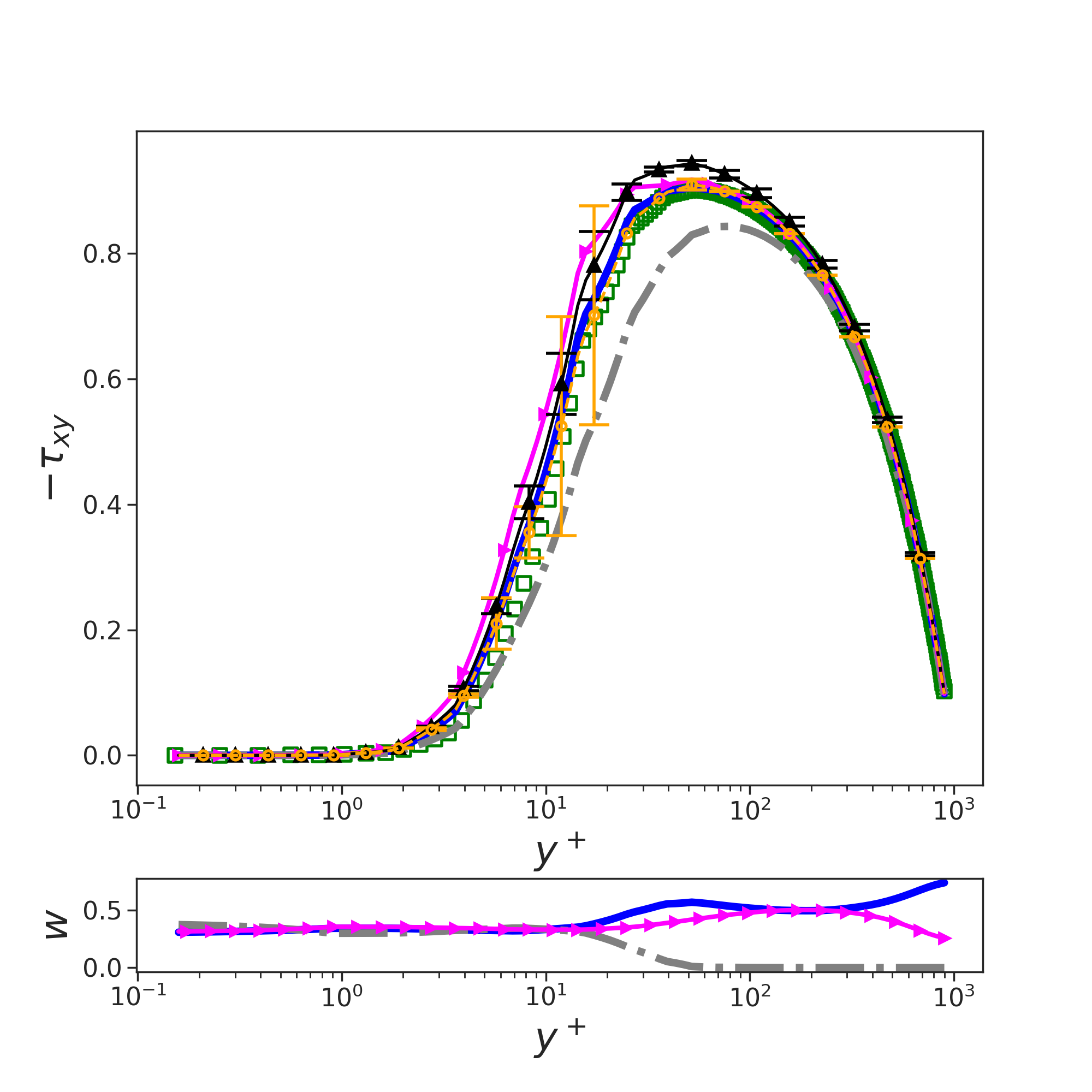}
\caption{$U^+$ vs. $y^+$ and $\tau_{xy}^+$ vs. $y^+$ and the corresponding model weights $w$ for the periodic channel flow case at $Re_\tau = 1000$. Baseline $k-\omega$ SST  (---); $\bm{M}^{(ANSJ)}$ ($-\cdot -$); $\bm{M}^{(SEP)}$ ($-\triangle-$); High-fidelity data {\color{green} ($\square$)}; Non-intrusive X-MA (orange with error bars); Intrusive X-MA (black with error bars).}
    \label{fig:CHAN_plots}
\end{figure}

\begin{figure}
    \centering
        \includegraphics[width=\linewidth, trim=0 15 0 15, clip]{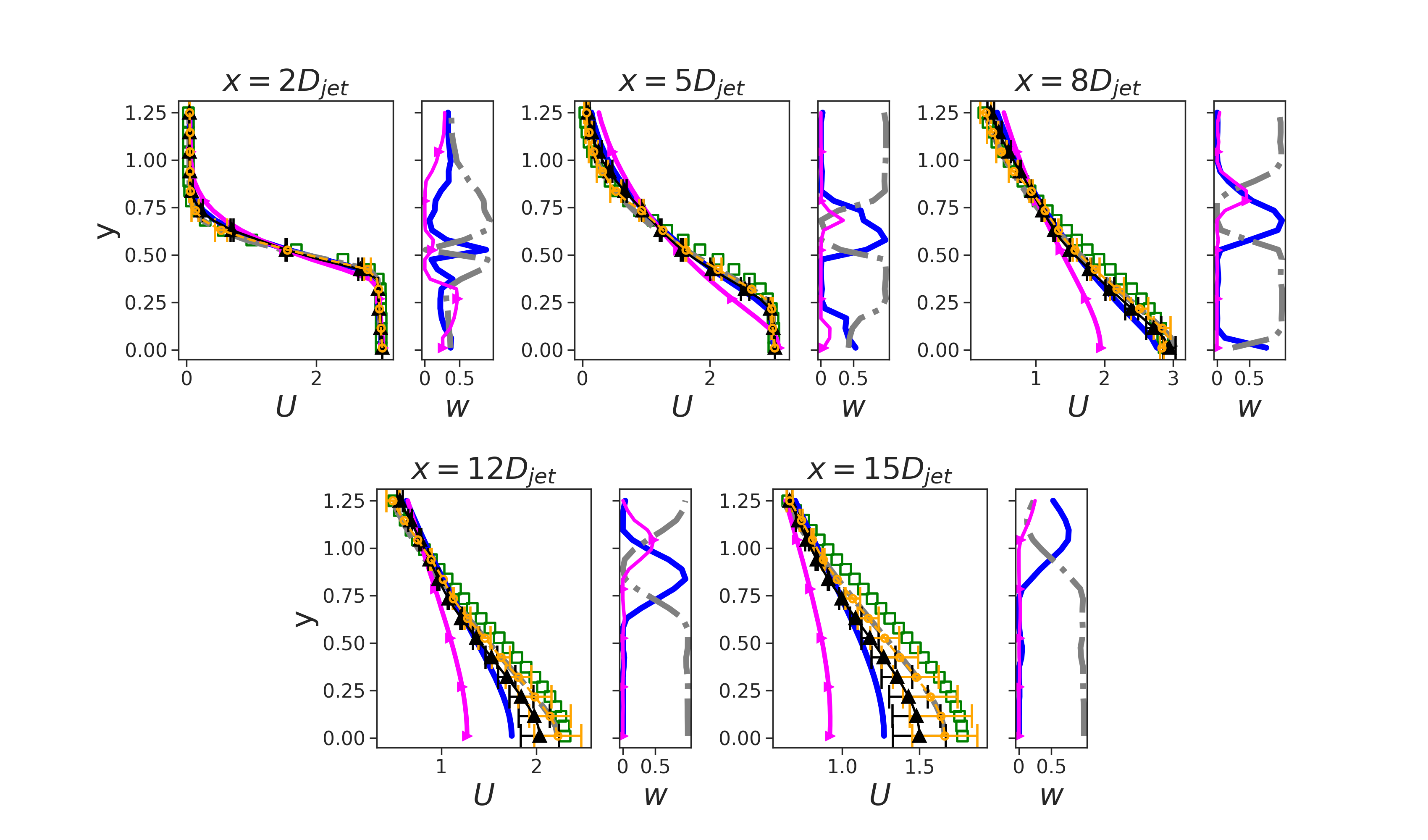}
        \includegraphics[width=\linewidth, trim=0 15 0 15, clip]{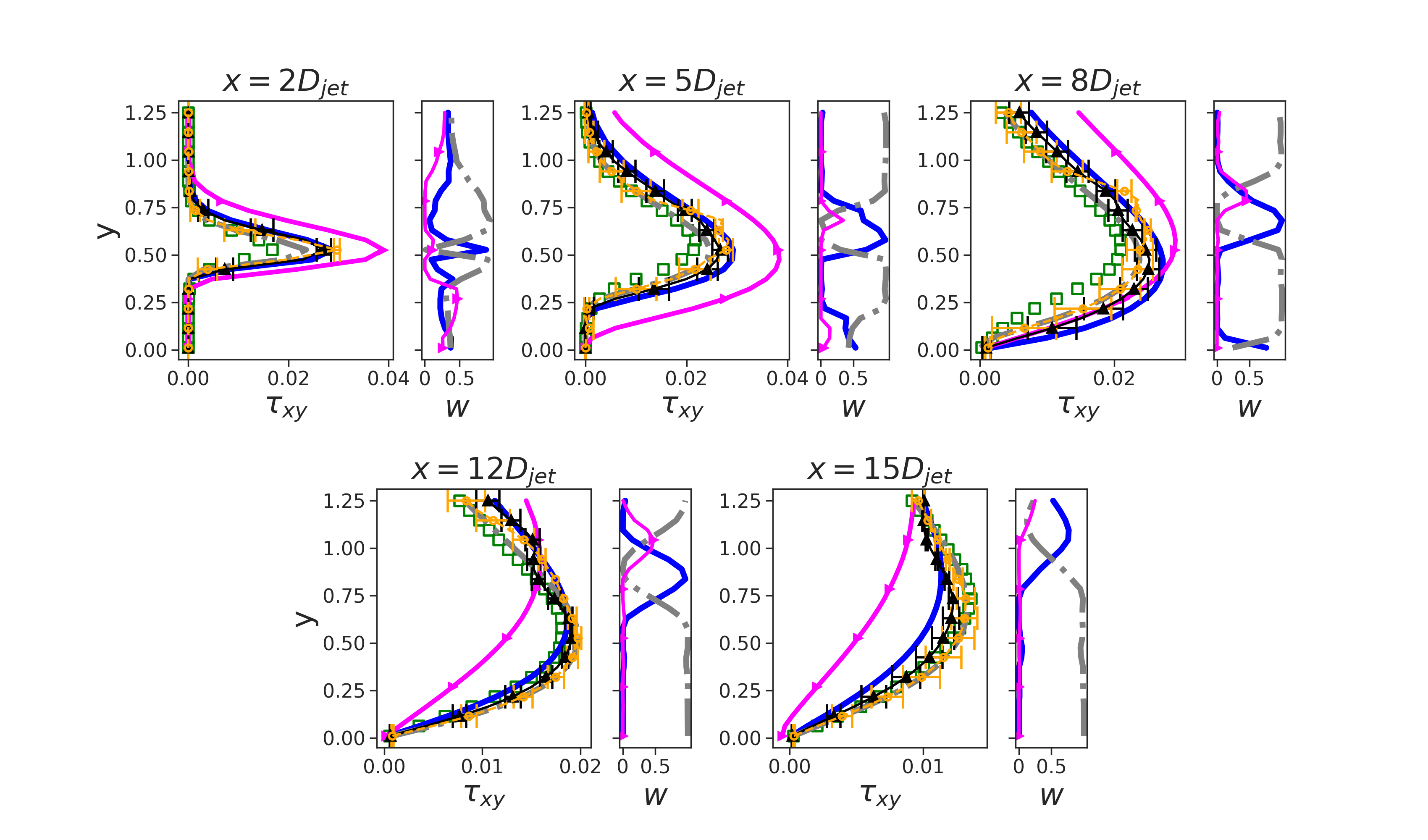}
    \caption{Horizontal velocity $U$ and Reynolds shear stresses $\tau_{xy}$ at various $x/D_{jet}$ positions for the ANSJ flow case. Baseline $k-\omega$ SST  (---); $\bm{M}^{(ANSJ)}$ ($-\cdot -$); $\bm{M}^{(SEP)}$ ($-\triangle-$); High-fidelity data {\color{green} ($\square$)}; Non-intrusive X-MA (orange with error bars); Intrusive X-MA (black with error bars).}\label{fig:ANSJ_plots}
\end{figure}

\begin{figure}
    \centering
        \includegraphics[width=1.\linewidth, trim= 75 0 110 0, clip]{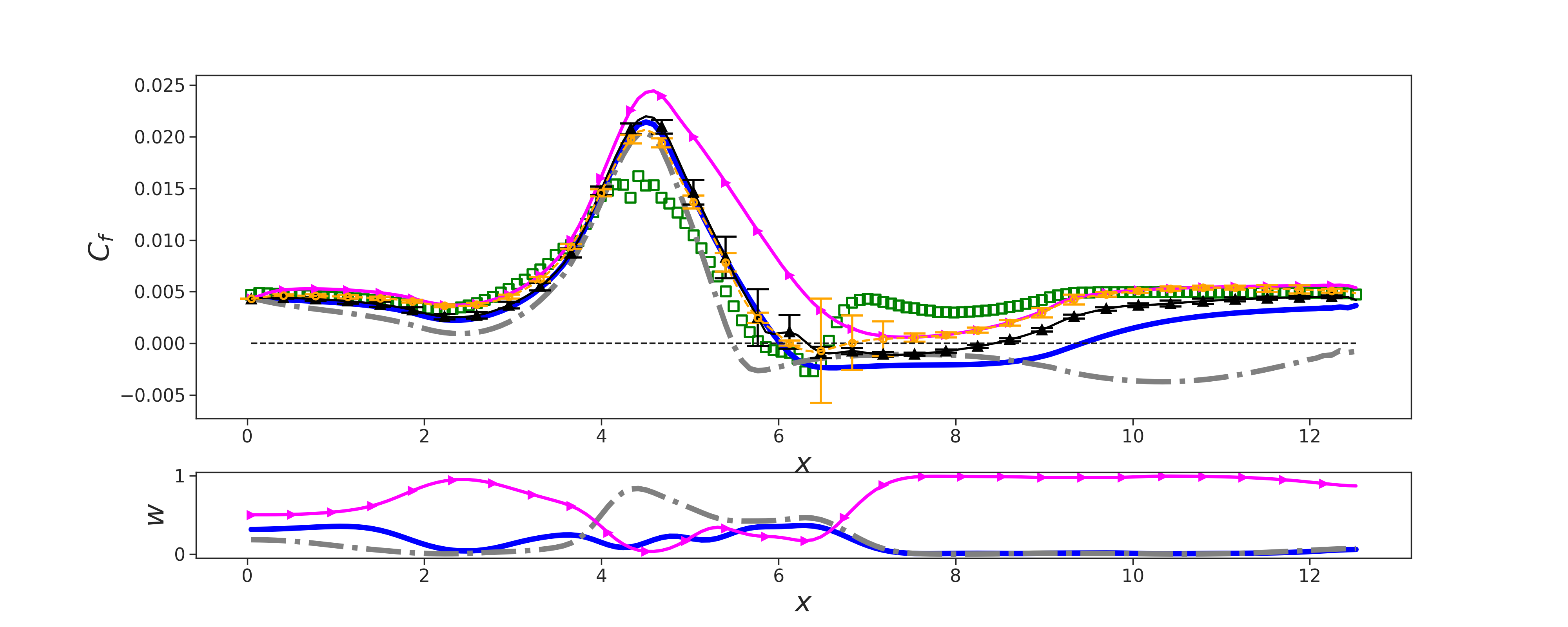}
        \includegraphics[width=\linewidth]{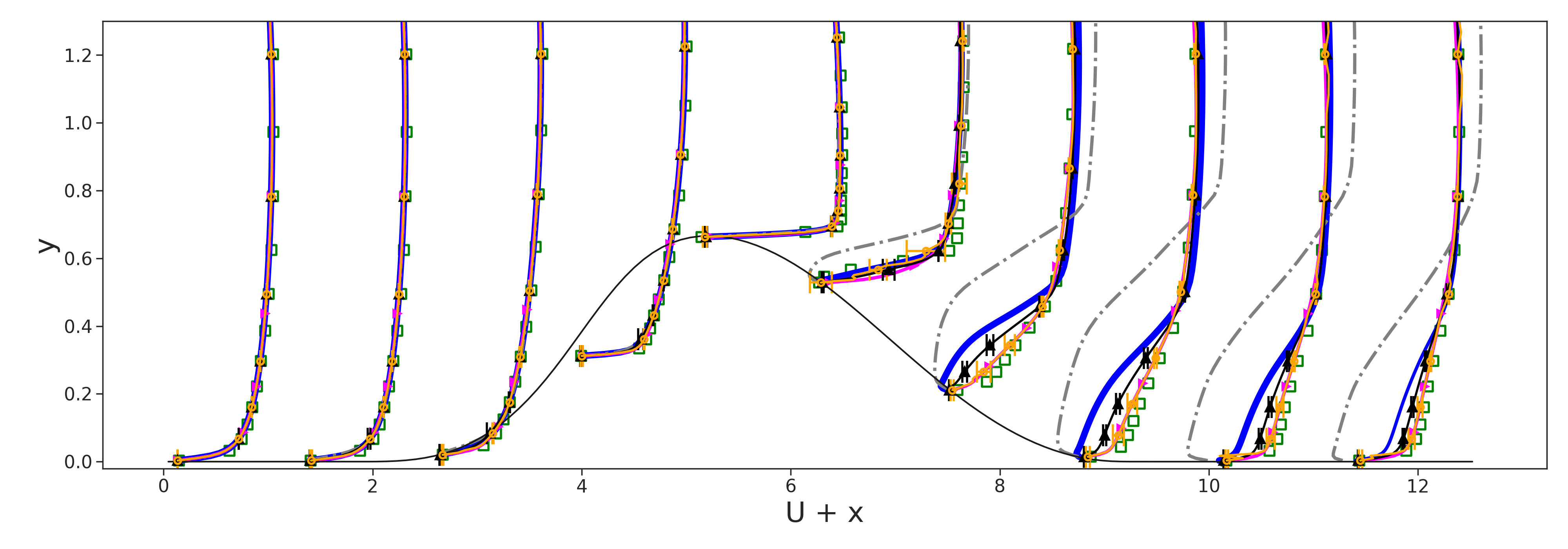}
        \includegraphics[width=\linewidth]{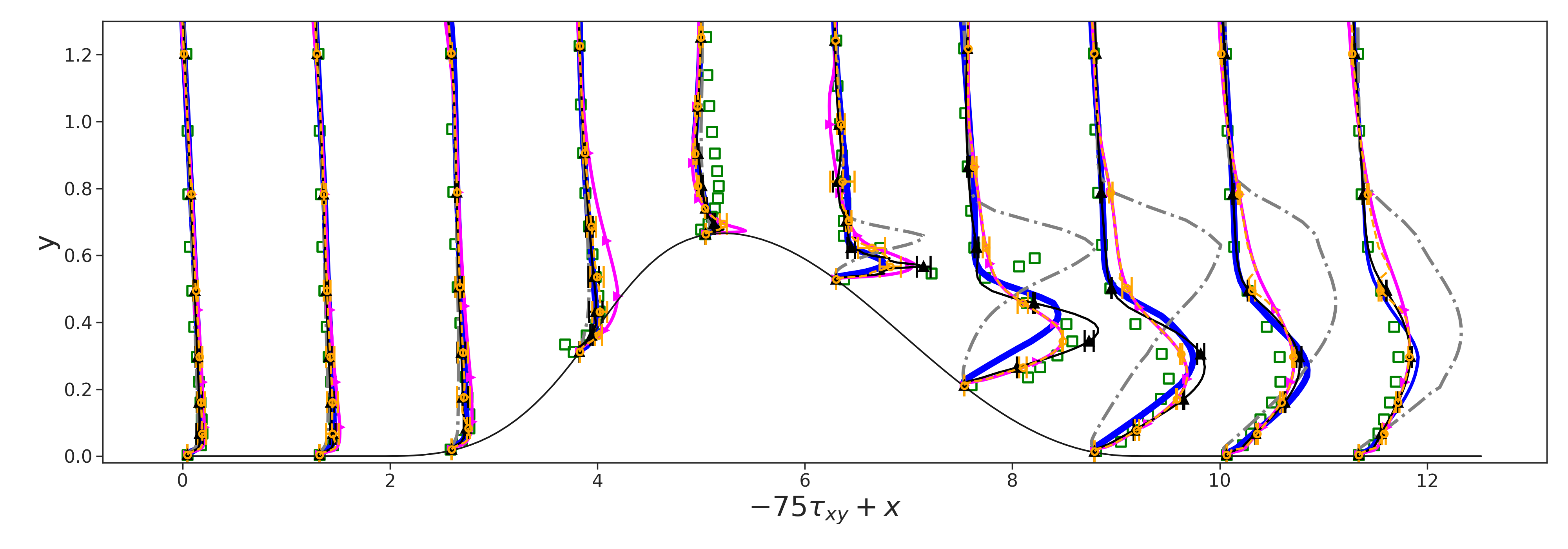}
 \caption{Horizontal velocity $U$ and Reynolds shear stresses $\tau_{xy}$ at various $x$ positions for the CD flow case. Baseline $k-\omega$ SST  (---); $\bm{M}^{(ANSJ)}$ ($-\cdot -$); $\bm{M}^{(SEP)}$ ($-\triangle-$); High-fidelity data {\color{green} ($\square$)}; Non-intrusive X-MA (orange with error bars); Intrusive X-MA (black with error bars).}\label{fig:SEP_plots}
\end{figure}

\begin{table}
 \centering
 {\footnotesize
\begin{tabular}{ c | c | c}
Case & Description & Source \\
\hline
ZPG & DNS of a zero pressure gradient turbulent boundary layer &\cite{schlatter2011progress} \\
& $ 670 \leq Re_{\theta} \leq 4060 $& \\
\hline
CHAN & DNS of turbulent channel flows & \cite{moser1999direct} \\
& $ 180 \leq Re_{\tau} \leq 5000 $& \cite{lee2019data} \\
\hline
APG &LES of adverse pressure-gradient TBL & \cite{bobke2017history} \\
& $Re_{\theta} \leq 4000$, $\beta \leq 4$, 5 different pressure gradients & \\
\hline
ANSJ & PIV of near sonic axisymmetric jet & \cite{bridges2010establishing} \\
\hline
SEP & LES of Periodic Hills (PH) at $Re=10595$ & \cite{breuer2009flow}\\
&DNS of converging-diverging channel (CD) at $Re=13600$ & \cite{laval2011direct} \\
&LES of curved backward facing step (CBFS) at $Re = 13700$ & \cite{bentaleb2012large} \\
\end{tabular}}
\caption{List of flow cases used to train customized SBL-SpaRTA corrections.} \label{tab:training_cases}
\end{table}
Results for the above-mentioned cases are reported in the figures \ref{fig:CHAN_plots}, \ref{fig:ANSJ_plots}, \ref{fig:SEP_plots}.
The specialized models learned for the various datasets are applied to the the incompressible turbulent channel flow, the near-sonic jet flow, and a separated flow, namely, the convergent-divergent channel (refer to \cite{cherroud2023space} for more information about the test cases).
The results show that the specialized models perform well for the flow classes they were trained for, but they are not better or worst than the baseline for radically different flows.
In the machine learning language, we say that the learned models do not generalize well outside their respective training sets.


Various authors have tried to improve the generalization capability of data-driven RANS, in the quest for a "universal" RANS model.
\cite{holland2019field} proposed a variant of the FIML approach where they trained their model simultaneously against multiple datasets using model-consistent training. The approach was applied to a class of airfoil flows with various angles of attack.
A similar idea has been recently explored by \cite{fang2023toward} within the GEP approach, where the learning requires the solution of a multiobjective optimization problem, each objective being the error with respect to a different dataset. The authors also enrich the set of features used to evaluate the function coefficients in Pope's representation.
Recently, \cite{rincon2023progressive} have a proposed a methodology for the progressive augmentation of data-driven models, along with a model-consistent training strategy. The approach successfully corrects model discrepancies for flows in noncircular ducts or for separated flows while preserving the attached boundary layer. However, such flow configurations do not have competing requirements. It would be interesting to see if the approach can simultaneously preserve accuracy for separated and jet flows simultaneously.

In the following, we discuss an approach using machine learning to generate a data-driven model mixture \citep{MZB2023}, with the intent of applying the best model to the right place in the flow.

\paragraph{Model aggregation methods}
In the attempt of learning a more general data-driven model that performs well over a diverse set of flow configurations, \cite{cherroud2023space} proposed using mixture of expert techniques, based on the space-dependent model aggregation (XMA) methodology of \cite{MZB2023}. The general idea consists in training so-called "weighting" or "gating" functions alongside the specialized models, which attribute high-scores to well-performing models and lower scores to ill-performing ones, based on a set of local flow features. Ideally, the weights should promote specialized models in their region of expertise, and penalizing them in regions where they have to extrapolate outside their training set. An aggregated solution is obtained as a convex linear combination of a set of competing experts (i.e. the sum of the weights is equal to one). The weights play the role of local model probabilities, reminiscent of Bayesian averaging methods. The strategy can be implemented in a non-intrusive or non-intrusive manner.
In the first case, predictions of a new flow are first produced by using a set of competing data-driven RANS models, and then the solutions are aggregated using XMA. Discrepancies among models can be used to estimate local uncertainty bounds on the reconstructed solution.
In the second case, a set of competing data-driven corrections for the RST and turbulent kinetic energy equation are first aggregated using XMA. Afterwards, the aggregated model (or blended model) is used to predict a new case.
The first option tends to provide more accurate results, but is more expensive (although it provides information on predictive uncertainties), and the aggregated prediction is not a solution of the NS equations.
The first option is cheaper and satisfies the conservation equations. However, due to transport effects (the corrections applied at some point are convected downstream), the solution is slightly less accurate than in non-intrusive XMA.

Let us consider $K$ SBL-SpaRTA models, learned in different environments, and let $d(\mathbf{x})$ be any Quantity of Interest (QoI) predictable as an output of a RANS flow solver at some spatial location $\mathbf{x}$ (\emph{e.g.} the predicted velocity or pressure fields, the skin friction distribution, etc.). 
In order to make predictions of $d$ that are robust to the choice of the data-driven turbulence model for an unseen flow scenario, we leverage the "Mixture-of-Experts" concept \citep{yuksel_twenty_2012} used in ensemble machine learning, and we build an ensemble solution by aggregating the individual solutions $d_k$ of the $K$ SBL-SpaRTA models:
\begin{equation}
d_\text{XMA}(\mathbf{x}) = \sum_{k=1}^K w_k(\mathbf{x})d_k(\mathbf{x})
\end{equation}
In the above, $w_k(\mathbf{\mathbf{x}})$ is the weighting function assigned to the $k^{th}$ component model, and $d_\text{XMA}(\mathbf{x})$ is the model ensemble or aggregated prediction. Following the approach of \cite{MZB2023}, we look for weighting functions satisfying the conditions:
\begin{equation}
\begin{dcases}
0\leq w_k(\mathbf{x})\leq 1\quad\forall k=1,...,K \\
\sum_{k=1}^K w_k(\mathbf{x})=1\quad\forall \mathbf{x}
\end{dcases}
\end{equation}
Given the preceding properties, they can be interpreted as the probability of model $k$ to contribute to the aggregated prediction $d_\text{XMA}(\mathbf{x})$ at location $\mathbf{x}$. Such an approach, called non-intrusive XMA, can be

On the other hand, the intrusive formulation of the space-dependent model aggregation algorithm (blended data-driven model) writes:
\begin{equation*}
  \begin{dcases}
   - \tau_{ij}    =     2k    \left(    \frac{1}{3}    \delta_{ij}     +b_{ij}^0    +    \sum_{k=1}^K
      w_k(\boldsymbol{\eta}(\mathbf{x})) b_{ij}^{\Delta^{(k)}}\right) \\ 
    \displaystyle P_k  = \displaystyle \min{\left(2\nu_t S^2  -2k\left(\sum_{k=1}^K w_k(\boldsymbol{\eta}(\mathbf{x}))
          b_{ij}^{\Delta^{(k)}}\right) \frac{\partial U_i}{\partial x_j}, 10 \beta^{*} \omega k \right)} \\ 
    R   =   2k   \left(\sum_{k=1}^K  w_k(\boldsymbol{\eta}(\mathbf{x}))   b_{ij}^{R^{(k)}}   \right)
    {\partial U_i}{\partial x_j} 
  \end{dcases}
\end{equation*}
where  $b_{ij}^0 $ is the baseline Boussinesq model, and  $b_{ij}^{\Delta^{(k)}}$   and  $b_{ij}^{R^{(k)}}$  are  data-driven   corrections  to  the $k^{\text{th}}$ SBL-SpaRTA model.  The weighting functions are learned offline in both the non-intrusive and the intrusive methods:
\begin{enumerate}
\item First, the exact values  of weight functions are calculated at certain points within
  the  domain where  high-fidelity  data are  available by  using  Equations \eqref{Eq:weights} rewritten below:
  \begin{equation}
    \begin{dcases}
      w_k(\delta^{(k)}(\bm{x});\bar{\delta}(\bm{x});\sigma_{w})                                            =
      \frac{g_k(\delta^{(k)}(\bm{x});\bar{\delta}(\bm{x});\sigma_{w})}{\displaystyle \sum_{l=1}^{K}
        g_l(\delta^{(l)}(\bm{x});\bar{\delta}(\bm{x});\sigma_{w})} \\
      g_k\bigl(\delta^{(k)}(\mathbf{x});\bar{\delta}(\mathbf{x});\sigma_{w}\bigr)  =   \exp  \left(-
        \frac{\Bigl(\delta^{(k)}(\mathbf{x})                                                       -
          \bar{\delta}(\mathbf{x})\Bigr)^T.\Bigl(\delta^{(k)}(\mathbf{x})                          -
          \bar{\delta}(\mathbf{x})\Bigr)}{\sqrt{   Var(  \bar{\delta}   )  }   \quad  \times   \quad
          2\sigma_w^2} \right)
          \label{Eq:weights}
    \end{dcases}
  \end{equation}
\item Second, a machine learning algorithm is used to regress the relationship
  \begin{equation}
    \underbrace{\boldsymbol{\eta}(\mathbf{x})                                                      =
      \bigl(\eta_1(\mathbf{x}),...,\eta_{11}(\mathbf{x})\bigr)}_{\text{local     flow     features}}
    \xrightarrow[\mathcal{W}]{ML} 
    \underbrace{\left(w_1\bigl(\delta^{(1)}(\mathbf{x});\bar{\delta}(\mathbf{x});\sigma_{w}\bigr),...,w_K
        \bigl(\delta^{(K)}(\mathbf{x});\bar{\delta}(\mathbf{x});\sigma_{w}\bigr)
      \right)}_{\text{local models weights}} \label{eq:MLweights}
  \end{equation}
  based on the values computed in the first stage, $\boldsymbol{\eta}$ still being a set of features taken from \cite{ling2015evaluation}.
\end{enumerate}
The features at each point in the flow domain are computed using the baseline $k-\omega$ SST model. The model weights depend on the hyperparameter $\sigma_w$. Both the local flow features and the model weights are computed for the training flow cases presented in Table \ref{tab:training_cases_ML}:
\begin{table}[h]
{\footnotesize
 \centering
 \begin{tabular}{ c | c | c}
  Case & Description & Source \\
  \hline
  CHAN & DNS of turbulent channel flows & \cite{moser1999direct} \\
  & $ 180 \leq Re_{\tau} \leq 5000 $  & \cite{lee2019data} \\
  \hline
  ANSJ & PIV of near sonic axisymmetric jet & \cite{bridges2010establishing} \\
  \hline
  SEP & LES of Periodic Hills (PH) at $Re=10595$ & \cite{breuer2009flow}  \\
  &  DNS of converging-diverging channel (CD) at $Re=13600$ & \cite{laval2011direct} \\
  &  LES of curved backward facing step (CBFS) at $Re = 13700$ & \cite{bentaleb2012large}   \\
 \end{tabular}}
 \caption{List of data used to train the ML regressor of model weights. \label{tab:training_cases_ML}}
\end{table}

Finally, we also show some results for a complex flow outside the training set, namely, the NASA Wall-Mounted hump Reynolds $80 \times 10^6$ (based on the hump height). This values is much higher than the characteristic values encountered for separated flows in the training set (which are order $10^4$). Therefore, this test case entails an extrapolation in both geometry and Reynolds number. Additionally, the flow features a significant portion with attached boundary layer upstream of the hump peak.

\begin{figure}
    \centering
        \includegraphics[width=0.48\linewidth]{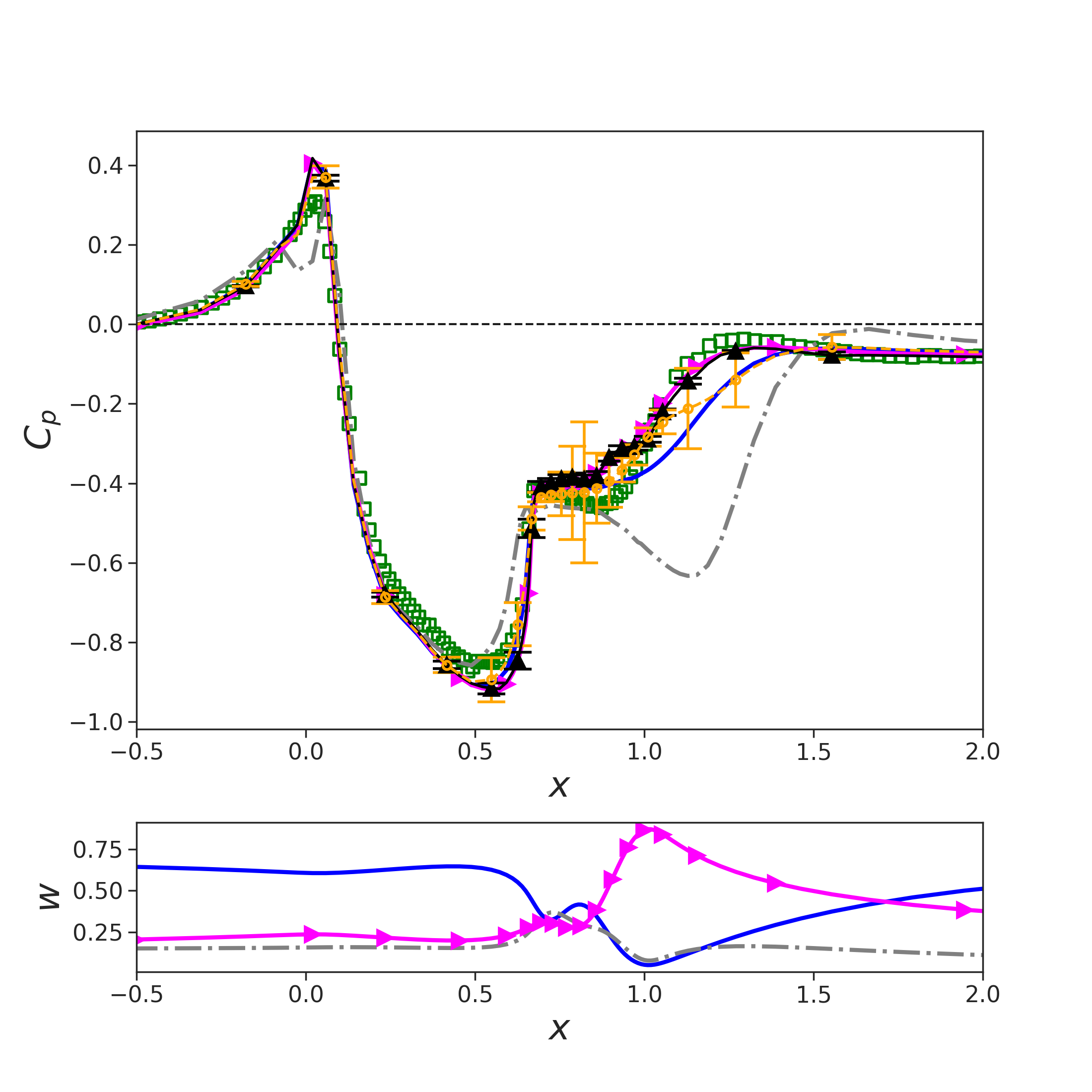}
     \includegraphics[width=0.48\linewidth]{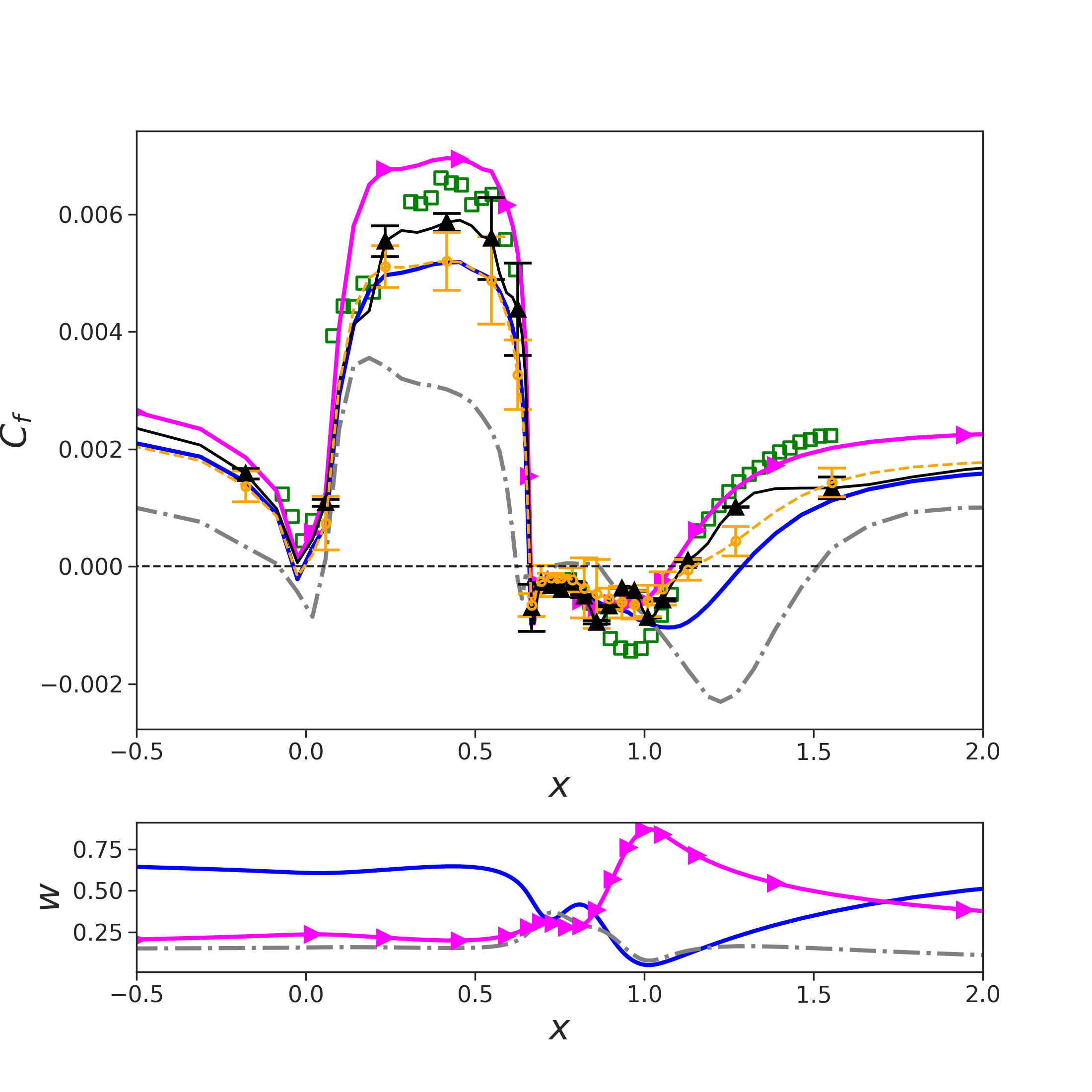}
    \caption{Pressure coefficient $C_p$ and friction coefficient $C_f$ along $x$ axis for the WMH flow case.  Baseline $k-\omega$ SST  (---); $\bm{M}^{(ANSJ)}$ ($-\cdot -$); $\bm{M}^{(SEP)}$ ($-\triangle-$); High-fidelity data {\color{green} ($\square$)}; Non-intrusive X-MA (orange with error bars); Intrusive X-MA (black with error bars).}\label{fig:WMH_Cp_Cf_GPR}
\end{figure}

The pressure and friction coefficients are displayed in Figure \ref{fig:WMH_Cp_Cf_GPR}, along with the model weights across the wall. The regions preceding separation are predominantly influenced by the baseline model, while the $\bm{M}^{(SEP)}$ and $\bm{M}^{(ANSJ)}$ models are assigned similar moderate weights. In this case, this benefits to the intrusive XMA prediction, where the locally received turbulent kinetic energy production is improved.

Downstream of the reattachment point, the $\bm{M}^{(SEP)}$ correction becomes predominant. Consequently, both  intrusive and non-intrusive XMA predictions of skin friction accurately depict the size and location of the recirculation bubble. Moving further downstream, the baseline model regains its influence, while the impact of $\bm{M}^{(SEP)}$ diminishes at the expense of the baseline model. This decrease in $\bm{M}^{(SEP)}$'s contribution is not immediately reflected in the intrusive X-MA predictions, primarily due to transport effects. 

\section{Conclusions and research trends} 
\label{sec:conclusion}

This course note summarized some of the most largely used techniques for data-driven turbulence modeling. We focused on RANS models, because they are expected to remain the workhorse tool for industrial CFD simulations in decades to come.  The ambitious goal of data-driven modeling would be to radically improve the predictive accuracy of RANS without incurring the significant computational overcost of scale-resolving simulations (LES or hybrid RANS/LES).

The field has been rapidly evolving field in the past decade. 

A number of tools is now available and working reasonably well, at least for simple flow problem and relatively narrow classes of flows. However, generalizability of the learned models is still an open question.
Several remedies have been explored, including extended sets of features, multi-case training, and multi-model ensemble learning. 
Some of them exhibit promise, and may contribute progress toward generalizable data-driven models that consistently perform better (or not worst) than classic physics-based models.

However, several issues are still open. Some of them are:
\begin{itemize}
\item inexpensive and robust model-consistent (with RANS solver in the training loop) training
\item choice and normalization of the features used to train the models
\item choice of the training data: kind, quantity
\item extension to fully 3D cases, for which standard RANS models have not been fine-tuned
\item benchmarking and documentation of the existing approaches
\item generalization to a sufficiently wide range of flows of interest for engineering applications
\item quantification of uncertainties associated with the data-driven models, especially when they are used in a purely predictive setting
\end{itemize}

\clearpage

\section*{References}

\bibliographystyle{apalike}
\bibliography{rans-uq-review}

\begin{thebibliography}{}

\bibitem[Amarloo et~al., 2023]{amarloo2023data}
Amarloo, A., Cinnella, P., Iosifidis, A., Forooghi, P., and Abkar, M. (2023).
\newblock Data-driven reynolds stress models based on the frozen treatment of
  reynolds stress tensor and reynolds force vector.
\newblock {\em Physics of Fluids}, 35(7).

\bibitem[Amarloo et~al., 2022]{amarloo2022frozen}
Amarloo, A., Forooghi, P., and Abkar, M. (2022).
\newblock Frozen propagation of {Reynolds} force vector from high-fidelity data
  into {Reynolds}-averaged simulations of secondary flows.
\newblock {\em Phys. Fluids}, 34(11):115102.

\bibitem[Arnst et~al., 2010]{Arnst2010}
Arnst, M., Ghanem, R., and Soize, C. (2010).
\newblock Identification of bayesian posteriors for coefficients of chaos
  expansions.
\newblock {\em Journal of Computational Physics}, 229(9):3134--3154.

\bibitem[Avdonin and Polifke, 2018]{avdonin2018quantification}
Avdonin, A. and Polifke, W. (2018).
\newblock Quantification of the impact of uncertainties in operating conditions
  on the flame transfer function with non-intrusive polynomial chaos expansion.
\newblock In {\em ASME Turbo Expo: Power for Land, Sea, and Air, Volume 4A:
  Combustion, Fuels, and Emissions}, page V04AT04A030. ASME.
\newblock Paper GT2018-75476.

\bibitem[Baldwin and Lomax, 1978]{baldwin1978thin}
Baldwin, B. and Lomax, H. (1978).
\newblock Thin layer approximation and algebraic model for separated turbulent
  flows.
\newblock AIAA Paper 78-257.

\bibitem[Banerjee et~al., 2007]{banerjee2007presentation}
Banerjee, S., Krahl, R., Durst, F., and Zenger, C. (2007).
\newblock Presentation of anisotropy properties of turbulence, invariants
  versus eigenvalue approaches.
\newblock {\em Journal of Turbulence}, 8(32):N32.

\bibitem[Beetham and Capecelatro, 2020]{beetham2020formulating}
Beetham, S. and Capecelatro, J. (2020).
\newblock Formulating turbulence closures using sparse regression with embedded
  form invariance.
\newblock {\em Physical Review Fluids}, 5(8):084611.

\bibitem[Bentaleb et~al., 2012]{bentaleb2012large}
Bentaleb, Y., Lardeau, S., and Leschziner, M.~A. (2012).
\newblock Large-eddy simulation of turbulent boundary layer separation from a
  rounded step.
\newblock {\em Journal of Turbulence}, (13):N4.

\bibitem[Berrone and Oberto, 2022]{berrone2022invariances}
Berrone, S. and Oberto, D. (2022).
\newblock An invariances-preserving vector basis neural network for the closure
  of {Reynolds}-averaged {Navier}--{Stokes} equations by the divergence of the
  {Reynolds} stress tensor.
\newblock {\em Phys. Fluids}, 34(9):095136.

\bibitem[Bishop, 2006]{bishop2006}
Bishop, C.~M. (2006).
\newblock {\em Pattern Recognition and Machine Learning}.
\newblock Springer.

\bibitem[Bobke et~al., 2017]{bobke2017history}
Bobke, A., Vinuesa, R., {\"O}rl{\"u}, R., and Schlatter, P. (2017).
\newblock History effects and near equilibrium in adverse-pressure-gradient
  turbulent boundary layers.
\newblock {\em Journal of Fluid Mechanics}, 820:667--692.

\bibitem[Brener et~al., 2024]{brener2024highly}
Brener, B.~P., Cruz, M.~A., Macedo, M.~S., and Thompson, R.~L. (2024).
\newblock A highly accurate strategy for data-driven turbulence modeling.
\newblock {\em Computational and Applied Mathematics}, 43(1):59.

\bibitem[Brener et~al., 2021]{brener2021conditioning}
Brener, B.~P., Cruz, M.~A., Thompson, R.~L., and Anjos, R.~P. (2021).
\newblock Conditioning and accurate solutions of {R}eynolds average
  {N}avier--{S}tokes equations with data-driven turbulence closures.
\newblock {\em J. Fluid Mech.}, 915.

\bibitem[Breuer et~al., 2009]{breuer2009flow}
Breuer, M., Peller, N., Rapp, C., and Manhart, M. (2009).
\newblock Flow over periodic hills--numerical and experimental study in a wide
  range of \text{Reynolds} numbers.
\newblock {\em Computers \& Fluids}, 38(2):433--457.

\bibitem[Bridges and Wernet, 2010]{bridges2010establishing}
Bridges, J. and Wernet, M. (2010).
\newblock Establishing consensus turbulence statistics for hot subsonic jets.
\newblock In {\em 16th AIAA/CEAS aeroacoustics conference}, page 3751.

\bibitem[Cabot and Moin, 2000]{cabot2000approximate}
Cabot, W. and Moin, P. (2000).
\newblock Approximate wall boundary conditions in the large-eddy simulation of
  high {R}eynolds number flow.
\newblock {\em Flow, Turbulence and Combustion}, 63(1-4):269--291.

\bibitem[Chaouat, 2017]{chaouat2017state}
Chaouat, B. (2017).
\newblock The state of the art of hybrid rans/les modeling for the simulation
  of turbulent flows.
\newblock {\em Flow, Turbulence and Combustion}, 99:279--327.

\bibitem[Cherroud et~al., 2022]{cherroud2022sparse}
Cherroud, S., Merle, X., Cinnella, P., and Gloerfelt, X. (2022).
\newblock Sparse bayesian learning of explicit algebraic reynolds-stress models
  for turbulent separated flows.
\newblock {\em International Journal of Heat and Fluid Flow}, 98:109047.

\bibitem[Cherroud et~al., 2023]{cherroud2023space}
Cherroud, S., Merle, X., Cinnella, P., and Gloerfelt, X. (2023).
\newblock Space-dependent aggregation of data-driven turbulence models.
\newblock {\em arXiv preprint arXiv:2306.16996}.

\bibitem[Cheung et~al., 2011]{cheung2011bayesian}
Cheung, S.~H., Oliver, T.~A., Prudencio, E.~E., Prudhomme, S., and Moser, R.~D.
  (2011).
\newblock Bayesian uncertainty analysis with applications to turbulence
  modeling.
\newblock {\em Reliability Engineering \& System Safety}, 96(9):1137--1149.

\bibitem[Chien, 1982]{chien1982predictions}
Chien, K.-Y. (1982).
\newblock Predictions of channel and boundary-layer flows with a
  low-{Reynolds}-number turbulence model.
\newblock {\em AIAA Journal}, 20(1):33--38.

\bibitem[Cinnella et~al., 2016]{cinnella2016review}
Cinnella, P., Dwight, R., and Edeling, W.~N. (2016).
\newblock Review of uncertainty quantification in turbulence modelling to date.
\newblock Minisymposium ``UQ in turbulence modelling'', SIAM Uncertainty
  Quantification conference, Lausanne, Switzerland. 5-8 April.
\newblock doi: 10.13140/RG.2.1.4512.5523.

\bibitem[Cruz et~al., 2019]{cruz2019use}
Cruz, M.~A., Thompson, R.~L., Sampaio, L.~E., and Bacchi, R.~D. (2019).
\newblock The use of the {R}eynolds force vector in a physics informed machine
  learning approach for predictive turbulence modeling.
\newblock {\em Comput. Fluids}, 192:104258.

\bibitem[de~Zordo-Banliat et~al., 2020]{MZB2023}
de~Zordo-Banliat, M., Dergham, G., Merle, X., and Cinnella, P. (2020).
\newblock Space-dependent turbulence model aggregation using machine learning.
\newblock {\em Journal of Computational Physics}, 497:112628.

\bibitem[{de Zordo-Banliat} et~al., 2020]{MZB_2020}
{de Zordo-Banliat}, M., Merle, X., Dergham, G., and Cinnella, P. (2020).
\newblock Bayesian model-scenario averaged predictions of compressor cascade
  flows under uncertain turbulence models.
\newblock {\em Computers \& Fluids}, 201:104473.

\bibitem[{de Zordo-Banliat} et~al., 2022]{MZB_2022}
{de Zordo-Banliat}, M., Merle, X., Dergham, G., and Cinnella, P. (2022).
\newblock Estimates of turbulence modeling uncertainties in naca65 cascade flow
  predictions by bayesian model-scenario averaging.
\newblock {\em International Journal of Numerical Methods for Heat \& Fluid
  Flow}, 32:1398--1414.

\bibitem[Diomede et~al., 2008]{diomede2008discharge}
Diomede, T., Davolio, S., Marsigli, C., Miglietta, M., Moscatello, A., Papetti,
  P., Paccagnella, T., Buzzi, A., and Malguzzi, P. (2008).
\newblock Discharge prediction based on multi-model precipitation forecasts.
\newblock {\em Meteorology and Atmospheric Physics}, 101(3-4):245--265.

\bibitem[Draper, 1995]{draper1995assessment}
Draper, D. (1995).
\newblock Assessment and propagation of model uncertainty.
\newblock {\em Journal of the Royal Statistical Society. Series B
  (Methodological)}, pages 45--97.

\bibitem[Duan et~al., 2007]{duan2007multi-model}
Duan, Q., Ajami, N.~K., Gao, X., and Sorooshian, S. (2007).
\newblock Multi-model ensemble hydrologic prediction using {Bayesian} model
  averaging.
\newblock {\em Advances in Water Resources}, 30(5):1371--1386.

\bibitem[Duraisamy, 2021a]{duraisamy2020perspectives}
Duraisamy, K. (2021a).
\newblock Perspectives on machine learning-augmented {R}eynolds-{A}veraged and
  {L}arge {E}ddy {S}imulation models of turbulence.
\newblock {\em Physical Review Fluids}, 6:050504.

\bibitem[Duraisamy, 2021b]{duraisamy2021perspectives}
Duraisamy, K. (2021b).
\newblock Perspectives on machine learning-augmented reynolds-averaged and
  large eddy simulation models of turbulence.
\newblock {\em Physical Review Fluids}, 6(5):050504.

\bibitem[Duraisamy et~al., 2019]{duraisamy2019turbulence}
Duraisamy, K., Iaccarino, G., and Xiao, H. (2019).
\newblock Turbulence modeling in the age of data.
\newblock {\em Annual Review of Fluid Mechanics}, 51.
\newblock doi: 10.1146/annurev-fluid-010518-040547.

\bibitem[Durbin, 1991]{durbin1991}
Durbin, P. (1991).
\newblock Near-wall turbulence closure modelling without damping functions.
\newblock {\em Theoretical and Computational Fluid Dynamics}, 3:1--13.

\bibitem[{Durbin}, 2018]{durbin2018some}
{Durbin}, P.~A. (2018).
\newblock Some recent developments in turbulence closure modeling.
\newblock {\em Annual Review of Fluid Mechanics}, 50(50):77--103.

\bibitem[Edeling et~al., 2014a]{edeling2014predictive}
Edeling, W., Cinnella, P., and Dwight, R.~P. (2014a).
\newblock Predictive {RANS} simulations via {Bayesian} model-scenario
  averaging.
\newblock {\em Journal of Computational Physics}, 275:65--91.

\bibitem[Edeling et~al., 2014b]{edeling2014bayesian}
Edeling, W.~N., Cinnella, P., Dwight, R.~P., and Bijl, H. (2014b).
\newblock Bayesian estimates of parameter variability in the $k$--$\varepsilon$
  turbulence model.
\newblock {\em Journal of Computational Physics}, 258:73--94.

\bibitem[Edeling et~al., 2018]{edeling2018bayesian}
Edeling, W.~N., Schmelzer, M., Cinnella, P., and Dwight, R.~P. (2018).
\newblock {Bayesian} predictions of {Reynolds}-averaged {Navier}--{Stokes}
  uncertainties using maximum a posteriori estimates.
\newblock {\em {AIAA} Journal}, pages 1--12.

\bibitem[Emory et~al., 2013]{emory2013modeling}
Emory, M., Larsson, J., and Iaccarino, G. (2013).
\newblock Modeling of structural uncertainties in {Reynolds}-averaged
  {Navier}-{Stokes} closures.
\newblock {\em Physics of Fluids}, 25(11):110822.

\bibitem[Emory et~al., 2011]{emory2011modeling}
Emory, M., Pecnik, R., and Iaccarino, G. (2011).
\newblock Modeling structural uncertainties in {Reynolds}-averaged computations
  of shock/boundary layer interactions.
\newblock In {\em 49th AIAA Aerospace Sciences Meeting including the New
  Horizons Forum and Aerospace Exposition}.
\newblock Orlando, FL, 2011 (AIAA, Reston, VA, 2017), paper 2011-479.

\bibitem[Evensen, 2003]{evensen2003ensemble}
Evensen, G. (2003).
\newblock The ensemble {Kalman} filter: theoretical formulation and practical
  implementation.
\newblock {\em Ocean Dynamics}, 53(4):343--367.

\bibitem[Evensen, 2009]{evensen2009data}
Evensen, G. (2009).
\newblock {\em Data Assimilation:The Ensemble {Kalman} Filter}.
\newblock Springer.

\bibitem[Fang et~al., 2023]{fang2023toward}
Fang, Y., Zhao, Y., Waschkowski, F., Ooi, A.~S., and Sandberg, R.~D. (2023).
\newblock Toward more general turbulence models via multicase
  computational-fluid-dynamics-driven training.
\newblock {\em AIAA Journal}, 61(5):2100--2115.

\bibitem[Ferreira, 2001]{ferreira2001gep}
Ferreira, C. (2001).
\newblock Gene expression programming: A new adaptive algorithm for solving
  problems.
\newblock {\em Complex Systems}, 13:87--129.

\bibitem[Ferrero et~al., 2020]{ferrero2020field}
Ferrero, A., Iollo, A., and Larocca, F. (2020).
\newblock Field inversion for data-augmented \text{RANS} modelling in
  turbomachinery flows.
\newblock {\em Computers \& Fluids}, 201:104474.

\bibitem[Fr{\"o}hlich and von Terzi, 2008]{frohlich2008hybrid}
Fr{\"o}hlich, J. and von Terzi, D. (2008).
\newblock Hybrid {LES/RANS} methods for the simulation of turbulent flows.
\newblock {\em Progress in Aerospace Sciences}, 44(5):349--377.

\bibitem[Gatski and Speziale, 1993]{gatski1993on}
Gatski, T. and Speziale, C. (1993).
\newblock On explicit algebraic stress models for complex turbulent flows.
\newblock {\em Journal of Fluid Mechanics}, 254:59--79.

\bibitem[Gerolymos et~al., 2012]{gerolymos2012term-by-term}
Gerolymos, G., Lo, C., Vallet, I., and Younis, B. (2012).
\newblock Term-by-term analysis of near-wall second-moment closures.
\newblock {\em AIAA journal}, 50(12):2848--2864.

\bibitem[Girimaji, 2006]{girimaji2006partially-averaged}
Girimaji, S.~S. (2006).
\newblock Partially-averaged {N}avier-{S}tokes model for turbulence: A
  {R}eynolds-averaged {N}avier--{S}tokes to direct numerical simulation
  bridging method.
\newblock {\em Journal of Applied Mechanics}, 73(3):413--421.

\bibitem[Gorl\'e et~al., 2015]{gorle2015quantifying}
Gorl\'e, C., Garcia-Sanchez, C., and Iaccarino, G. (2015).
\newblock Quantifying inflow and {RANS} turbulence model form uncertainties for
  wind engineering flows.
\newblock {\em Journal of Wind Engineering and Industrial Aerodynamics},
  144:202--212.

\bibitem[Gorl{\'e} and Iaccarino, 2013]{gorle2013framework}
Gorl{\'e}, C. and Iaccarino, G. (2013).
\newblock A framework for epistemic uncertainty quantification of turbulent
  scalar flux models for {Reynolds}-averaged {Navier}-{Stokes} simulations.
\newblock {\em Physics of Fluids}, 25(5):055105.

\bibitem[Hastings, 1970]{hastings1970MCMC}
Hastings, W.~K. (1970).
\newblock Monte carlo sampling methods using markov chains and their
  applications.
\newblock {\em Biometrika}, 57(1):97--109.

\bibitem[He et~al., 2022]{he2022explainability}
He, X., Tan, J., Rigas, G., and Vahdati, M. (2022).
\newblock On the explainability of machine-learning-assisted turbulence
  modeling for transonic flows.
\newblock {\em International Journal of Heat and Fluid Flow}, 97:109038.

\bibitem[Hoeting et~al., 1999]{hoeting1999bayesian}
Hoeting, J.~A., Madigan, D., Raftery, A.~E., and Volinsky, C.~T. (1999).
\newblock Bayesian model averaging: a tutorial.
\newblock {\em Statistical science}, 14(4):382--401.

\bibitem[Holland et~al., 2019a]{holland2019field}
Holland, J.~R., Baeder, J.~D., and Duraisamy, K. (2019a).
\newblock Field inversion and machine learning with embedded neural networks:
  Physics-consistent neural network training.
\newblock In {\em AIAA Aviation 2019 Forum}, page 3200.

\bibitem[Holland et~al., 2019b]{holland2019towards}
Holland, J.~R., Baeder, J.~D., and Duraisamy, K. (2019b).
\newblock Towards integrated field inversion and machine learning with embedded
  neural networks for rans modeling.
\newblock In {\em AIAA Scitech 2019 forum}, page 1884.

\bibitem[Iglesias et~al., 2013]{iglesias2013ensemble-kalman}
Iglesias, M.~A., Law, K. J.~H., and Stuart, A.~M. (2013).
\newblock {Ensemble Kalman} methods for inverse problems.
\newblock {\em Inverse Problems}, 29(4):045001 (20pp).

\bibitem[James et~al., 2013]{james2013introduction}
James, G., Witten, D., Hastie, T., and Tibshirani, R. (2013).
\newblock {\em An introduction to statistical learning}, volume 112.
\newblock Springer.

\bibitem[Jiang et~al., 2021]{jiang2021interpretable}
Jiang, C., Vinuesa, R., Chen, R., Mi, J., Laima, S., and Li, H. (2021).
\newblock An interpretable framework of data-driven turbulence modeling using
  deep neural networks.
\newblock {\em Physics of Fluids}, 33(5):055133.

\bibitem[Kaandorp and Dwight, 2020]{kaandorp2020data}
Kaandorp, M.~L. and Dwight, R.~P. (2020).
\newblock Data-driven modelling of the reynolds stress tensor using random
  forests with invariance.
\newblock {\em Computers \& Fluids}, 202:104497.

\bibitem[Kato and Obayashi, 2013]{kato2013approach}
Kato, H. and Obayashi, S. (2013).
\newblock Approach for uncertainty of turbulence modeling based on data
  assimilation technique.
\newblock {\em Computers \& Fluids}, 85:2--7.

\bibitem[Kawai and Larsson, 2012]{kawai2012wall}
Kawai, S. and Larsson, J. (2012).
\newblock Wall-modeling in large eddy simulation: Length scales, grid
  resolution, and accuracy.
\newblock {\em Phys. Fluids}, 24(1):015105.

\bibitem[Kennedy and O'Hagan, 2001]{kennedy2001bayesian}
Kennedy, M.~C. and O'Hagan, A. (2001).
\newblock Bayesian calibration of computer models.
\newblock {\em Journal of the Royal Statistical Society: Series B (Statistical
  Methodology)}, 63(3):425--464.

\bibitem[Kline et~al., 1969]{kline1969computation}
Kline, S.~J., Coles, D.~E., and Hirst, E. (1969).
\newblock {\em Computation of turbulent boundary layers -- 1968
  {AFOSR-IFP-Stanford} Conference: proceedings held at Stanford University,
  August 18-25, 1968}.
\newblock Thermosciences Division, Stanford University.

\bibitem[Laval and Marquillie, 2011]{laval2011direct}
Laval, J.-P. and Marquillie, M. (2011).
\newblock Direct numerical simulations of converging--diverging channel flow.
\newblock In {\em Progress in wall turbulence: understanding and modeling},
  pages 203--209. Springer.

\bibitem[Lee and You, 2019]{lee2019data}
Lee, S. and You, D. (2019).
\newblock Data-driven prediction of unsteady flow over a circular cylinder
  using deep learning.
\newblock {\em Journal of Fluid Mechanics}, 879:217--254.

\bibitem[Ling et~al., 2016a]{ling2016machine}
Ling, J., Jones, R., and Templeton, J. (2016a).
\newblock Machine learning strategies for systems with invariance properties.
\newblock {\em Journal of Computational Physics}, 318:22--35.

\bibitem[Ling et~al., 2016b]{ling2016reynolds}
Ling, J., Kurzawski, A., and Templeton, J. (2016b).
\newblock Reynolds averaged turbulence modelling using deep neural networks
  with embedded invariance.
\newblock {\em Journal of Fluid Mechanics}, 807:155--166.

\bibitem[Ling and Templeton, 2015]{ling2015evaluation}
Ling, J. and Templeton, J. (2015).
\newblock Evaluation of machine learning algorithms for prediction of regions
  of high {Reynolds averaged Navier Stokes} uncertainty.
\newblock {\em Physics of Fluids (1994-present)}, 27(8):085103.

\bibitem[Liu et~al., 2017]{liu2017quantification}
Liu, D., Litvinenko, A., Schillings, C., and Schulz, V. (2017).
\newblock Quantification of airfoil geometry-induced aerodynamic
  uncertainties---comparison of approaches.
\newblock {\em SIAM/ASA Journal on Uncertainty Quantification}, 5(1):334--352.

\bibitem[Liu et~al., 2021]{liu2021iterative}
Liu, W., Fang, J., Rolfo, S., Moulinec, C., and Emerson, D.~R. (2021).
\newblock An iterative machine-learning framework for rans turbulence modeling.
\newblock {\em International Journal of Heat and Fluid Flow}, 90:108822.

\bibitem[Lumley, 1978]{lumley1978computational}
Lumley, J.~L. (1978).
\newblock Computational modeling of turbulent flows.
\newblock {\em Advances in applied mechanics}, 18(123):213.

\bibitem[Mandler and Weigand, 2022]{mandler2022realizable}
Mandler, H. and Weigand, B. (2022).
\newblock A realizable and scale-consistent data-driven non-linear eddy
  viscosity modeling framework for arbitrary regression algorithms.
\newblock {\em International Journal of Heat and Fluid Flow}, 97:109018.

\bibitem[Margheri et~al., 2014]{margheri2014epistemic}
Margheri, L., Meldi, M., Salvetti, M., and Sagaut, P. (2014).
\newblock Epistemic uncertainties in {RANS} model free coefficients.
\newblock {\em Computers \& Fluids}, 102:315--335.

\bibitem[Mariotti et~al., 2016]{mariotti2016freestream}
Mariotti, A., Salvetti, M., Omrani, S., and Witteween, J. (2016).
\newblock Stochastic analysis of the impact of freestream conditions on the
  aerodynamics of a rectangular 5:1 cylinder.
\newblock {\em Computers and Fluids}, 136:170--192.

\bibitem[McConaghy, 2011]{McConaghy2011fast}
McConaghy, T. (2011).
\newblock {\em FFX: Fast, Scalable, Deterministic Symbolic Regression
  Technology}, pages 235--260.
\newblock Springer New York, New York, NY.

\bibitem[Menter, 1992]{menter1992improved}
Menter, F.~R. (1992).
\newblock Improved two-equation k-$\omega$ turbulence models for aerodynamic
  flows.
\newblock {\em Nasa Sti/recon Technical Report N}, 93:22809.

\bibitem[{Mons} et~al., 2016]{mons2016reconstruction}
{Mons}, V., {Chassaing}, J.-C., {Gomez}, T., and {Sagaut}, P. (2016).
\newblock Reconstruction of unsteady viscous flows using data assimilation
  schemes.
\newblock {\em Journal of Computational Physics}, 316:255--280.

\bibitem[Moser et~al., 1999]{moser1999direct}
Moser, R.~D., Kim, J., and Mansour, N.~N. (1999).
\newblock Direct numerical simulation of turbulent channel flow up to re
  $\tau$= 590.
\newblock {\em Physics of fluids}, 11(4):943--945.

\bibitem[Oliver and Moser, 2011]{oliver2011bayesian}
Oliver, T.~A. and Moser, R.~D. (2011).
\newblock Bayesian uncertainty quantification applied to {RANS} turbulence
  models.
\newblock {\em Journal of Physics: Conference Series}, 318:042032.

\bibitem[Parish and Duraisamy, 2016]{parish2016paradigm}
Parish, E.~J. and Duraisamy, K. (2016).
\newblock A paradigm for data-driven predictive modeling using field inversion
  and machine learning.
\newblock {\em Journal of Computational Physics}, 305:758--774.

\bibitem[Parussini and Pediroda, 2007]{parussini2007fictitious}
Parussini, L. and Pediroda, V. (2007).
\newblock Fictitious domain with least-squares spectral element method to
  explore geometric uncertainties by non-intrusive polynomial chaos method.
\newblock {\em Computer modeling in engineering and science}, 22(1):41--63.

\bibitem[Perot and Moin, 1996]{perot1996new}
Perot, B. and Moin, P. (1996).
\newblock A new approach to turbulence modeling.
\newblock Technical report, Proceedings of Summer Research Program, Center of
  Turbulence Research, Stanford University, Stanford, CA, USA.

\bibitem[Piomelli and Balaras, 2002]{piomelli2002wall}
Piomelli, U. and Balaras, E. (2002).
\newblock Wall-layer models for large-eddy simulations.
\newblock {\em Ann. Rev. Fluid Mech.}, 34(1):349--374.

\bibitem[Platteeuw et~al., 2008]{platteeuw2008uncertainty}
Platteeuw, P. D.~A., Loeven, G. J.~A., and Bijl, H. (2008).
\newblock Uncertainty quantification applied to the $k$--$\epsilon$ model of
  turbulence using the probabilistic collocation method.
\newblock In {\em 10th AIAA Non-Deterministic Approaches Conference}.
\newblock Paper no.: 2008-2150.

\bibitem[Pope, 1975]{pope1975more}
Pope, S. (1975).
\newblock A more general effective-viscosity hypothesis.
\newblock {\em Journal of Fluid Mechanics}, 72(2):331--340.

\bibitem[Pope, 2000]{pope2000turbulent}
Pope, S.~B. (2000).
\newblock {\em Turbulent Flows}.
\newblock Cambridge University Press, Cambridge.

\bibitem[Poroseva et~al., 2006]{poroseva2006improving}
Poroseva, S.~V., Hussaini, M.~Y., and Woodruff, S.~L. (2006).
\newblock Improving the predictive capability of turbulence models using
  evidence theory.
\newblock {\em AIAA Journal}, 44(6):1220--1228.

\bibitem[Prudencio and Cheung, 2012]{prudencio2012parallel}
Prudencio, E. and Cheung, S.~H. (2012).
\newblock Parallel adaptive multilevel sampling algorithms for the bayesian
  analysis of mathematical models.
\newblock {\em International Journal for Uncertainty Quantification},
  2(3):215--237.

\bibitem[Ray et~al., 2018]{ray2018robust}
Ray, J., Dechant, L., Lefantzi, S., Ling, J., and Arunajatesan, S. (2018).
\newblock Robust {B}ayesian calibration of k--$\varepsilon$ model for
  compressible jet-in-crossflow simulations.
\newblock {\em AIAA Journal}.
\newblock In press.

\bibitem[Rinc{\'o}n et~al., 2023]{rincon2023progressive}
Rinc{\'o}n, M.~J., Amarloo, A., Reclari, M., Yang, X.~I., and Abkar, M. (2023).
\newblock Progressive augmentation of reynolds stress tensor models for
  secondary flow prediction by computational fluid dynamics driven surrogate
  optimisation.
\newblock {\em International Journal of Heat and Fluid Flow}, 104:109242.

\bibitem[{Rodi}, 1976]{rodi1976algebraic}
{Rodi}, W. (1976).
\newblock {A new algebraic relation for calculating the Reynolds stresses}.
\newblock {\em Gesellschaft Angewandte Mathematik und Mechanik Workshop Paris
  France}, 56:219.

\bibitem[Rumsey et~al., 2022]{rumsey2022search}
Rumsey, C.~L., Coleman, G.~N., and Wang, L. (2022).
\newblock In search of data-driven improvements to rans models applied to
  separated flows.
\newblock In {\em AIAA Scitech 2022 Forum}, page 0937.

\bibitem[Sagaut, 2006]{sagaut2006large}
Sagaut, P. (2006).
\newblock {\em Large Eddy Simulations for Incompressible Flows: An
  Introduction}.
\newblock Springer.

\bibitem[Sagaut et~al., 2013]{sagaut2006multiscale}
Sagaut, P., Deck, S., and Terracol, M. (2013).
\newblock {\em Multiscale and Multiresolution Approaches in Turbulence}.
\newblock Imperial College Press, second edition.

\bibitem[Sa{\"\i}di et~al., 2022]{saidi2022}
Sa{\"\i}di, I. B.~H., Schmelzer, M., Cinnella, P., and Grasso, F. (2022).
\newblock {CFD}-driven symbolic identification of algebraic reynolds-stress
  models.
\newblock {\em Journal of Computational Physics}, 457:111037.

\bibitem[Sandberg and Zhao, 2022]{sandberg2022review}
Sandberg, R.~D. and Zhao, Y. (2022).
\newblock Machine-learning for turbulence and heat-flux model development: A
  review of challenges associated with distinct physical phenomena and progress
  to date.
\newblock {\em International Journal of Heat and Fluid Flow}, 95:108983.

\bibitem[Schlatter et~al., 2011]{schlatter2011progress}
Schlatter, P., Orlu, R., Li, Q., Brethouwer, G., Johansson, A.~V., Alfredsson,
  P.~H., and Henningson, D.~S. (2011).
\newblock Progress in simulations of turbulent boundary layers.
\newblock In {\em Seventh International Symposium on Turbulence and Shear Flow
  Phenomena}. Begel House Inc.

\bibitem[Schmelzer et~al., 2020]{schmelzer2020discovery}
Schmelzer, M., Dwight, R.~P., and Cinnella, P. (2020).
\newblock Discovery of algebraic \text{Reynolds-stress} models using sparse
  symbolic regression.
\newblock {\em Flow, Turbulence and Combustion}, 104(2):579--603.

\bibitem[Schmitt, 2007]{schmitt2007boussinesq}
Schmitt, F.~G. (2007).
\newblock About {B}oussinesq's turbulent viscosity hypothesis: historical
  remarks and a direct evaluation of its validity.
\newblock {\em Comptes Rendus M{é}canique}, 335:617--627.

\bibitem[Simonsen and Krogstad, 2005]{simonsen2005turbulent}
Simonsen, A. and Krogstad, P.-{\AA}. (2005).
\newblock Turbulent stress invariant analysis: {Clarification} of existing
  terminology.
\newblock {\em Physics of Fluids}, 17(8):088103.

\bibitem[Singh and Duraisamy, 2016]{singh2016using}
Singh, A.~P. and Duraisamy, K. (2016).
\newblock Using field inversion to quantify functional errors in turbulence
  closures.
\newblock {\em Physics of Fluids}, 28:045110.

\bibitem[Singh et~al., 2017]{singh2017machine-learning-augmented}
Singh, A.~P., Medida, S., and Duraisamy, K. (2017).
\newblock Machine-learning-augmented predictive modeling of turbulent separated
  flows over airfoils.
\newblock {\em AIAA Journal}, 55(7):2215--2227.

\bibitem[Spalart, 2000]{spalart2000}
Spalart, P. (2000).
\newblock Strategies for turbulence modelling and simulations.
\newblock {\em International Journal of Heat and Fluid Flow}, 21:252--263.

\bibitem[Spalart, 2023]{spalart2023oldfashioned}
Spalart, P. (2023).
\newblock An old-fashioned framework for machine learning in turbulence
  modeling.

\bibitem[Spalart and Shur, 1997]{spalart_shur1997}
Spalart, P. and Shur, M. (1997).
\newblock On the sensitization of simple turbulence models to rotation and
  curvature.
\newblock {\em Aerospace Science and Technology}, 1:297--302.

\bibitem[Spalart, 2009]{spalart2009detached-eddy}
Spalart, P.~R. (2009).
\newblock Detached-eddy simulation.
\newblock {\em Annual Review of Fluid Mechanics}, 41:181--202.

\bibitem[Spalart, 2015]{spalart2015philosophies}
Spalart, P.~R. (2015).
\newblock Philosophies and fallacies in turbulence modeling.
\newblock {\em Progress in Aerospace Sciences}, 74:1--15.

\bibitem[Spalart and Allmaras, 1992]{spalart1992one-equation}
Spalart, P.~R. and Allmaras, S.~R. (1992).
\newblock A one-equation turbulence model for aerodynamic flows.
\newblock AIAA Paper 1992-439.

\bibitem[Speziale, 1995]{speziale1995}
Speziale, C. (1995).
\newblock A review of {R}eynolds {S}tress models for {T}urbulent shear flows.
\newblock Technical Report NASA-CR-195054, NASA.

\bibitem[Speziale, 1987]{speziale1987on}
Speziale, C.~G. (1987).
\newblock On nonlinear $k$-$l$ and $k$-$\varepsilon$ models of turbulence.
\newblock {\em Journal of Fluid Mechanics}, 178:459--475.

\bibitem[St{\"o}cker et~al., 2023]{stocker2023dns}
St{\"o}cker, Y., Golla, C., Jain, R., Fr{\"o}hlich, J., and Cinnella, P.
  (2023).
\newblock Dns-based turbulent closures for sediment transport using symbolic
  regression.
\newblock {\em Flow, Turbulence and Combustion}, pages 1--25.

\bibitem[Taghizadeh et~al., 2020]{taghizadeh2020turbulence}
Taghizadeh, S., Witherden, F.~D., and Girimaji, S.~S. (2020).
\newblock Turbulence closure modeling with data-driven techniques: physical
  compatibility and consistency considerations.
\newblock {\em New Journal of Physics}, 22(9):093023.

\bibitem[Tebaldi and Knutti, 2007]{tebaldi2007use}
Tebaldi, C. and Knutti, R. (2007).
\newblock The use of the multi-model ensemble in probabilistic climate
  projections.
\newblock {\em Philosophical Transactions of the Royal Society of London A:
  Mathematical, Physical and Engineering Sciences}, 365(1857):2053--2075.

\bibitem[Thompson et~al., 2019]{thompson2019eigenvector}
Thompson, R.~L., Mishra, A.~A., Iaccarino, G., Edeling, W., and Sampaio, L.
  (2019).
\newblock Eigenvector perturbation methodology for uncertainty quantification
  of turbulence models.
\newblock {\em Physical Review Fluids}, 4(4):044603.

\bibitem[Tibshirani, 1996]{tibshirani1996regression}
Tibshirani, R. (1996).
\newblock Regression shrinkage and selection via the lasso.
\newblock {\em Journal of the Royal Statistical Society. Series B
  (Methodological)}, 58(1):267--288.

\bibitem[Tinoco et~al., 2018]{tinoco2018summary}
Tinoco, E.~N., Brodersen, O.~P., Keye, S., Laflin, K.~R., Feltrop, E.,
  Vassberg, J.~C., Mani, M., Rider, B., Wahls, R.~A., Morrison, J.~H., Hue, D.,
  Roy, C.~J., Mavriplis, D.~J., and Murayama, M. (2018).
\newblock Summary data from the sixth {AIAA} {CFD} drag prediction workshop:
  {CRM} cases.
\newblock {\em Journal of Aircraft}, 55(4):1352--1379.

\bibitem[Tipping, 2001]{tipping2001sparse}
Tipping, M.~E. (2001).
\newblock Sparse \text{Bayesian} learning and the relevance vector machine.
\newblock {\em Journal of {M}achine {L}earning research}, 1:211--244.

\bibitem[Tracey et~al., 2013]{tracey2013application}
Tracey, B., Duraisamy, K., and Alonso, J. (2013).
\newblock Application of supervised learning to quantify uncertainties in
  turbulence and combustion modeling.
\newblock In {\em 51st AIAA Aerospace Sciences Meeting}.
\newblock {Dallas}, TX, paper 2013-0259.

\bibitem[Virgolin and Pissis, 2022]{virgolin2022symbolic}
Virgolin, M. and Pissis, S.~P. (2022).
\newblock Symbolic regression is {NP}-hard.
\newblock {\em Transactions on Machine Learning Research}.

\bibitem[Volpiani et~al., 2022]{volpiani2022neuralIJHFF}
Volpiani, P.~S., Bernardini, R.~F., and Franceschini, L. (2022).
\newblock Neural network-based eddy-viscosity correction for rans simulations
  of flows over bi-dimensional bumps.
\newblock {\em International Journal of Heat and Fluid Flow}, 97:109034.

\bibitem[Volpiani et~al., 2021]{volpiani2021machinePRF}
Volpiani, P.~S., Meyer, M., Franceschini, L., Dandois, J., Renac, F., Martin,
  E., Marquet, O., and Sipp, D. (2021).
\newblock Machine learning-augmented turbulence modeling for rans simulations
  of massively separated flows.
\newblock {\em Physical Review Fluids}, 6(6):064607.

\bibitem[Wallin and Johansson, 2000]{wallin2000explicit}
Wallin, S. and Johansson, A.~V. (2000).
\newblock An explicit algebraic {R}eynolds stress model for incompressible and
  compressible turbulent flows.
\newblock {\em Journal of Fluid Mechanics}, 403:89--132.

\bibitem[Wang et~al., 2017]{wang2017physics-informed}
Wang, J.-X., Wu, J.-L., and Xiao, H. (2017).
\newblock Physics-informed machine learning approach for reconstructing
  {Reynolds} stress modeling discrepancies based on {DNS} data.
\newblock {\em Physical Review Fluids}, 2(3):034603.

\bibitem[Weatheritt and Sandberg, 2016]{weatheritt2016novel}
Weatheritt, J. and Sandberg, R. (2016).
\newblock A novel evolutionary algorithm applied to algebraic modifications of
  the {RANS} stress--strain relationship.
\newblock {\em Journal of Computational Physics}, 325:22--37.

\bibitem[Weatheritt and Sandberg, 2017]{weatheritt2017development}
Weatheritt, J. and Sandberg, R.~D. (2017).
\newblock The development of algebraic stress models using a novel evolutionary
  algorithm.
\newblock {\em International Journal of Heat and Fluid Flow}, 68:298--318.

\bibitem[Wilcox, 2006]{wilcox2006turbulence}
Wilcox, D.~C. (2006).
\newblock {\em Turbulence modeling for CFD}.
\newblock DCW Industries, 3 edition.

\bibitem[Wu et~al., 2018a]{wu2018pde-informed}
Wu, J.-L., Str{\"o}fer, C.~M., and Xiao, H. (2018a).
\newblock {PDE}-informed construction of covariance kernel in uncertainty
  quantification of random fields.
\newblock Manuscript in preparation. Bibliography details to be available while
  this article is under review.

\bibitem[Wu et~al., 2016]{wu2016bayesian}
Wu, J.-L., Wang, J.-X., and Xiao, H. (2016).
\newblock A {Bayesian} calibration--prediction method for reducing model-form
  uncertainties with application in {RANS} simulations.
\newblock {\em Flow, Turbulence and Combustion}, 97(3):761--786.

\bibitem[Wu et~al., 2018b]{wu2018data-driven}
Wu, J.-L., Xiao, H., and Paterson, E. (2018b).
\newblock Physics-informed machine learning approach for augmenting turbulence
  models: A comprehensive framework.
\newblock {\em Physical Review Fluids}, 3(7):074602.

\bibitem[Wu et~al., 2018c]{wu2018rans}
Wu, J.-L., Xiao, H., Sun, R., and Wang, Q. (2018c).
\newblock {RANS} equations with {Reynolds} stress closure can be
  ill-conditioned.
\newblock Submitted. arXiv preprint arXiv:1803.05581.

\bibitem[Xiao and Cinnella, 2019]{xiao2019review}
Xiao, H. and Cinnella, P. (2019).
\newblock Quantification of model uncertainty in {RANS} simulations: A review.
\newblock {\em Progress in Aerospace Sciences}, 108:1--31.

\bibitem[Xiao and Jenny, 2012]{xiao2012consistent}
Xiao, H. and Jenny, P. (2012).
\newblock A consistent dual-mesh framework for hybrid {LES/RANS} modeling.
\newblock {\em Journal of Computational Physics}, 231(4):1848--1865.

\bibitem[Xiao et~al., 2016]{xiao2016quantifying}
Xiao, H., Wu, J.-L., Wang, J.-X., Sun, R., and Roy, C. (2016).
\newblock Quantifying and reducing model-form uncertainties in
  {Reynolds-averaged} {Navier--Stokes} simulations: A data-driven,
  physics-informed {Bayesian} approach.
\newblock {\em Journal of Computational Physics}, 324:115--136.

\bibitem[Yang et~al., 2015]{yang2015integral}
Yang, X. I.~A., Sadique, J., Mittal, R., and Meneveau, C. (2015).
\newblock Integral wall model for large eddy simulations of wall-bounded
  turbulent flows.
\newblock {\em Phys. Fluids}, 27(2):025112.

\bibitem[Yuksel et~al., 2012]{yuksel_twenty_2012}
Yuksel, S.~E., Wilson, J.~N., and Gader, P.~D. (2012).
\newblock Twenty {Years} of {Mixture} of {Experts}.
\newblock {\em IEEE transactions on neural networks and learning systems},
  23(8):1177--1193.

\bibitem[Zhao et~al., 2020]{zhao2020rans}
Zhao, Y., Akolekar, H.~D., Weatheritt, J., Michelassi, V., and Sandberg, R.~D.
  (2020).
\newblock \text{RANS} turbulence model development using \text{CFD}-driven
  machine learning.
\newblock {\em Journal of Computational Physics}, 411:109413.

\bibitem[Zhong et~al., 2017]{zhong2017gep}
Zhong, J., Feng, L., and Ong, Y.-S. (2017).
\newblock Gene expression programming: A survey.
\newblock {\em IEEE Computational Intelligence Magazine}, 12(3):54--72.

\bibitem[Zou and Hastie, 2005]{zou2005regularization}
Zou, H. and Hastie, T. (2005).
\newblock {Regularization and variable selection via the elastic net}.
\newblock {\em Journal of the royal statistical society: series B (statistical
  methodology)}, 67(2).

\end{thebibliography}
\end{document}